%% file: dissertacao.tex
\pdfoutput=1

\documentclass[
	12pt,					% tamanho da fonte
	openright,				% capítulos começam em pág ímpar (insere página vazia caso preciso)
	twoside,					% para impressão em verso e anverso. Oposto a oneside
	a4paper,					% tamanho do papel. 
	brazil,			% idioma adicional para hifenização
	english,           % o último idioma é o principal do documento
 ,%draft
	]{abntex2}

\usepackage{times}				% Usa a fonte Times New Roman			
\usepackage[T1]{fontenc}			% Selecao de codigos de fonte.
\usepackage[utf8]{inputenc}		% Codificacao do documento (conversão automática dos acentos)
\usepackage{lastpage}			% Usado pela Ficha catalográfica
\usepackage{indentfirst}			% Indenta o primeiro parágrafo de cada seção.
\usepackage{color}				% Controle das cores
\usepackage[table]{xcolor}
\usepackage{graphicx}	% Inclusão de gráficos
\usepackage{epsfig,subfig}		% Inclusão de figuras
\usepackage{microtype} 			% Melhorias de justificação
\usepackage{pdfpages}           %Permite incluir pdf
\usepackage{background}

\usepackage{makeidx} %inserção do index remissivo
\makeindex

\usepackage{lipsum}						% Geração de dummy text
\usepackage{amsmath,amssymb,mathrsfs,arydshln}	% Comandos matemáticos avançados 
\usepackage{setspace}  					% Para permitir espaçamento simples, 1 1/2 e duplo
\usepackage{verbatim}					% Para poder usar o ambiente "comment"
\usepackage{tabularx} 					% Para poder ter tabelas com colunas de largura auto-ajustável
\usepackage{afterpage} 					% Para executar um comando depois do fim da página corrente
\usepackage{url} 						% Para formatar URLs (endereços da Web)
\usepackage{titlesec}
\usepackage{multirow}
\usepackage{makecell}
\usepackage{array}
\usepackage{arydshln}
\usepackage{float}

\usepackage{titlesec}
\titleformat{\chapter}[hang]{\normalfont\huge\bfseries}{\thechapter.}{1em}{}

\usepackage{pdfpages}

\newtheorem{definition}{Definition}[section]

\providecommand{\aver}[1]{\langle #1 \rangle}
\DeclareMathOperator{\diag}{diag}

\DeclareUnicodeCharacter{2212}{-}

\usepackage[a4paper, margin=2.5cm]{geometry}
\usepackage{titlesec}

\include{extras/dados}

\include{extras/conf_pdf}
\usepackage[style=phys, articletitle=true, biblabel=brackets, chaptertitle=true, pageranges=false, eprint=true, date=year]{biblatex} 
\graphicspath{ {./figs/} }

\makeindex
\backgroundsetup{contents={}}

\begin{document}
\selectlanguage{english}
% Configurações de CITAÇÕES para abntex2
%\include{extras/conf_citacoes}

% Retira espaço extra obsoleto entre as frases.
\frenchspacing

% ----------------------------------------------------------
% ELEMENTOS PRÉ-TEXTUAIS (Capa, Resumo, Abstract, etc.)
% ----------------------------------------------------------
\pretextual

% Capa
\include{pretextual/capa}

% Folha de rosto (o * indica que haverá a ficha bibliográfica)
\imprimirfolhaderosto*

% Imprimir Ficha Catalografica
\include{pretextual/catalografica}

% Inserir Folha de Aprovação
\include{pretextual/assinaturas}

% Dedicatória
\include{pretextual/dedicatoria}

% Agradecimentos
\include{pretextual/agradecimentos}

% Epígrafe
\include{pretextual/epigrafe}

% Resumo e Abstract
\include{pretextual/resumos}

% Lista de ilustrações
%\pdfbookmark[0]{\listfigurename}{lof}
%\listoffigures*
%\cleardoublepage

%Lista de tabelas
\pdfbookmark[0]{\listtablename}{lot}
\listoftables*
\cleardoublepage

\begin{siglas}
 \item[CP] Charge-Parity
 \item[C] Conjugation Charge
 \item[P] Parity 
 \item[VLQ] Vector-like-Quarks
 \item[SM] Standard Model
 \item[WBT] Weak basis transformation
 \item[CKM] Cabibbo-Kobayashi-Maskawa
 \item[GSW] Glashow-Salam-Weinberg
 \item[SSB] Spontaneous symmetry breaking
 \item[VEV] Vacuum expectation value
 \item[FCNC] Flavor changing neutral currents
 \item[HS] Hilbert Series 
 \item[PE] Plethystic Exponential
 \item[PL] Plethystic Logarithmic
 \item[LHC] Large Hadron Collider
 \item[SCPV] Spontaneous CP Violation
 \item[GUT] Grand Unification Theory
 \item[BAu] Baryon Asymmetry of the Universe
\end{siglas}

% Lista de símbolos
%\begin{simbolos}
%  \item[$ \Gamma $] Letra grega Gama
%  \item[$ \Lambda $] Lambda
%  \item[$ \zeta $] Letra grega minúscula zeta
%  \item[$ \in $] Pertence
%\end{simbolos}

% Inserir o SUMÁRIO
%\pdfbookmark[0]{\contentsname}{toc}
\tableofcontents*
\cleardoublepage

% ----------------------------------------------------------
% ELEMENTOS TEXTUAIS (Capítulos)
% ----------------------------------------------------------
\textual
% Elementos textuais com numeração arábica
\pagenumbering{arabic}
% Reinicia a contagem do número de páginas
\setcounter{page}{1}

% Inclui cada capitulo da Dissertação
\include{capitulos/1-introducao}

\include{capitulos/2-SM}

\include{capitulos/3-SH}

\include{capitulos/4-VLQS}

\include{capitulos/5-Resultados}

\include{capitulos/6-Conclusion}
% PARTE - Define a divisão do documento em partes (Não é obrigatório)
%\part{Preparação da pesquisa}
%\include{capitulos/referencias}
%\include{capitulos/estadodaarte}
%\include{capitulos/ferramentas}

% PARTE
%\part{Proposta}
%\include{capitulos/proposta}

% PARTE
%\part{Parte Final}
%\include{capitulos/resultados}
%\include{capitulos/conclusao}

% ----------------------------------------------------------
% ELEMENTOS PÓS-TEXTUAIS (Referências, Glossário, Apêndices)
% ----------------------------------------------------------
\postextual

% Referencias bibliograficas
\printbibliography
% Glossario (Consulte o manual)
%\glossary

% Apêndices
\include{postextual/apendices}

% Anexos
 \include{postextual/anexos}

% Índice remissivo (Consultar manual)
%\phantompart
%\printindex

\printindex

\end{document}

%% file: extras/dados.tex
\titulo{Hilbert series for quark flavor invariants and CP violation}
\autor{Eduardo Lourenço Fabio de Lima}
\local{Santo André - SP}
\data{2024}
\orientador{Celso Chikahiro Nishi}
\instituicao{%
  Universidade Federal do ABC -- UFABC
  \par
  Programa de Pós-Graduação em Física}
\tipotrabalho{Dissertação (Mestrado)}
\preambulo{\textbf{Dissertação de Mestrado} realizada sob  orientação do Prof. Dr. Celso Chikahiro Nishi, apresentada ao Programa de Pós-Graduação em Física, como parte dos requisitos necessários para a obtenção do título de Mestre em Física.}

%% file: pretextual/capa.tex
% ---
% Impressão da Capa
% ---
  \begin{capa}%
    \begin{figure}[h!]%
        \centering%
        \includegraphics[scale=1.2]{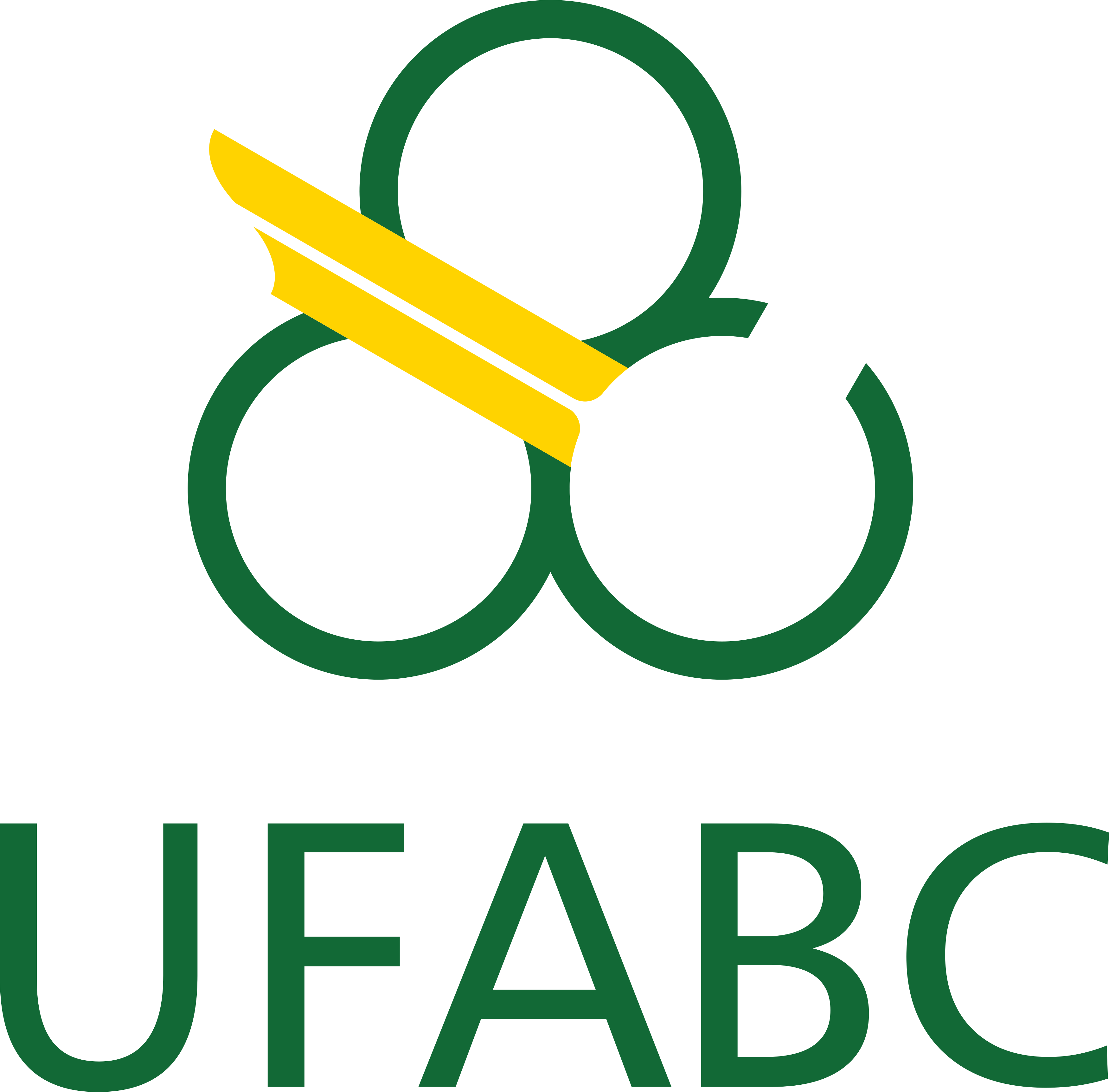}%
      \end{figure}%
    \center
	\large{Universidade Federal do ABC \\ Programa de Pós-Graduação em Física}
	%\vspace{1.5cm}

    \vfill
    \bfseries\LARGE\imprimirtitulo
    \vfill

	%\vfill
	\large\imprimirautor
	\vfill
%
	% Número de Ordem : XXXX
	
    \large\imprimirlocal,\\
    \large\imprimirdata

    \vspace*{1cm}
  \end{capa}
% ---

%% file: pretextual/catalografica.tex
\includepdf[pages=-]{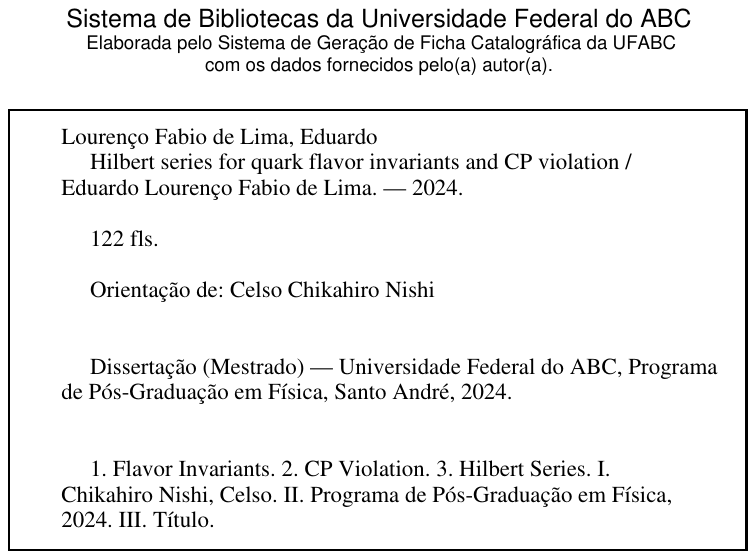}

%% file: pretextual/assinaturas.tex
\includepdf[pages=-]{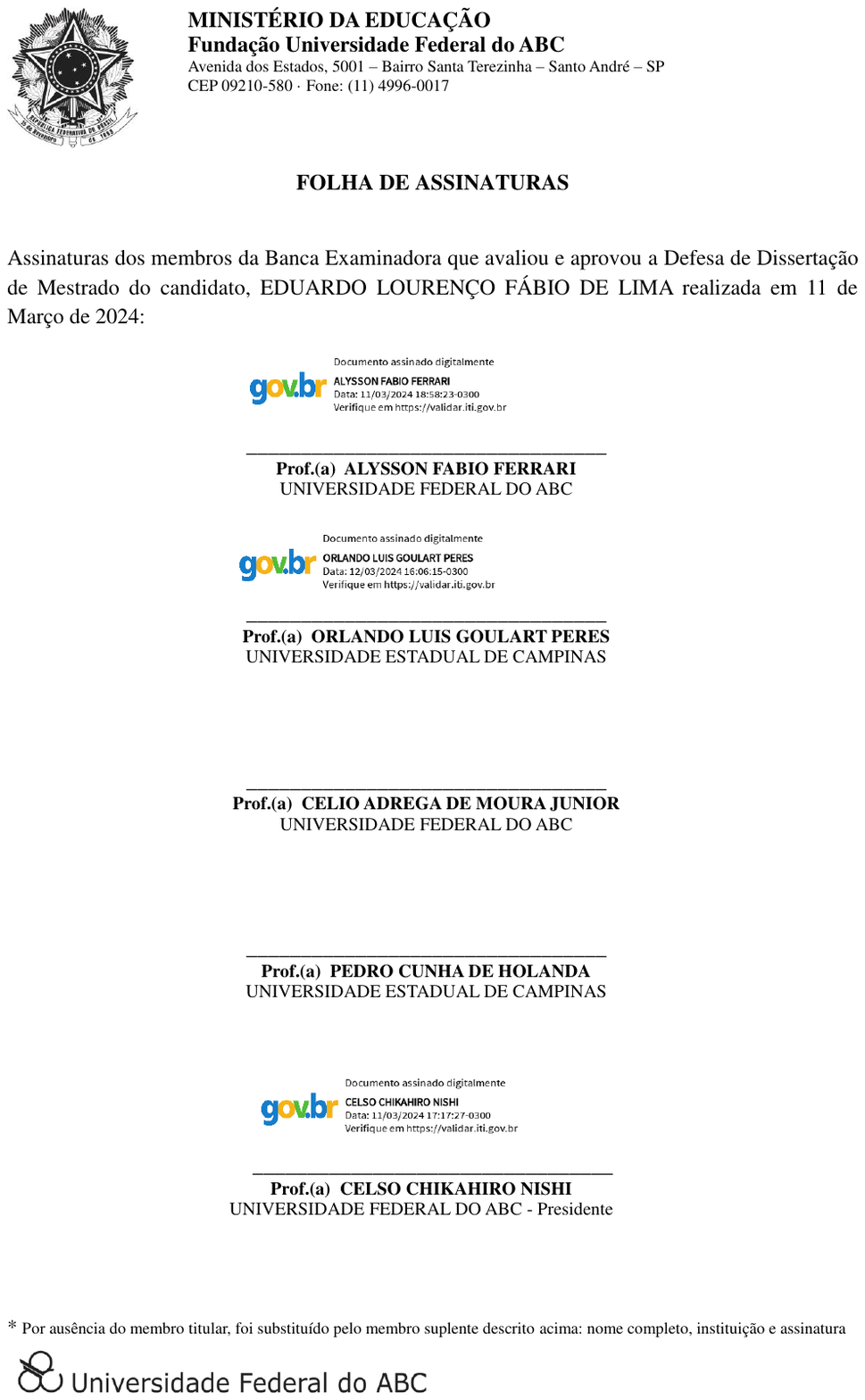}

%% file: pretextual/dedicatoria.tex
% ---
% Dedicatória
% ---
\begin{dedicatoria}
   \vspace*{\fill}
   \centering
   \noindent
   \textit{} \vspace*{\fill}
\end{dedicatoria}
% ---

%% file: pretextual/agradecimentos.tex
% ---
% Agradecimentos
% ---
\begin{agradecimentos}

I would like to express my heartfelt gratitude to my parents, Arnaldo and Benedita, for their unwavering support and dedication to my well-being throughout my journey.

I extend my sincere appreciation to my advisor, Prof. Celso Nishi, for his invaluable guidance, patience, and assistance during this period.

This research was conducted with the financial support of the Coordenação de Aperfeiçoamento de Pessoal de Nível Superior – Brasil (CAPES), under Funding Code 001.

\end{agradecimentos}
%% ---

%% file: pretextual/epigrafe.tex
% ---
% Epígrafe
% ---
\begin{epigrafe}
    \vspace*{\fill}
	\begin{flushright}
		\textit{"Abane a cabeça Leitor"}\\
  Joaquim Maria Machado de Assis
	\end{flushright}
\end{epigrafe}
% ---

%% file: pretextual/resumos.tex
% ---
% RESUMOS
% ---

% RESUMO em português
 \setlength{\absparsep}{18pt} % ajusta o espaçamento dos parágrafos do resumo
 \begin{resumo}[Resumo]

 Este trabalho desenvolve um estudo dos invariantes de sabor e, em especial, dos invariantes capazes de detectar a violação CP (Paridade-Carga). Por meio da ferramenta matemática chamada série de Hilbert, enumeramos e exploramos sistematicamente os invariantes de sabor que permanecem inalterados sob transformações de base fracas. Após revisar as séries de Hilbert e os invariantes de sabor do setor de quarks do Modelo Padrão (SM), aplicamos a série de Hilbert ao SM estendido por um Vector-Like quark (VLQ) do tipo down. A introdução dessas partículas hipotéticas leva a uma extensão simples do SM que pode ser motivada por muitos problemas, incluindo a necessidade de novas fontes de violação CP para explicar a assimetria observada entre matéria e antimatéria no universo. Tivemos sucesso ao calcular as séries de Hilbert para a extensão VLQ na base de massa do VLQ, onde as transformações de spurion são mais simples. Com base na série de Hilbert, construímos e enumeramos os invariantes de sabor básicos com os quais todos os invariantes podem ser construídos. Para uma base genérica, onde as transformações de spurion envolvem um grupo maior, conseguimos obter apenas os primeiros termos das séries de Hilbert.

  \textbf{Palavras-chaves}: Violação de CP, Invariantes de Sabor, Série de Hilbert. 
 \end{resumo}

% ABSTRACT in english
\begin{resumo}[Abstract]
 \begin{otherlanguage*}{english}

This work delves into the study of flavor invariants and, in special, invariants capable of detecting CP (Charge-Parity) violation. Through the mathematical tool of the Hilbert series, we systematically enumerate and explore flavor invariants that are unchanged under weak basis transformations. After reviewing the Hilbert series and the flavor invariants of the SM quark sector, we apply the tool of Hilbert series to the SM extended by a singlet vector-like quark (VLQ) of down-type. The introduction of these hypothetical particles leads to a simple extension of the SM that can be motivated by many problems, including the need for new sources of CP violation to explain the observed matter-antimatter asymmetry in the universe. We were successful in calculating the Hilbert series for the VLQ extension in the mass basis of the VLQ, where the spurion transformations are simpler. Based on the Hilbert series, we build and enumerate the basic flavor invariants with which all invariants can be constructed. For a generic basis, where the spurion transformations involve a larger group, we could only get the first few terms of the Hilbert series.

\vspace{\onelineskip}
 
   \noindent 
   \textbf{\emph{Keywords}}: CP Violation, Flavor Invariants, Hilbert Series. 
 \end{otherlanguage*}
\end{resumo}

%% file: capitulos/1-introducao.tex
% ----------------------------------------------------------
% Introdução 
% Capítulo sem numeração, mas presente no Sumário
% ----------------------------------------------------------

\chapter*[Introduction]{Introduction}
\addcontentsline{toc}{chapter}{Introduction}

The Standard Model (SM) of particle physics is a remarkably successful theory that describes the fundamental particles and their interactions through the electromagnetic, weak, and strong forces, excluding gravity. It encapsulates our understanding of the subatomic world, providing a framework that has been confirmed by numerous experiments over decades \cite{ParticleDataGroup:2022pth}. Despite its triumphs, the SM is not complete. It does not incorporate the gravitational force, it does not account for the dark matter or dark energy, constituting most of the mass-energy content of the universe, and it does not explain the matter-antimatter asymmetry observed in the universe \cite{gaillard_standard_1999, altarelli_standard_2005}.

One of the intriguing aspects of the SM is the phenomenon of CP violation, which refers to the difference in behavior between particles and their antiparticles. CP violation is essential for explaining the matter-antimatter asymmetry in the universe,  yet the explanation provided by the SM, through the complex phase in the Cabibbo-Kobayashi-Maskawa (CKM) matrix, is insufficient to account for the observed imbalance. This inadequacy has led to extensive research into physics beyond the SM, seeking mechanisms that could provide additional sources of CP violation \cite{jarlskog1989cp}.

Invariants play a crucial role in the study of CP violation and flavor physics. They are quantities that remain unchanged under basis transformations in flavor space, such as those involving quark fields in the SM\,\cite{jarlskog_commutator_1985}. The analysis of flavor invariants, especially in the context of extensions to the SM, provides deep insights into the nature of CP violation and the possible existence of new physics. These invariants are tools for exploring the parameter space of extended models in a basis-independent manner, allowing for a clear comparison with experimental data\cite{grossman_just_2018}.

The Hilbert series is a mathematical instrument that facilitates the formulation of 
group invariants. It elucidates the algebraic structure of a set by depicting it as a power series, enabling the quantification of invariants and their corresponding  degrees. By employing the Hilbert series, it becomes possible to construct the Plethystic Logarithm \cite{xiao_standard_2019}. This tool reveals the number of basic invariants, i.e., invariants that remain unchanged after symmetry transformations, and their degrees, but also identifies the  interrelationships of them. 

An application is described in \cite{jenkins_algebraic_2009}, where flavor invariants for quarks and leptons are constructed. Invariants were systematically classified for the quark sector of both two and three generations, and for the seesaw model  incorporating only two generations. In \cite{lehman_hilbert_2015}, it was utilized to calculate the number of operators with a specified mass dimension for a given field content. In particle physics, the series aids in the analysis of flavor physics and consequently in CP violation, providing a systematic approach to identifying and classifying flavor invariants.

% \red{Melhor inverter: diga o que é o serie de Hilbert, e para que serve.
% Mencione primeiro a aplicação no estudo de invariantes de sabor. Cite o Manohar.
% A aplicação na construção de operadores/Lagrangiana, você menciona depois.
% (É muito importante também.)
% }
Vector-like quarks (VLQ) are hypothetical quarks of spin $1/2$, whose left and right-handed components transforms identically under the SM gauge group  $SU(3)_C \times SU(2)_L \times U(1)_Y$. VLQs introduce additional fermionic states beyond those predicted by the SM, providing opportunities to explain phenomena such as flavor mixing, CP violation, and the matter-antimatter asymmetry observed in the universe. These extensions often involve the introduction of new symmetry-breaking mechanisms and coupling structures, which can significantly impact the predictions of the model.  VLQs also play a crucial role in scenarios aiming to unify fundamental forces, as their presence can modify the running of coupling constants and stabilize the Higgs mass \cite{aguilar-saavedra_handbook_2013, dermisek_unification_2013}.

% \red{Ponha uma frase inicial dizendo o que são VLQs, escreva "vector-like quark (VLQ)" pela primeira vez.}

% \red{Acho que pode fazer um parágrafo sobre aplicação de Hilbert series em física de partículas.}

% \red{E outro parágrafo sobre extensões com VLQs.}

% This dissertation delves into the realm of VLQs within the framework of the SM. 
This dissertation delves into the study of quark flavor invariants, aided with the tool of Hilbert series, which we apply to the SM and its extension with a singlet VLQ of down type. Chapter 2 provides a comprehensive review of the SM, focusing on the flavor aspects and CP violation. In this chapter, we introduce the concept of Weak Basis Transformations (WBT). It elaborates on how WBTs are used to simplify the Lagrangian without altering physical observables, facilitating the analysis of CP violation and flavor physics. In Chapter 4, we examine the Hilbert series and its application in identifying and classifying flavor invariants in the SM. This mathematical tool enables the systematic exploration of the algebra of invariants, revealing its underlying structure. In chapter 5, we compute the Hilbert series in the VLQ mass basis. We also construct the basic invariants, aided by the information
from the Hilbert series.
We perform an analysis regarding the basic invariants and categorize them into CP odd and CP even types. The dissertation concludes with an analysis of the invariants found in Ref.\,\cite{albergaria_cp-odd_2023}.

%% file: capitulos/2-SM.tex
 \chapter[The Standard Model and CP Violation]{The Standard Model \\
 and CP Violation}

   In this chapter, we will provide a brief review of the SM and then analyze CP violation through the CKM matrix mechanism, where quark masses, quark mixing, and CP violation share a common origin. We will demonstrate how, in the SM, the appearance of the CP violation phase depends on the chosen basis. Finally, we will analyze the model to identify quantities that are independent of convention and, therefore, more suitable for adoption as observables of the theory. In the case of the SM with three families, there is a single rephasing invariant, the Jarlskog invariant. We will also describe the weak basis invariants.

\section{The Standard Model}

    The SM of particle physics is a quantum field theory in four dimensions, i.e., $3+1$ space-time. It describes the strong, weak and electromagnetic interactions with local gauge symmetry \cite{glashow_partial-symmetries_1961,weinberg_model_1967},
    \begin{equation}
        G_{SM} = SU(3)_C \times SU(2)_L \times U(1)_Y, \label{sym_gauge}
    \end{equation}
    where the indices $C,L$ and $Y$ correspond to  color, left-handed and weak hypercharge, respectively. 

    This group determines interactions with vector bosons corresponding to each group's generators. The group (\ref{sym_gauge}) has $8+3+1 = 12 \quad \text{generators}$. The group $SU(3)$ has eight generators, corresponding to eight massless gluons mediating strong interactions. The group $SU(2) \times U(1)$ has four generators, corresponding  to four gauge bosons, three massive ($W^{\pm}$, $Z^0$) and one massless (the photon $\gamma$), responsible for mediating the electroweak interaction. The groups $SU(3)$, $SU(2)$, $U(1)$ are generated by the Gellmann matrices, the Pauli matrices and one complex phase, respectively. 

    The electroweak sector is commonly referred to as the Glashow–Salam–Weinberg (GSW) model. In SM, the fermions are organized into three families. Each family consists of left-handed (L) fermions, which belong to weak isodoublets, and right-handed (R) fermions, which belong to  weak isosinglets: 
        \begin{equation}
         q_{iL}=
        \begin{pmatrix}
            u_{iL} \\
            d_{iL}
        \end{pmatrix} ;       
        \quad u_{iR} ; \quad d_{iR} ;  \quad
        \ell_i= \begin{aligned} 
        \begin{pmatrix}
            \nu_{iL} \\
            e_{iL}         
        \end{pmatrix}; \quad e_{iR}. \label{representations}
         \end{aligned}
        \end{equation}
    Regarding these conventions, the SM is characterized as a chiral theory, in which the fermion fields are $\psi_{R,L} = \frac{1}{2} (1 \pm \gamma_5)\psi$. Here, the fields $\psi$ represent any of the quarks. There are five types of fields in each generation, transforming under \eqref{sym_gauge} as
    \begin{eqnarray}
    q_L \sim (3,2)_{1/6} \hspace{0.5em} ; \hspace{0.5em} u_R \sim (3,1)_{2/3} \hspace{0.5em} ; \hspace{0.5em} d_R \sim (3,1)_{-1/3} \hspace{0.5em} ; \hspace{0.5em} \ell \sim (1,2)_{-1/2} \hspace{0.5em} ; \hspace{0.5em} e_R \sim (1,1)_{-1}. 
    \end{eqnarray} 
    The numerical notations in these representations indicate their respective transformations under the $SU(3)$ and $SU(2)$ groups. Moreover, the subscript numbers within each representation denote the hypercharge values. The comprehensive SM field content is presented in Table \ref{Fields:SM}.
     
     Spontaneous symmetry breaking (SSB) in the context of equation \eqref{sym_gauge} occurs when the Higgs scalar field acquires the vacuum expectation value (VEV), a phenomenon known as the Higgs mechanism. This leads to the breaking of the electroweak symmetry, denoted as $SU(2)_L \times U(1)_Y \to U(1)_{\text{em}}$ (see Appendix \ref{SSB:A}). It is important to note that the subgroup $U(1)_{\text{em}}$, which has dimension of 1, remains a symmetry of the vacuum. After this symmetry breaking, masses for particles are generated through the Yukawa couplings of fermions to the Higgs field. In the SM, the different fermion generations are distinguished solely by their Yukawa couplings \cite{higgs_spontaneous_1966, hocker_cp_2006}. In the Yukawa sector, the CKM matrix emerges as one 
     of the sources 
     of CP violation, describing the mixing of quark generations \cite{cabibbo_unitary_1963, kobayashi_cp_1973} and will be seen in the following.
   
    % The spontaneous breaking in the symmetry (\ref{sym_gauge}) occurs when the Higgs scalar field acquires the vacuum expectation value (VEV) (Higgs mechanism), such that, $SU(2)_L \times U(1)_Y \to U(1)_{em}$ \footnote{The subgroup $U(1)_em$, with dimension of 1, should continue to be a symmetry of the vacuum}, and this breaking is called  ElectroWeak Spontaneous Breaking (EWSB)  (vide appendix A). The masses are generated after this symmetry breaking, and this is due to the Yukawa couplings of fermions to the Higgs field, and in the SM,the fermion generations are differentiated exclusively by the Yukawa couplings.\cite{higgs_spontaneous_1966, hocker_cp_2006}.  In the Yukawa sector arise CKM (Cabibbo-Kobayashi-Maskawa) matrix, which is one of the sources of CP violation, which describes the mixing of quark generations \cite{cabibbo_unitary_1963, kobayashi_cp_1973} and will be seen following.  
    
    The quark fields, namely $q_L$, $u_R$, and $d_R$, are triplets under $SU(3)_C$, signifying their strong force interactions. Additionally, the left-handed chiral representations, $q_L$ and $l_L$, are doublets under $SU(2)_L$ and interact with the gauge bosons of this group. The electroweak interactions occurs through the charged, neutral and electromagnetic currents,
    \begin{align}
    J^+_\mu &= \Bar{u}_L \gamma_\mu d_L + \Bar{e}_L \gamma_\mu \nu_L ;\\
    J^3_\mu &= \frac{1}{2} (\Bar{u}_L \gamma_\mu u_L + \Bar{d}_L \gamma_\mu d_L + \Bar{\nu}_L \gamma_\mu \nu_L - \Bar{e}_L \gamma_\mu e_L) ;\\
    J^{\text{em}}_\mu &= \frac{2}{3} \Bar{u}_L \gamma_\mu u_L - \frac{1}{3} \Bar{d}_L \gamma_\mu d_L - \Bar{e}_L \gamma_\mu e_L;\\
    J^Z_\mu &= J^3_\mu - \sin^2{\theta_W}\cdot J^{\text{em}}_\mu.  
    \end{align}
    Here, $\theta_W$ is known as the Weinberg angle or the weak mixing angle. It plays a crucial role in determining the relative strengths of the weak and electromagnetic interactions. This angle is pivotal in defining the mixing between these two interactions, which are unified under the electroweak theory. The Weinberg angle is defined as the angle by which the neutral weak force carrier particles (W and Z bosons) must be rotated to match the observed particles (the photon and the Z boson),
    \begin{equation}
    \cos \theta_W = \frac{M_W}{M_Z},
    \end{equation}
    where $M_W$ and $M_Z$ are the masses of the W and Z bosons, respectively.
    
    Therefore, quarks are particles that engage in all interactions, including the strong, weak, electromagnetic, and gravitational forces. In contrast, leptons do not partake in the strong interaction, and among the leptons, neutrinos do not carry electric charges \cite{giunti_fundamentals_2007}. 

     The Lagrangian of the Standard Model can be separated, as
    \begin{equation}
        \mathcal{L}_{SM} = \mathcal{L}_{\text{Kin + gauge}} + \mathcal{L}_{\text{Higgs}} + \mathcal{L}_{\text{Yukawa}}.
    \end{equation}
    The last term refers to the Yukawa Lagrangian, which describes the coupling of the Higgs field with fermions. It also contains all flavor physics within the Yukawa interactions of the quarks \cite{isidori_flavor_2010}
    \begin{equation}
             -\mathcal{L}_{Yukawa} = Y^d_{ij} \Bar{q}_{iL} H d_{jR} + Y^u_{ij}\Bar{q}_{iL}\Tilde{H} u_{jR} + h.c., \label{yukawa} 
    \end{equation}
    where the indices $i$ and $j$ range from 1 to $n=3$, the number of families, $H$ is the Higgs field,  $\Tilde{H} = i \sigma_2 H^*$ and $i \sigma_2$ is the $2 \times 2$ antisymmetric tensor. As mentioned in \eqref{representations}, $\overline{q}$ is a doublet, and the terms $u_L H u_R$,$d_L H u_R$, $u_L H d_R$, $d_L H d_R$ are because of $SU(2)$ symmetry. In general, $Y^f$ ($f= u,d$) are complex $3 \times 3$ matrices in flavor space, and initially contain 36 parameters. In this sector, CP violation occurs due to the fact that Yukawa matrices are arbitrary complex matrices. This violation is characterized by the presence of a complex phase in the CKM matrix, which arises from the diagonalization of the Yukawa matrices for up-type and down-type quarks \cite{bigi_cp_2009}. When the Higgs field $H$ acquires a VEV, $\aver{H} = (0, v/\sqrt{2})$, the equation \eqref{yukawa} results in Dirac mass terms for quarks, expressed as $3 \times 3$ mass matrices.
    \begin{equation}
        M^u = \frac{v Y^u}{\sqrt{2}} ; \quad M^d = \frac{v Y^d}{\sqrt{2}}.
    \end{equation}
    
\begin{table}[!htbp]
    \centering
    \begin{tabular}{|c|c|c|c|c|}
        \hline
         & Names & Fields & $G_{SM}$ & $U(1)_{EM}$ \\
        \cline{2-5}
        \multirow{6}{*}{\makecell[c]{Matter\\(spin 1/2)}} & \multirow{3}{*}{\makecell[c]{Quarks\\}} & $q_L= (u_L \quad d_L)$ & $(3,2)_{1/6}$ & $2/3 \quad -1/3$ \\
        & & $u_R$ & $(3,1)_{+2/3}$ & $+2/3$ \\
        & & $d_R$ & $(3,1)_{-1/3}$ & $-1/3$ \\
        \cline{2-5}
        & \multirow{2}{*}{\makecell[c]{Leptons\\}} &$l_L = (\nu_l \quad e_L)$ & $(1,2)_{-1/2}$ & $0 \quad -1$\\
        &  & $e_R$ & $(1,1)_{-1}$ & $-1$ \\
        \hline
        \multirow{2}{*}{\makecell[c]{Higgs\\(spin 0)}}  & \multirow{2}{*}{\makecell[c]{Higgs\\}} &$H= (H^+ \quad H^0)$ & $(1,2)_{1/2}$ & $1 \quad 0$\\
        &  & & & \\
        \hline
        \multirow{3}{*}{\makecell[c]{Gauge\\(spin 1)}}  & Gluon & $g$ & $(8,1)_0$ & $0$ \\
        & Boson  & $W^{\overset{+}{-}} \quad W^0$ & $(1,3)_0$ & $\overset{+}{-}1 \quad 0$ \\
        & Boson  & $Z^0$ & $(1,1)_0$ & $0$ \\
        \hline 
    \end{tabular} 
    \caption{This table categorizes the Standard Model fields \cite{giunti_fundamentals_2007}.}
    \label{Fields:SM}
\end{table}

\section{Weak Bases} \label{weakbases}

     In the SM, gauge invariance does not impose restrictions on the flavor structure of Yukawa couplings. The Yukawa matrices, $Y^u$ and $Y^d$ are $3 \times 3$ arbitrary complex matrices, containing 36 parameters in the most general case. Thus, one can perform weak basis transformations (WBT):
    \begin{equation}
        q_{iL} = (U_{L})_{ij} q'_{jL}, \quad d_{iR} = (U^d_{R})_{ij} d'_{jR}, \quad u_{iR} = (U^u_{R})_{ij} u'_{jR},
    \label{WB}
    \end{equation}
    in which $U^q_{L},U^u_{R}$, and $U^d_{R}$ are $3 \times 3$ unitary matrices in flavor space, and the indices represent the number of families. Physical properties of the theory are independent of WBT and an appropriate choice allows counting the number of independent physical parameters. This is useful for reducing the number of parameters and expressing them in terms of physical parameters. Additionally, these transformations preserve the electroweak currents invariant. This WBT keeps the entire Lagrangian invariant, except the Yukawa sector, and will be useful in the analysis of CP violation. 

    Applying WBT to the Lagragian \eqref{yukawa}, we obtain 
    \begin{equation}
        -\mathcal{L}_{Yukawa} = \Bar{q'_L} U^{\dagger}_L Y^d U^d_R d'_R H +\Bar{q'_L} U^{\dagger}_L Y^u U^u_R d'_R \tilde{H}, \label{Lyukawa}
    \end{equation}
    where we can identify the transformed Yukawa matrices
    \begin{equation}
      Y'^d = U^{\dagger}_L Y^d U^d_R; \quad Y'^u = U^{\dagger}_L Y^u U^u_R. 
    \label{WBT}
    \end{equation}
    The Yukawa matrices can be decomposed through singular value decomposition as \cite{quigg1985gauge}
   \begin{equation}
       Y^u = W^u_L \hat{Y^u} W^{u \dagger}_R; \quad Y^d = W^d_L \hat{Y^d} W^{d \dagger}_R.
    \end{equation}
    where the matrices $W$ are also unitary matrices and $\hat{Y^u}$, $\hat{Y^d}$ are diagonal matrices. Then, we obtain
    \begin{equation}
         Y'^u = \hat{Y}^{u} \quad Y'^d = W^{u\dagger}_L W^d_L \hat{Y}^d\label{diagonalize}
    \end{equation}
    after choosing
    \begin{equation}
        U_L= W^u_L; \quad U^u_R= W^u_R; \quad U^d_R= W^d_R.
    \end{equation}
    The Lagragian becomes
    \begin{eqnarray}
    -\mathcal{L}_{Yukawa} = \Bar{q'}_{L} H (W^{u\dagger}_L W^d_L) \hat{Y}^d  d'_{R} + \Bar{q'}_{L}\Tilde{H} \hat{Y^u} u'_{R} + h.c.,  \label{Lckm}
    \end{eqnarray}
    where $\hat{Y}^d = \diag(y_d, y_s, y_b)$ and $\hat{Y}^u = \diag(y_u, y_c, y_t)$. In terms of quark masses, we have
    \begin{eqnarray}
       \hat{Y}^u = \frac{\sqrt{2}}{v} \diag(m_u, m_c, m_t); \quad \hat{Y}^d = \frac{\sqrt{2}}{v} \diag(m_d, m_s, m_b).
    \end{eqnarray}  

    From \eqref{Lckm} we can also define the unitary CKM matrix,\footnote{Note that, in the mass eigenstate basis, the neutral-current part of the Lagrangian remains unchanged, meaning that there are no flavor-changing neutral currents (FCNC) at the tree level.} 
    \begin{equation}
       V = \begin{pmatrix}
                V_{ud} & V_{us} & V_{ub} \\
                V_{cd} & V_{cs} & V_{cb}\\
                V_{td} & V_{ts} & V_{tb}        
            \end{pmatrix} = W^{u\dagger}_L W^d_L\,. 
   \end{equation}
   The CKM matrix is a $n \times n$ unitary matrix. Furthermore, the CKM matrix is a quark mixing matrix that incorporates the physical effects of mixing between different generations and contributes significantly to the coupling of the charged current in weak interactions \cite{giunti_fundamentals_2007}. Since $V$ is a unitary matrix, it satisfies the condition $V V^\dagger = I$. Therefore, considering this property and the ability to arbitrarily choose the global phases of the quark fields, the original nine complex elements of $V$ are reduced to three real numbers and one phase. This phase is responsible for CP violation. 
   
   Let us explain the counting of parameters. A general complex $n \times n$ matrix has $2n^2$ real parameters. The unitary relations
   \begin{equation}
       \sum_j V_{ij}V^*_{jk}= \delta_{ik}, \label{unitaryrelations}
   \end{equation}
   lead to $n$ real equations and $n(n-1)/2$ complex equations. Thus, the unitary matrix contains $n^2$ independent real parameters. 
   
   Keeping the same structure, the quark fields can be rotated freely, such that
    \begin{equation}
        q_{iL} \rightarrow e^{i \alpha_i} q_{iL}; \quad u_{iR} \rightarrow e^{i \beta_i} u_{iR}; \quad d_{iR} \rightarrow e^{i \gamma_i} d_{iR} \label{rephasing}
    \end{equation}
    If $\beta_i=\alpha_i$, the second term of \eqref{Lckm} is already invariant while the first term changes in such a way that
        \begin{equation}
            V \rightarrow 
            \begin{pmatrix}
                e^{-i \alpha_1} &   & \\
                & e^{-i \alpha_2} & \\
                & & e^{-i \alpha_3}        
            \end{pmatrix}
        \cdot V \cdot
            \begin{pmatrix}
                e^{i \gamma_1} &  & \\
                & e^{i \gamma_2} & \\
                & & e^{i \gamma_3}        
            \end{pmatrix} \label{repahsing2}
       \end{equation}
    The elements in the two diagonal above are pure phases. Choosing these phases appropriately, we can remove $2n -1$ relative phases from $V$. Then the number of parameters reduce from $2n^2$ to  $(n-1)^2$ independent parameters. As  $V$ contains $(n-1)^2 =4$ parameters, 3 angles and 1 phase. The Lagrangian (\ref{Lckm}) now has 4 parameters that arise from the CKM matrix, and the other 6 from the diagonalized $Y^u$ and $Y^d$ matrices. The flavor sector of the SM in the Lagrangian \eqref{Lyukawa} contains 10 parameters, 3 in $\hat{Y}^u$, 3 in $\hat{Y}^d$ and 4 in $V$.
    
    Considering that a $n \times n$ real orthogonal matrix can be parametrized with $n(n-1)/2$ angles, we are left with $(n-1)^2 - n(n-1)/2 = (n-1)(n-2)/2$ phases. Hence, irremovable phases that lead to CP violation only appear for $n \ge 3$. For $n=3$ we have just one phase.

    \section{CP violation}
    
    % To better understand Charge-Parity (CP) transformations in quantum field theory and their impact on Yukawa interactions, 
    % % (CP-violation), 
    % we need to break down the process step by step. We start with the Dirac spinor \footnote{$\psi_{R,L} \equiv \gamma_{R,L}\psi \quad \gamma_{R,L} \equiv \frac{1 \pm \gamma_5}{2}$}.
    To better understand Charge-Parity (CP) transformations in quantum field theory and their impact on Yukawa interactions, we need to consider the transformations of parity (P) and charge conjugation (C) on the Dirac spinor $\psi$.
    
    \subsection{Parity Transformation (P)} 
    
   Parity transformation (P) essentially inverts spatial coordinates, acting like a “mirror” in physics equations. For a Dirac field $\psi(\mathbf{x}, t)$, the parity transformation is expressed as\,\cite{Branco:1999fs}
    \begin{align}
        \psi(\mathbf{x},t) \to e^{i\beta} \gamma^0 \psi(-\mathbf{x},t), \nonumber \\
        \overline{\psi}(\mathbf{x},t) \to e^{-i\beta}  \overline{\psi}(-\mathbf{x},t) \gamma^0,
    \end{align}
   where $\gamma^0$ is one of the Dirac matrices\footnote{Note that $\gamma^0 \gamma_\mu \gamma^0 = \gamma^{\mu}$, $(\gamma^0)^2= \mathbb{I}$, $(\gamma^i)^2 = -\mathbb{I}$ ~for~ $i=1,2,3$, $\gamma^{0\dagger}= \gamma^0$.} and $\beta$ is an arbitrary phase. 

    \subsection{Charge conjugation (C)}

% \red{Se tiver mais algum "charge transformation", troque por "charge conjugation".}

    Charge conjugation (C) exchanges particles for their antiparticles. For Dirac fields, this transformation is expressed as,
    \begin{align}
        \psi(x,t) \to e^{i\beta} \mathcal{C} \overline{\psi(x,t)}^T, \nonumber \\
        \overline{\psi}(x,t) \to e^{-i\beta}  \overline{\psi(x,t)}^T \mathcal{C},
    \end{align}
    where $\mathcal{C}$ is the charge conjugation matrix and $\overline{\psi}$  denotes the Dirac conjugate of the field $\psi$. 
    % \red{Se não definiu $\bar{\psi}$, defina aqui.}
    % Notably, $\mathcal{C}$ plays a pivotal role in Dirac's field equations.
    The matrix $\mathcal{C}$ obeys the properties
\begin{align}
            \mathcal{C}\gamma^T_\mu \mathcal{C}^{-1} = -\gamma_\mu, \nonumber\\
            \gamma^0 \mathcal{C}= \mathcal{C}^{-1}\gamma^0, \nonumber\\
            \mathcal{C}^{\dagger} = \mathcal{C}^{-1}, \nonumber \\ 
            \mathcal{C}^T = -\mathcal{C}, \nonumber \\
            \mathcal{C} \mathcal{C}^{-1} = \mathbb{I}
        \end{align}
        Furthermore, we have
        \begin{align}
            \mathcal{C}(\gamma^5)^T \mathcal{C}^{-1}= \gamma^5, \\
            \mathcal{C}(\sigma^{\mu \nu})^T \mathcal{C}^{-1}= -\sigma^{\mu \nu}.
        \end{align}
    
\subsection{Combined CP Transformation}

In quantum field theory, the combined Charge-Parity transformation for an arbitrary Dirac field $\psi(x,t)$ is:
    \begin{align}
         \psi(x,t) \to e^{i \beta_{\psi}}\gamma^0 \mathcal{C} \overline{\psi}^T (-x, t). \label{cp:psi}
    \end{align}
    The corresponding transformation for the Dirac conjugate field $\overline{\psi}(x,t)$ is expressed as:
    \begin{align}
       \overline{\psi}(x,t) \to -e^{-i \beta_{\psi}} \psi^T (-x, t) \mathcal{C}^{-1}\gamma^0, 
    \end{align}
    where we first apply the parity transformation and then the charge conjugation. This transformation combines the effects of both parity and charge conjugation.

    The CP transformation \eqref{cp:psi} is also valid for the chiral fields defined by
        \begin{equation}
       \psi_{R,L} \equiv \gamma_{R,L}\psi \quad \gamma_{R,L} \equiv \frac{1 \pm \gamma_5}{2},
    \end{equation}
    and
    \begin{equation}
        \overline{\psi}_{R,L} = \overline{\psi} \gamma_{L,R}.
    \end{equation}
    It is evident that
    \begin{align}
        \psi =& \psi_R + \psi_L = \gamma_R \psi + \gamma_L \psi, \nonumber\\
             =&(\gamma_L+\gamma_R)\psi = \left(\frac{1+\gamma_5}{2} + \frac{1-\gamma_5}{2} \right)\psi = \psi\,. 
    \end{align}
        
    The Yukawa interactions for quarks in the SM are given by eq.\,\eqref{yukawa}. The CP transformation \eqref{cp:psi} for the quark fields and the Higgs doublet is given by
\begin{equation}
\begin{aligned} 
    \overline{q}_{L} \to -e^{-i\beta_q} q_L^T \mathcal{C}^{-1}\gamma^0,  \\
    d_{R} \to e^{i\beta_d} \gamma^0 \mathcal{C} \overline{d_R}^T,\\ 
    H \to H^*,
\label{cp.transf}
\end{aligned}
\end{equation}
    where $e^{-i\beta_q}=\diag(e^{-i\beta_{q_1}},e^{-i\beta_{q_2}},e^{-i\beta_{q_3}})$ contains three phases and $e^{i\beta_d}$ is similar. Then the first Yukawa term in \eqref{Lckm} transforms as,     
    \begin{equation}
    \begin{aligned} 
    \overline{q_L} H V \hat{Y}^d d_R \overset{\text{CP}}{\to}& - q_L^T \mathcal{C}^{-1} \gamma^0 H^* e^{-i\beta_q} V \hat{Y}^d e^{i\beta_d} \gamma^0 \mathcal{C}\overline{d_R}^T
    \cr
    &=-q_L^T H^* e^{-i\beta_q}Ve^{i\beta_d} \hat{Y}^d \overline{d_R}^T,
    \cr
    &=+\overline{d_R}H^\dag \hat{Y}^d e^{i\beta_d}V^T e^{-i\beta_q} q_L,
    \label{CPtranformation}
    \end{aligned}
    \end{equation}
    while the hermitian conjugate of the original term is
    \begin{equation}\begin{aligned}
    (\overline{q_L} H V \hat{Y}^d d_R )^\dag= \overline{d_R} \hat{Y}^d (V)^\dagger H^\dagger q_L. \label{conjugate}
    \end{aligned}\end{equation}
    The rest of the SM Lagrangian is invariant if the similar phases $\beta_u$ for the $u_R$ fields obey $e^{i\beta_u}=e^{i\beta_q}$. Comparison of \eqref{CPtranformation} with \eqref{conjugate} leads us to the conclusion that for the CP transformations \eqref{cp.transf} to be a symmetry, the CKM matrix $V$ must satisfy the condition
    \begin{equation}
        V = e^{i\beta_q}V^* e^{-i\beta_d}\,. \label{V:V^*}
    \end{equation}
    Therefore, for CP conservation, $V$ must be real after some rephasing transformations. Equation \eqref{V:V^*} mandates that all rephasing-invariant functions of the CKM matrix must be real, thus uncovering the core of CP violation in the SM \cite{Branco:1999fs}. The possibility of rephasing transformations makes the use of rephasing-invariant functions of the CKM matrix very convenient in determining the presence or absence of CP violation within the Standard Model. Ultimately, the complex phases of $V$ comes from the complex entries of the Yukawa matrices $Y^u,Y^d$. If these matrices were purely real, CP symmetry would be preserved.

\section{Parametrization of the CKM Matrix}
    \subsection{Standard Parametrization}

     There are infinite ways to represent the elements of the matrix $V$ using three rotation Euler angles  and one phase. However, we will use the standard parametrization of the Particle Data Group, given by 3 Euler angles ($\theta_{12}, \theta_{23}, \theta_{13}$) and a phase $\delta_{13}$ \cite{ParticleDataGroup:2022pth}:
    \begin{equation}
    V = 
    \begin{pmatrix}
    c_{12}c_{13} & s_{12}c_{13} & s_{13}e^{-i\delta_{13}} \\
    -s_{12}c_{23} - c_{12}s_{23}s_{13}e^{i\delta_{13}} & c_{12}c_{23} - s_{12}s_{23}s_{13}e^{i\delta_{13}} & s_{23}c_{13} \\
    s_{12}s_{23} - c_{12}c_{23}s_{13}e^{i\delta_{13}} & -c_{12}s_{23} - s_{12}c_{23}s_{13}e^{i\delta_{13}} & c_{23}c_{13}
    \end{pmatrix}
    \end{equation}
    where $c_{ij}= \cos \theta_{ij}$ and $s_{ij}= \sin \theta_{ij}$. These angles can be chosen in the range $\theta_{ij} \in [0, \pi/2]$, so that $s_{ij}$, $c_{ij}\geq 0$ and $\delta \in [0, 2 \pi]$. The presence of a physical complex phase generally implies that $V^* \neq V$. These parameters are experimentally determined as follows, $\theta_{12} = 12.9^\circ$, $\theta_{23} = 2.4^\circ$, $\theta_{13} = 0.22^\circ$, $\delta = 69.4^\circ$ \cite{ParticleDataGroup:2022pth}. 

    \subsection{Wolfenstein Parametrization}

   The Wolfenstein parametrization is an approximation of the CKM matrix, which describes the mixing of quarks in the SM. This parametrization is particularly useful because it simplifies the representation of the CKM matrix by considering the hierarchy of the experimentally determined mixing angles $\theta_{12} \gg \theta_{23} \gg \theta_{13}$ \cite{wolfenstein_parametrization_1983}. In this parametrization, the independent parameters $\lambda, A, \rho, \eta$ re defined in terms of the mixing angles and the CP-violating phase $\delta_{13}$
    \begin{align}
        \sin{\theta_{12}} =& \lambda ; \\
        \sin{\theta_{23}} =& A \lambda^2; \\
        \exp{(-i \delta_{13})}\sin{\theta_{13}} =& A\lambda^3 (\rho -i\eta);
    \end{align}
 where $\lambda \approx 0.225$, $A \approx 0.82$, $\rho \approx 0.14$, $\eta \approx 0.35$. This parametrization is an expansion series around the parameter $\lambda$. Up to order, $\lambda^3$ the CKM is
    \begin{equation}
        V = \begin{pmatrix}
1 - \frac{\lambda^2}{2} & \lambda & A\lambda^3(\rho - i\eta) \\
-\lambda & 1 - \frac{\lambda^2}{2} & A\lambda^2 \\
A\lambda^3(1 - \rho - i\eta) & -A\lambda^2 & 1
\end{pmatrix} + \mathcal{O}(\lambda^4).
    \end{equation}
    The parameter $\eta$ in the Wolfenstein parametrization represents the imaginary part of the CKM matrix, which implies CP violation in weak interactions. CP violation is a phenomenon where the laws of physics change when particles are replaced by their antiparticles and their spatial coordinates are reversed. To quantify CP violation independently of phase conventions, the Jarlskog invariant $J$ is defined. In the Wolfenstein parametrization, this invariant is
    \begin{equation}
        J \sim A^2 \lambda^6 \eta, 
    \end{equation}
   The value of the Jarlskog invariant is of the order of $10^{-5}$,  indicating the level of CP violation in the quark sector of the SM.
    
\section{Unitarity Triangle}

    The unitarity of the CKM matrix leads to several important relations, both real and complex. These relations are fundamental in understanding the CP violation in the quark sector of the SM. The real equations of the unitary relations \eqref{unitaryrelations} are,
    \begin{equation}
        \sum^3_{i=1} |V_{ij}|^2 =1; \quad j=1,2,3,
    \end{equation}
   where $V_{ij}$ are the elements of the CKM. These relations assert that the sum of the squares of the absolute values of the elements in any column or row of the CKM matrix is 1. While the complex unitary relations are,
    \begin{equation}
    \sum^3_{i=1}  V_{ij} V^*_{ik} = \sum^3_{i=1}  V_{ji} V^*_{ki} = 0; \quad j,k=1,2,3; \quad j\neq k.   \label{TU}  
    \end{equation}
    These relations imply that the product of any row with the complex conjugate of another one must be zero. The same is true for columns.
    
    Explicitly, the $j \neq k$ cases give rise to six mutually orthogonal constraints:
    \begin{align}
            V_{ud} V^*_{ub} + V_{cd} V^*_{cb} + V_{td} V^*_{tb} = 0, \\
            V_{us} V^*_{ub} + V_{cs} V^*_{cb} + V_{ts} V^*_{tb} = 0, \\
            V_{ud} V^*_{us} + V_{cd} V^*_{cs} + V_{td} V^*_{ts} = 0, \\
            V^*_{ud} V_{cd} + V^*_{us} V_{cs} + V^*_{ub} V_{cb} = 0, \\
            V^*_{ud} V_{td} + V^*_{us} V_{ts} + V^*_{ub} V_{tb} = 0, \\
            V^*_{cd} V_{td} + V^*_{cs} V_{ts} + V^*_{cb} V_{tb} = 0. 
        \end{align}
    These six equations can be represented geometrically as triangles in the complex plane. From these relationships, it is possible to derive what we refer to as the “unitarity triangles”. The information about CP violation is embedded in the equations above, since the CKM matrix contains complex elements. One of these triangles can be represented as follows:  
    \begin{figure}[!h]
        \centering
        \includegraphics[width= 0.8\linewidth]{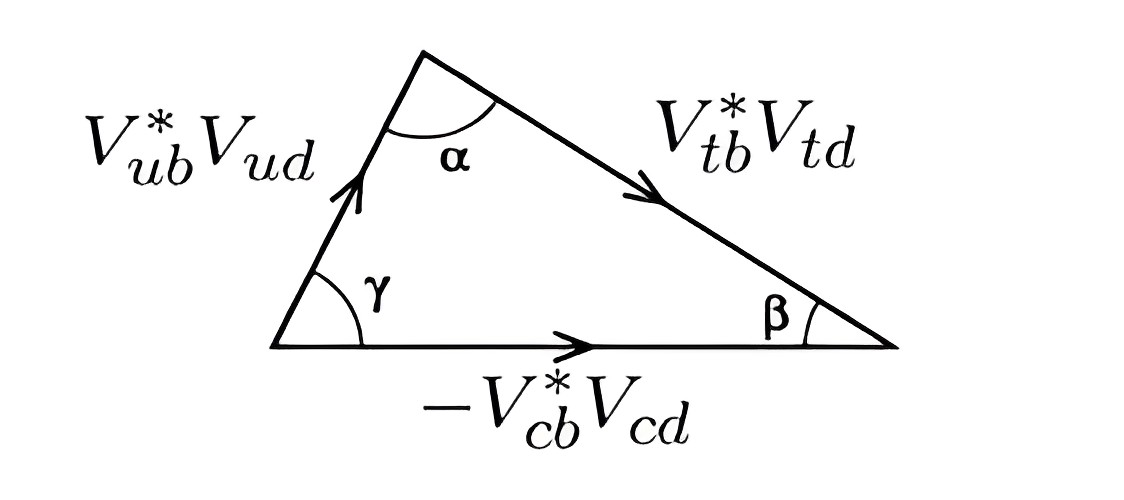}
        \caption{The unitarity triangle}
        \label{figure1}
    \end{figure}
    
    The angles $\alpha, \beta$ and $\gamma$ in the unitarity triangle are defined as
    \begin{align}
       \alpha \equiv \arg \left[-\frac{V_{td}V^*_{tb}}{V_{ud}V^*_{ub}} \right], \\
       \beta \equiv \arg \left[-\frac{V_{cd}V^*_{cb}}{V_{td}V^*_{tb}} \right], \\
       \gamma \equiv \arg \left[-\frac{V_{ts}V^*_{tb}}{V_{cb}V^*_{cb}} \right].
   \end{align}
    These definitions refer to Figure~\ref{figure1}. It is important to note that these definitions remains independent of conventions as long as phase transformations correspond to rotations in the complex plane and do not impact the angles. The equation corresponding to Figure~\ref{figure1} is
    \begin{equation}
    V_{ud} V^*_{ub} + V_{cd} V^*_{cb} + V_{td} V^*_{tb} = 0. \label{tu}
    \end{equation}
    Multiplying equation \eqref{tu} by $V_{ub} V_{ud}^*$,    
    \begin{equation}
        \lvert V_{ub} V_{ud}\rvert^2 + V_{ub}V^*_{ud} V_{cd}V^*_{cb} + V_{ub}V^*_{ud} V_{td}V^*_{tb} = 0. 
    \end{equation}
    Thus,
    \begin{equation}
            Im [V_{ub}V^*_{ud} V_{cd}V^*_{cb}] = - Im[V_{ub}V^*_{ud} V_{td}V^*_{tb}].
    \end{equation}
    If now we multiply \eqref{tu} by $V^*_{cd} V_{cb}$ instead, we get \cite{bigi_cp_2009, Jarlskog:1985cw}
    \begin{equation}
         Im [V_{ub}V^*_{ud} V_{cd}V^*_{cb}] = - Im[V_{cd}V^*_{cb} V_{td}V^*_{tb}].
    \end{equation}
    So, repeating this procedure for the other relations, we conclude that the absolute value of all quartic invariants
    \begin{equation}
     |Im V^*_{km}V_{lm} V_{kn}V^*_{ln}| = \lvert Im V^*_{mk}V_{ml} V_{nk}V^*_{nl} \rvert, 
    \end{equation}
    are equal irrespective of the indices $k, l, m, n$, where $k\neq l$ and $m \neq n$. Hence, we only need to define one of them as the Jarlskog invariant
    \begin{equation}
    J=Im [V_{ub}V^*_{ud} V_{cd}V^*_{cb}] \label{Jarlskog}.
    \end{equation}
    It is a fundamental parameter that characterizes CP violation in the CKM matrix. It arises due to the presence of a complex phase in the unitary CKM matrix, which is represented by the phase $\delta$. This phase is responsible for introducing CP violation into the Standard Model. In essence, $\delta$ determines the extent to which certain physical processes involving quarks violate the combined symmetries of charge conjugation (C) and parity (P), collectively known as CP symmetry.

    Therefore, the Jarlskog invariant in \eqref{Jarlskog}, typically on the order of $10^{-5}$, quantifies CP violation in the SM. For CP violation to occur, the phase cannot be null, hence $J \neq 0$. This parameter is commonly expressed using the standard parametrization formula,
    \begin{equation}
    J= c_{12} c^2_{13} c_{23} s_{12} s_{13} s_{23} \sin(\delta),
    \label{CKM.J}
    \end{equation}
    and in the Wolfenstein parametrization,
    \begin{equation}
       J= A^2 \lambda^6 \eta. 
    \end{equation} 
    Moreover, the phase $\delta$ involved in $V$ undergoes sign reversal under CP transformations, i.e., $\delta \to -\delta$, and as a result, $J$
    % , being CP odd, 
    also changes sign.

\section{Weak Basis Invariants}

    In the last section, we observed that CP violation occurs through the CKM matrix via a complex phase. For 3 generations, there is one physical phase that cannot be eliminated by quark field rephasing.

    In this section, we will describe the invariants under the WBT in \eqref{WBT}. One CP odd invariant indicates whether CP violation occurs or not, and quantify CP violation in a basis invariant way. As any choice of weak basis does not affect the physical properties of the theory, the transformed mass matrices $Y'^{u,d}$ in \eqref{WBT} contain the same physical information as the original $Y^{u,d}$. By definition, we can calculate invariants by WBT in any basis, and we can use $Y^{u,d}$ to denote the Yukawa matrices in any basis.

    The information on quark mixing is encoded in the $q_L$ space, and it is convenient to define the hermitian matrices 
     \begin{equation}
         X_u \equiv Y^u Y^{u\dagger}; \quad X_d \equiv Y^d Y^{d\dagger}.
    \label{Xu.Xd}
     \end{equation}
    In the u-diagonal basis of \eqref{diagonalize}, these quantities becomes \cite{branco_texture_2000}
    \begin{align}
    X_u &\equiv \diag(y^2_u,y^2_c,y^2_t), \label{diag1} \\
    X_d &\equiv V \diag(y^2_d,y^2_s,y^2_b) V^\dagger. \label{diag2}
    \end{align}
    Another possibility is
     \begin{align}
    X_u &\equiv V^\dagger \diag(y^2_u,y^2_c,y^2_t) V, \\
    X_d &\equiv \diag(y^2_d,y^2_s,y^2_b), 
    \end{align}
    in the d-diagonal representation.
    
    Under change of WBT, the Lagrangian must retain the same form. The outcome of any physical process depends solely on WB-invariant quantities, which are crucial for analyzing CP violation \cite{Branco:1999fs}.

    To the ten physical parameters of the SM in the Yukawa sector correspond ten algebraically independent invariants, which can be constructed as traces of powers and products of the matrices $X_u$ and $X_d$\,\cite{trautner_systematic_2019,jenkins_algebraic_2009, branco_weak-basis_1986, branco_rephasing_1988, bernabeu_cp_1986}. Six invariants correspond to the masses or Yukawas, and the other four correspond to 3 angles and one phase of the CKM. For example, using equations \eqref{diag1} and \eqref{diag2}, the Yukawas $(y_u,y_c,y_t)$ and $(y_d,y_s,y_b)$ can be reconstructed from
    \begin{align}
    Tr(X^n_u) &= y^n_u + y^n_c + y^n_t, \label{inv:Xu} \\
    Tr(X^m_d) &= y^m_d + y^m_s + y^m_b, \label{inv:Xd}
    \end{align}
    with $n,m=1,2,3$.
    % In the same way, we have the product of both
    The parameters of the CKM can be obtained from 
    \begin{equation}
        Tr (X^n_u X^m_d) = \sum_{\alpha = u,c,t} \sum_{\beta=d,s,b} y^{2n}_{\alpha} V_{\alpha \beta} y^{2m}_{\beta} V^{\dagger}_{\beta \alpha}. \label{inv:XuXd}
        \end{equation}
    In the u-diagonal basis, for three families, we have 
    \begin{equation}
    Tr[X_uX_d]=Tr[\hat{X}_u V\hat{X}_d V^\dag]
    =\sum_{\alpha,\beta}y_\alpha^2 V_{\alpha\beta} y_\beta^2 V^\dag_{\beta\alpha} \quad \text{for} \quad n,m=1        
    \end{equation}
    \begin{equation}
    Tr[X_u^2 X_d^2]=Tr[\hat{X}_u^2 V\hat{X}_d V^\dag V\hat{X}_d V^\dag]
    =\sum_{\alpha,\beta}y_\alpha^4 V_{\alpha\beta} y_\beta^4 V^\dag_{\beta\alpha} \quad \text{for} \quad n,m=2        
    \end{equation}
    \begin{equation}
    Tr[X_u^3 X_d^3]=Tr[\hat{X}_u^3 V\hat{X}_d V^\dag V\hat{X}_d V^\dag V\hat{X}_d V^\dag]
    =\sum_{\alpha,\beta}y_\alpha^6 V_{\alpha\beta} y_\beta^6 V^\dag_{\beta\alpha} \quad \text{for} \quad n,m=3.        
    \end{equation}

\subsection{Invariants for CP violation}

    The invariants in \eqref{inv:Xu}, \eqref{inv:Xd} and \eqref{inv:XuXd} are CP even. The information about CP violation is indicated by the CP odd invariants. So we can consider the following trace of the powers of commutator
    \begin{equation}
        Tr[X_u, X_d]^r
        % = 0 
        \quad \text{for $r$ odd.}
    \end{equation}
    The case $r=1$ is trivial. The nontrivial relations start with $r=3$. As previously mentioned, CP violation arises in systems with three or more families due to the emergence of complex phases. For three families, there is one physical phase and the unique quantity with $r=3$ can be associated with CP violation \cite{bigi_cp_2009}:
    \begin{align}
        Tr[X_u, X_d]^3 &= 6i Im\,Tr(X^2_u X^2_d X_u X_d), \nonumber\\
        &= 6i\frac{2^6}{v^{12}}(m^2_c - m^2_t)(m^2_t - m^2_u)(m^2_c - m^2_u)(m^2_b - m^2_s)(m^2_b - m^2_d)(m^2_d - m^2_s) J,\quad  \label{Invariant3}
    \end{align}
    % \red{Lembre-se que os $X$ dependem dos Yukawas, não das massas.}
    where $J$ is the Jarlskog invariant in \eqref{CKM.J}. We show the first equality in Appendix \ref{Appendix trace}. 
    See that this invariant vanishes, i.e., CP is a symmetry, unless 
    \begin{eqnarray}
        m_u \neq m_c \neq m_t \text{~~and~~} m_d \neq m_s \neq m_b \nonumber \\
        \theta_{12},\theta_{13},\theta_{23} \neq 0, \pi/2 \nonumber\\
        \delta \neq 0.
    \end{eqnarray} 
    Thus CP violation occurs only if the quark of same charge are non-degenerate. 
    % The masses cannot be zero, because If any quark had zero mass, it would simplify equations in ways that are inconsistent with physical observations. Such a zero mass for a quark would have significant implications for the CKM matrix. The structure of this matrix, being unitary, depends on the mass differences between quarks. A zero mass would alter how we calculate the elements of the CKM matrix, and consequently, how we understand quark mixing and CP violation.
   
   We can formulate the same condition in terms of a hermitian matrix $\mathcal{C}$ defined as \cite{jarlskog_commutator_1985}:
    \begin{equation}
    i\mathcal{C} \equiv [X_u, X_d].
    \end{equation}
    Then, the invariant in  \eqref{Invariant3} can be written as
    \begin{equation}
    \det \mathcal{C} = 
    \det (-i[X_u, X_d])
    =-\frac{1}{3} Tr[X_u, X_d]^3
    \,.
    % \neq 0
    \end{equation} 
    A nonzero $\det \mathcal{C}$ is a requisite for CP violation. 
   % \blue{However, the interpretation of $\det \mathcal{C}$, a parameter with a high mass dimension, is not straightforward. It is value, proportional to $M^{12}$, can vary significantly based on the relevant scale $M$ for a specific process. For instance, if we consider baryogenesis at the electroweak scale, characterized by $M \sim 100$ GeV, CKM dynamics appear irrelevant. CP violation is observable in $K$ and $B$ decays, primarily because the CP asymmetries in these decays are ratios of CKM parameters. Despite the numerators, governed by $\det \mathcal{C}$, being small, the denominators are similarly small, which is particularly evident in B decays \cite{bigi_cp_2009}}.
   This invariant is highly suppressed and has a numerical value of the order of $\det\mathcal{C}\sim 10^{-17}$. Therefore, if we consider baryogenesis at the electroweak scale, characterized by $M\sim v \sim 100$ GeV, CKM dynamics appears irrelevant. On the other hand, CP violation is observable in $K$ and $B$ decays, primarily because the CP asymmetries in these decays are ratios with similarly small denominators. This is particularly evident in $B$ decays \cite{bigi_cp_2009}.

%% file: capitulos/3-SH.tex
\chapter[Invariants and the Hilbert series]{Invariants \\
and the Hilbert series}

    The Hilbert series (HS) is a mathematical tool used to explain the algebraic structure of a set by representing it as a power series in suitable variables. Its principal utility resides in the quantification of invariants within a given theoretical framework.  This chapter is dedicated to delving into the Hilbert series, particularly emphasizing its application to the study of flavor invariants within the quark sector. 

\section{The Hilbert Series}

   The Hilbert series serves as an instrumental tool in the field of theoretical physics, particularly in the context of characterizing all physical information encapsulated within invariants under a group $G$. 
   It can be defined as \cite{xiao_standard_2019, lehman_hilbert_2015, jenkins_algebraic_2009}
   \begin{equation}
       H(q) \equiv \sum_{n=0}^{\infty} c_n q^n, \label{Hilbert}
   \end{equation}
   where $q$ is an arbitrary complex variable with $|q|< 1$. We will often denote the variable $q$ as a \emph{spurion}, in reference to fields or couplings with definite transformation properties under $G$. The coefficient  $c_n$  is the number  of invariants of degree $n$. The Hilbert series, therefore, provides a comprehensive method to encapsulate and enumerate the invariants associated with a given symmetry group in a theory, playing a crucial role in understanding the underlying structure of the theory and its physical implications.
   
   The Hilbert series can generally be written as the ratio of two polynomials as\cite{jenkins_algebraic_2009}
    \begin{equation}
           H(q) = \frac{N(q)}{D(q)}. \label{definition}
   \end{equation}
    The numerator consists of a polynomial with non-negative integer coefficients, exhibiting a palindromic configuration    
    \begin{equation}
        N(q) = 1+ c_1 q + \cdots c_{d_N -1} q^{d_N -1} + q^{d_N}, \quad c_k= c_{d-k},
    \end{equation}
    and the denominator can be expressed in the general form 
    \begin{equation}
        D(q) =  \prod_{r=1}^{p} (1-q^{d_r})\,,
    \label{HS:deno}
    \end{equation}
    where $d_1\le d_2\le \cdots\le d_p$ and the degree is $d_D=\sum_r d_r$. A highly nontrivial result is that the denominator of the Hilbert series encodes the information of all algebraically independent invariants. The factors in the denominator $D(q)$ describe the degree and number of these invariants: each factor $(1-q^{d_r})$ corresponds to an invariant of degree $d_r$, while the number of denominator factors $p$ corresponds to the number of physical parameters that make up the invariants \cite{jenkins_algebraic_2009}. 

%%%%%%%%%%%%%%%%%%%%%%%%%% 
    \subsection{Example using definition} \label{Example-definition}
    
    Let us consider a theoretical framework with one coupling $m_1$, which transform under a $U(1)$ symmetry group as 
    \begin{equation}
       m_1 \to e^{i\phi} m_1, \quad m^*_1 \to  e^{-i\phi} m^*_1.
    \label{U1:m1-m1*}
   \end{equation}
    In this scenario, an  invariant can be constructed as $I = e^{i\phi} m_1 e^{-i\phi} m^*_1 = m_1 m^*_1$. In this simple example we can list all invariants which are the powers $(m_1m_1^*)^n$. So we can write the Hilbert series \eqref{Hilbert} directly as
\begin{equation}
       H = \sum_{n=1}^\infty c_n (m_1 m^*_1)^n = 1 + (m_1 m^*_1)+(m_1 m^*_1)^2+(m_1 m^*_1)^3+\cdots.        
   \end{equation}
% where $c$ denotes the number of distinct invariant possibilities.
    where $c_n=1$ for all powers $n$.
    % denotes the number of distinct invariant possibilities. Given that the dimension referring to the quantity of invariants is 1 in this case, the Hilbert series can be expanded as
    % \begin{equation}
    %     H= 1 + (m_1 m^*_1)+(m_1 m^*_1)^2+(m_1 m^*_1)^3+\cdots. 
    % \end{equation}
    % This series is nothing more than a geometric series\footnote{Proof regarding geometric series in Appendix C}, and taking into account that the invariant ($m_1 m^*_1$) is less than 1
    This series is nothing more than a geometric series\footnote{Proof regarding geometric series in Appendix \ref{Appendix geometric series}.}, and considering ($m_1 m^*_1$) as a variable which is less than unity,    
    % the Hilbert series takes the following form
    the Hilbert series can be rewritten in the form
    \begin{equation}
        H = \frac{1}{(1-m_1 m_1^*)}. \label{m1m1}
    \end{equation}
   
    The invariant is given by $I= m_1 m_1^*$, and since each coupling has degree 1, the overall invariant has degree 2. So, from the Hilbert series \eqref{m1m1} it becomes evident that there is only one invariant of degree 2 that can be constructed with $m_1$ and $m_1^*$. Since the entire power of the denominator is equal to 1, this indicates that there is only one invariant. This Hilbert series is termed \textit{refined} or \textit{multigraded} because it encompasses multiple variables, specifically $m_1$ and $m_1^*$. 

    % In this example, the condition when $N(q) = 1$,  indicates that there are no interrelations or dependencies among the identified invariant. 
    In this example, when comparing \eqref{m1m1} with \eqref{definition}, the trivial numerator $N(q) = 1$ indicates that there are no interrelations or dependencies among the invariants, because there is no invariant there. This dependency arises when terms in the numerator become negative. This will be addressed further ahead.       
    
    The number and degree of invariants become more apparent when we express the series in an \textit{unrefined} form, as in \eqref{Hilbert}, representing it in terms of a single variable. This is achieved by substituting both $m_1$ and $m_1^*$ with $q$ in the series. In this case, the unrefined Hilbert series becomes
    \begin{equation}
        H(q) = \frac{1}{(1-q^2)}.
    \end{equation}
    This form is particularly useful in invariant theory, where the focus is on understanding the properties and relationships of invariants that remain unaffected by certain transformations. The existence of exactly one invariant is evidenced by the power of $(1 - q^2)^1$  being 1, indicating a single invariant. Furthermore, the degree of this invariant is 2, as denoted by the exponent in $q^2$. This representation highlights the specific characteristics of the invariant within the theory, where the power of the term reflects the
    number of invariants and the exponent of $q$ indicates their degree.

    The unrefined Hilbert series must also satisfy Knop's theorem \cite{jenkins_algebraic_2009, knop_grad_1987} which tells us that
   \begin{equation}
       \dim V \geq d_D - d_N \geq p,
    \label{knop}
   \end{equation}
    where $d_D = \sum_r d_r$ and $\dim V$ represents the dimension of the vector space upon which the group transformations are applied. In this example, $\dim V = 2$ because of $m_1$ and $m_1^*$,  $d_D=2$ is the degree of the denominator and $p=1$ refers to the number of independent parameters in the model:
    only the modulus $|m_1|$ is physical as the phase can be removed by the symmetry transformation. Here, Knop’s theorem gives $2\geq 2-0 \geq 1$.
    
\section{Molien-Weyl Formula}

   In the previous section, we calculated the Hilbert series using its definition. However, for more complicated cases, computing it directly becomes nearly impossible due to the exponential increase in the number of linearly independent invariants with increasing degree.
   
   The Molien-Weyl formula is a pivotal tool in Invariant Theory. It streamlines the calculation of the Hilbert series by reducing it to the evaluation of a limited number of complex integrals, which can be efficiently computed using the residue theorem. This formula is extensively utilized in Invariant Theory, particularly in contexts involving the action of a symmetry group on mathematical objects. It facilitates the computation of invariant quantities under the specified symmetry.
   
   The enumeration of invariants for each degree, particularly in relation to quarks, can be succinctly represented through the Hilbert series.
   This provides crucial insights for determining the number of independent invariants at each degree.
   
    The Molien-Weyl formula consists of two primary components: the integration measure, which is the Haar measure of the symmetry group, and the integrand, represented by the Plethystic exponential.
    
    For a spurion $X$ transforming under a finite group $G$ in a representation $R$, the Hilbert series in terms of the variable\footnote{We will often use the same symbol for the spurion and for the variable.} $X$ is determined by the following Molien-Weyl formula \cite{wang_flavor_2021}:
    \begin{equation}
        H(X)= \frac{1}{|G|} \sum_{g \in G} \frac{1}{\text{Det} (\mathbb{I} - X R(g))}, \label{Hilb}
    \end{equation}
    where $|G|$  represents the order of  $G$, meaning the total number of elements in  $G$ . Here, \( \mathbb{I} \) denotes the identity matrix. 
   
   If $G$ is a reductive Lie group\footnote{Reductive Lie groups are characterized by their decomposable nature into a direct sum of a semisimple Lie group and an Abelian group.}, then the equation \eqref{Hilb} can be rewrite as
    \begin{equation}
    H(X) = \int d\mu_{G} \frac{1}{\text{Det} (\mathbb{I} - X R(g))}. \label{H(q)}
    \end{equation}
    Here, $d\mu_G$ is the Haar measure, and $[\text{Det} (\mathbb{I} - XR(g))]^{-1}$ is known as the Plethystic exponential \cite{hanany_hilbert_2010}, i.e.,
    \begin{equation*}
       PE[X,R] = \frac{1}{\text{Det} (\mathbb{I} - X R(g))},
    \end{equation*}
    the deduction is in the following section.

\subsection{Plethystic exponential}

    To construct the Hilbert series using the Molien-Weyl formula, it is necessary first to define a function known as the Plethystic Exponential (PE). The Plethystic exponential is a mathematical function that generates symmetrical functions corresponding to the representations of a Lie group. The Plethystic exponential is formulated in terms of a spurion $X$ which transforms with a representation $R$ of the Lie group $G$ \cite{xiao_standard_2019}. Then we can rewrite the expression \eqref{H(q)} as 
    \begin{align}
        \text{Det} (\mathbb{I} - XR(g))]^{-1} &= \prod_{j=1}^{d} (1- X z_j)^{-1} = \exp{\left(\sum_{j=1}^d \sum_{r=1}^\infty \frac{X^r z_j^r}{r}\right)} 
        \cr
        &= \exp{\left(\sum_{r=1}^\infty \frac{X^r \chi_R (z_1^r,...,z_d^r)}{r}\right)},
    \end{align} 
    where $\chi_R$ is the character function of representation $R$ of $G$, with $z_j$ being the eigenvalues of $R(g)$ and $d$ is the dimension of the representation; see \cite{feng_counting_2007} and \cite{benvenuti_counting_2007} for details.
    This allows us to define the $PE$ as follows   
    \begin{equation}
       PE[X,R] = \exp \left(\sum_{r=1}^\infty \frac{X^r \chi_R(z^r_j)}{r} \right). \label{PE}
   \end{equation}

    Depending on the context, the spurion $X$ can represent different types of fields. For instance, it can represent fermion fields in certain applications, highlighting the versatility of the Plethystic exponential in different physical contexts. In this case, the Plethystic exponential should be adapted to the fermionic case \cite{lehman_hilbert_2015}.

   The Plethystic exponential encompasses all possible products of spurions, enabling its extension to additional spurions and representations of diverse groups.   This formulation is a pivotal component in the analysis of group symmetries and their corresponding invariants, particularly in the field of theoretical physics.

\subsection{Characters}

    In this section, we will review the character of representations and  present the characters that will be used throughout the work. Starting from the definition \cite{tung1985group}:
 \begin{definition}
     The character $\chi^D(g)$ of a group element $g \in G$ in a representation $D$ of $G$, is defined to be the trace of the matrix representing it:
 \end{definition}
    \begin{equation}
        \chi(g) \equiv Tr(D(g)) = \sum_{i=1}^{\dim D} D_{ii}(g).
    \end{equation} 
    Then all group elements of a given conjugacy class have the same character, i.e.,\linebreak $Tr[D(p) D(g) D(p^{-1})]= Tr D(g)$. 

% \red{Use o símbolo $R(g)$ para denotar representações ao invés de $D(g)$. Já foi usado acima.}
    For example, for a matrix $U$ in the fundamental representation (representation $2$) of the $SU(2)$ group, its character  is 
    \begin{equation}
    \label{su2:chi:2}
                \chi_{2}(U)=\operatorname{Tr}(U) = \operatorname{Tr} \begin{pmatrix}
                e^{i \theta} & \\
                & e^{-i \theta}         
            \end{pmatrix} 
            = z + \frac{1}{z},
        \end{equation}
    where $e^{i \theta} = z$ and $e^{-i \theta} = z^{-1}$, so $z$ is a complex number with modulus 1. 
    
    As another example, we can consider the rotation group on the plane, the group $SO(2)$, represented by
   \begin{equation}
   \label{so2}
            R(\theta) = \begin{pmatrix}
                \cos{\theta} & \sin{\theta} \\
              - \sin{\theta}  & \cos{\theta}         
            \end{pmatrix}, 
    \end{equation}
    such that
    \begin{equation}
        R(\theta) \begin{pmatrix}
            x \\
            y
    \end{pmatrix}
        =  \begin{pmatrix}
            x \cos{\theta} + y \sin{\theta} \\
            - x \sin{\theta} + y \cos{\theta}
        \end{pmatrix}.
    \end{equation}
    It is easy to verify that $R(\theta)R(\phi) = R(\theta + \phi)$, so this group is Abelian and consequently, in complex space, any representation is reducible to one-dimensional representations. Explicitly,
    \begin{equation}
        M R(\theta) M^{-1} = \begin{pmatrix}
                e^{-i \theta} & \\
                & e^{i \theta}         
            \end{pmatrix}, 
    \end{equation}
    where
    \begin{equation}
        M = \frac{1}{\sqrt{2}}\begin{pmatrix}
               1&i \\
                i& 1         
            \end{pmatrix}. 
    \end{equation}
    The vector in the new basis is
    \begin{equation}
        M \begin{pmatrix}
            x \\
            y
    \end{pmatrix}
        = \frac{1}{\sqrt{2}}\begin{pmatrix}
            x+ iy \\
            ix +y
        \end{pmatrix}.
    \end{equation}
    The  character of the representation \eqref{so2} is
    \begin{equation}
        \chi=Tr[R(\theta)] 
        =2 \cos{\theta}
        = Tr[ M R(\theta) M^{-1}]
        = e^{i \theta} + e^{-i \theta} = z+\frac{1}{z}.
    \end{equation}

    For the $U(1)$ group, the character is trivial, since the matrix representation of $U(1)$ is a $1 \times 1$ matrix. For a spurion of charge $Q$,
    \begin{equation}
        \chi_{Q} = Tr[e^{i Q\theta}] = z^Q,
    \end{equation}
    where $z=e^{i\theta}$.

    The character can be extended to direct product of groups. Let $G$ and $H$ be groups, and let $\chi(g)$ and $\sigma(h)$ be characters of $g\in G$ and $h\in H$, respectively. The character $\chi \otimes \sigma$ of $(g,h)\in G\otimes H$ is defined by \cite{tung1985group}
    \begin{equation}
       (\chi \otimes \sigma)(g,h)=\chi(g) \sigma(h).
    \end{equation}

% \red{Está misturando produto tensorial de representações com produto de grupos.}

    Similarly, different representations of a group $G$ can be constructed from the tensorial product of a few basic representations. We can then calculate the character of the product representation from the character of these basic representations. If $R,R'$ are two representations of $G$, the character of the product $R\otimes R'$ is given by \cite{zee_group_2016} 
    \begin{equation}
    \chi_{R\otimes R'}(g)=\chi_R(g)\chi_{R'}(g).
    \end{equation}
    
    \textbf{Proof}: For a given $g\in G$, let us write $A=R(g)$ and $B=R'(g)$. The product representation $R\otimes R'$ is given by the Kronecker product $(R\otimes R')(g)=R(g)\otimes R'(g)=A\otimes B$. The latter is defined as the block matrix
   \begin{equation}
       A \otimes B = \begin{bmatrix}
    a_{11} B & \cdots & a_{1n} B \\
    \vdots & \ddots & \vdots \\
    a_{m1} B & \cdots & a_{mn} B
    \end{bmatrix}.
    \end{equation}
    Then
    \begin{align}
        Tr[A\otimes B] &= \sum_{i=j=1}^n [a_{11} b_{ij} + a_{22} b_{ij} + \cdots] \\ 
        &= [a_{11} b_{11} + a_{11} b_{22} + \cdots a_{22} b_{11}+ \cdots] \\
        &= (a_{11} + a_{22} + a_{33}+\cdots)(b_{11}+ b_{22}+b_{33} + \cdots) = Tr[A] Tr[B]\,.
    \end{align}
    
    With these results we can calculate the character of the product of groups and of the representations constructed from the product of other representations.
    For $SU(2)$, given the character of fundamental representation \eqref{su2:chi:2}, the character of the representation $2\times\bar{2}$ is given simply by
    \begin{equation}
       \chi_{2 \times \Bar{2}}(z) = 
       \chi_2(z)\chi_2(z)=
       \left(  z + \frac{1}{z} \right) \left(  z + \frac{1}{z} \right) = z^2 + \frac{1}{z^2} +2,
   \end{equation}
    considering that the antifundamental representation $\bar{2}$ is equivalent to the fundamental representation $2$. From the character above we can extract the character of the adjoint representation (3) by subtracting one since $2\otimes\bar{2}=3\oplus 1$ and the character of the singlet representation is trivially $\chi_1=1$. This property is valid for all $SU(n)$: $n \otimes \bar{n} = 1 \oplus \text{adj}$. 

    The procedure for other groups is analogous. We summarize the characters of the different representations and groups in table~\ref{tab:char}.
   \begin{table}[h]
    \centering
    \caption{\label{tab:char}%
    Character function for several groups \cite{lehman_hilbert_2015, sturmfels_algorithms_2008, hanany_hilbert_2010}.}.
    \begin{tabular}{|c|c|}
    \hline
    Representation & Character \\
    \hline
    SU(2): 2 & $z+\frac{1}{z}$ \\
    SU(2): $2 \otimes \Bar{2}$ & $z_2 + \frac{1}{z_2} +2$ \\
    SU(3): 3 & $z_1 + \frac{z_2}{z_1} + \frac{1}{z_2}$ \\
    SU(3): $\bar{3}$ & $z_2 + \frac{z_1}{z_2} + \frac{1}{z_1}$ \\
    SU(3): $3\otimes \bar{3}$ & 
    $z_1 z_2 + \frac{z^2_1}{z_2} + \frac{z^2_2}{z_1} \frac{z_1}{z^2_2}+\frac{z_2}{z^2_1} + \frac{1}{z_1 z_2} + 3$ \\
    SU(4): 4 & $z_1 + \frac{z_2}{z_1} + \frac{z_3}{z_2} + \frac{1}{z_3}$\\
    SU(4): $\Bar{4}$ & $\frac{1}{z_1} + \frac{z_1}{z_2} + \frac{z_2}{z_3} + z_3$ \\
     SU(4): $4 \otimes \Bar{4}$ & $\frac{z_1^2}{z_2}+ \frac{z_2}{z_1^2}+ \frac{z_1 z_2}{z_3}+ \frac{z_3}{z_1 z_2} + z_1 z_3 + \frac{1}{z_1z_3} + \frac{z_2^2}{z_1 z_3} + \frac{z_1 z_3}{z_2^2} + \frac{z_2 z_3}{z_1} + \frac{z_1}{z_2 z_3} + \frac{z_3^2}{z_2} + \frac{z_2}{z_3^2} +4$ \\
    U(1): charge Q & $z^Q$ \\
    \hline 
    \end{tabular}
\end{table}

    % Using the fundamental representation of SU(2), for example, the Plethystic exponential \eqref{PE} can be written as 
    We can now write, for example, the Plethystic exponential \eqref{PE} for $X$ in the fundamental representation of SU(2):
    \begin{equation}
       PE[X,R] = \exp \left(\sum_{r=1}^\infty \frac{X^r}{r} \left(z^r +\frac{1}{z^r} \right) \right) = \exp \left(\sum_{r=1}^\infty \left( \frac{(Xz)^r}{r} + \frac{X^r}{z^r r} \right)\right)\,.\label{PEX}
   \end{equation}
    Using the series expansion
    \begin{equation}
        \sum_{k=1}^\infty \frac{x^k}{k} = -\ln(1-x), \label{log1}
    \end{equation}
equation \eqref{PEX} becomes
   \begin{equation}
   PE[X,R] = \exp \left[ -\ln{(1-Xz)} - \ln{\left(1-\frac{X}{z}\right)} \right] = \frac{1}{(1-Xz) \left(1-\frac{X}{z}\right)}.       
   \end{equation}

\subsection{Plethystic Logarithm}

In theories dealing with flavor symmetry, a key characteristic is that all flavor invariants can be expressed as polynomials of a \emph{finite} number of \emph{basic} invariants.
% This leads to an interesting question: how do we determine the combination that exists between these invariants? The Plethystic Logarithm (PL) offers a solution, providing insights into the number of basic invariants and the dependencies (syzygies) among them.
This leads to an interesting question: how do we determine the number and degree of these basic invariants? The Plethystic Logarithm (PL) offers a solution, providing the number and degree of basic invariants and also the possible dependencies (syzygies) among them.

% \red{Esse parágrafo, você tem que mudar. Desse jeito, você vai ter que explicar o que é um ring, o que é finitely generated, etc. Dá para explicar isso de um jeito mais simples.
% Algo como "Nos casos de interesse, todos os invariantes podem ser escritos como polinômios de um número finito de invariantes.
% Esse invariantes são chamados de básicos." E o PL dá esse número.
% ... Estou em dúvida se não é melhor definir "polynomial ring" e etc.
% }
    
The Plethystic logarithm of the Hilbert series is defined as follows,
    \begin{equation}
   PL[H(q)]= \sum_{r=1}^\infty \frac{\mu(r) \log H(q^r)}{r}\,, \label{PL:log}
   \end{equation}
    where $H(q)$ is the Hilbert series and $\mu(r)$ is the M\"obius function. The M\"obius function is defined as
    \begin{equation}
    \begin{aligned}
    \mu(r) := &\begin{cases}
        0, &  \text{$r$ has repeated prime factors} \\
        1, &  r = 1 \\
        (-1)^n, & \text{ $r$ is a product of $n$ distinct primes}
    \end{cases}
    \end{aligned}
    \end{equation} 
    In this context, the initial positive terms of the Taylor expansion of this function correspond to the basic invariants (Irreducible flavor invariants) , which are the generators of the polynomial formed by the flavor invariants. These terms reveal both the number and the degrees of the basic invariants.
 
    The negative terms in the Plethystic logarithm function indicate the interdependencies among the basic invariants. These dependencies, known as \emph{syzygies}, show how certain invariants can be expressed in terms of other \cite{benvenuti_counting_2007}.

     By calculating the $PL$ of the Hilbert series, we can systematically understand the fundamental building blocks of the flavor invariants basics and their interrelations, i.e., their syzygies. This approach is crucial for delving deeper into the theoretical framework governing the interactions and symmetries in particle physics, particularly in understanding the structure of quark and lepton mixing patterns and CP violation phenomena. The phenomenon of CP violation arises from the characteristics of the invariants, which will be discussed in the subsequent chapters. 
    
\subsection{Haar measure}

    The Haar measure for a generic group $G$ is given by\cite{derksen2015computational,sturmfels_algorithms_2008}
\begin{equation}
        \int d\mu_{G} = \frac{1}{(2\pi i)^r} \oint_{|z_1| = 1} \frac{d z_1}{z_1} \cdots \oint_{z_{r}=1} \frac{d z_r}{z_r} \prod_{\alpha^+} \left( 1- \prod_{l=1}^r z_l^{\alpha^+} \right).
        \label{haar}
    \end{equation}
    Here, $\alpha^+$ represents the positive roots belonging to the Lie algebra of the group $G$, while $r$ indicates the rank of $G$ \cite{hanany_hilbert_2010}. This information can also be analyzed using the root diagram, which are used in the classification of Lie algebras. The roots appear in the $n \otimes \Bar{n}$  representation of the group, and we can extract the powers $z_l^{\alpha^+}$ from the character of this representation.
    These powers are summarized in table \ref{tab:haar}.
\begin{table}[h]
    \centering
    \caption{Representations of roots of the Lie groups.}
   \begin{tabular}{|c|c|c|}
    \hline
    Representation & Character & $z_l^{\alpha^+}$  \\
    \hline
    $SU(2):2 \otimes\Bar{2}$ & $z^2 + \frac{1}{z^2} +2$ & $z^2$ \\
    $SU(3):3 \otimes\Bar{3}$ & $z_1 z_2 + \frac{z^2_1}{z_2} + \frac{z^2_2}{z_1}+\frac{z_1}{z^2_2}+\frac{z_2}{z^2_1} + \frac{1}{z_1 z_2} + 3$ & $z_1 z_2, \frac{z^2_1}{z_2}, \frac{z^2_2}{z_1}$  \\
    \makecell{$SU(4):4 \otimes \Bar{4}$} & 
    \makecell{
    $\frac{z_1^2}{z_2}+ \frac{z_2}{z_1^2}+ \frac{z_1 z_2}{z_3}+ \frac{z_3}{z_1 z_2} + z_1 z_3 +$ \\ 
    $\frac{1}{z_1z_3} + \frac{z_2^2}{z_1 z_3} + \frac{z_1 z_3}{z_2^2} + \frac{z_2 z_3}{z_1} +$ \\
    $\frac{z_1}{z_2 z_3} + \frac{z_3^2}{z_2} + \frac{z_2}{z_3^2} +4$} & 
    \makecell{
    $\frac{z_1^2}{z_2}, \frac{z_1 z_2}{z_3}, z_1 z_3,$ \\ 
    $\frac{z_2^2}{z_1 z_3}, \frac{z_2 z_3}{z_1}, \frac{z_3^2}{z_2}$} \\ 
    U(1)- Charge Q & $z^Q$ & 0\\
    \hline
    \end{tabular} \label{tab:haar}
\end{table}
    
    Using eq.\,\eqref{haar} we can now write the Haar measure for the groups we will use:\footnote{Additional examples of Haar measures for various Lie groups and their proofs can be found in  \cite{hanany_hilbert_2010, chen_hilbert_2011}.}
    \begin{align}
        &\int_{U(1)} d\mu_{U(1)} = \frac{1}{2\pi i} \oint_{|z|=1} \frac{dz}{z}; \label{haar:u1}
\\
        &\int_{SU(2)} d\mu_{SU(2)} = \frac{1}{2\pi i} \oint_{|z|=1} \frac{dz}{z} (1-z^2); \label{su2}\\
        &\int_{SU(3)} d\mu_{SU(3)} = \frac{1}{(2\pi i)^2} \oint_{|z_1|=1} \frac{d z_1}{z_1} \oint_{|z_2|=1} \frac{d z_2}{z_2} (1-z_1 z_2) \left(1- \frac{z^2_1}{z_2} \right) \left(1- \frac{z^2_2}{z_1}\right). \label{haar su3}
    \end{align}
    \begin{multline}
        \int_{SU(4)} d\mu_{SU(4)} = \frac{1}{(2\pi i)^3} \oint_{|z_1|=1} \frac{d z_1}{z_1} \oint_{|z_2|=1} \frac{d z_2}{z_2} \oint_{|z_3|=1} \frac{d z_3}{z_3} (1-z_1 z_3) \left(1- \frac{z^2_1}{z_2} \right) \\
        \left(1- \frac{z_1 z_2}{z_3}\right) \left(1- \frac{z_2^2}{z_1 z_3} \right) \left(1- \frac{z_2 z_3}{z_1}\right) \left(1- \frac{z_3^2}{z_2}\right)  \label{haar su4}
    \end{multline}
    It can be observed that the $U(1)$ and $SU(2)$ groups have rank one, 
    meaning only one variable $z$ is necessary to describe conjugacy classes. The $SU(3)$ group, on the other hand, has rank 2. This means two variables $z_1$ and $z_2$ are necessary to describe the conjugacy classes. Furthemore, the $SU(4)$ group has rank 3, implying that three variables, denoted as $z_1$, $z_2$ and $z_3$ are necessary to describe the conjugacy classes.
    
    We can now recalculate the Hilbert series of Example \ref{Example-definition} using the Molien-Weyl Formula.

\subsection{Example \ref{Example-definition} using the Molien-Weyl Formula}

   Let us consider the same spurions $m_1,m_1^*$ transforming as in \eqref{U1:m1-m1*}. We use \eqref{PE},  where the summation extends over the different spurions $m_1$ and $m^*_1$. The character associated with $m_1$ is $z$ and with $m_1^* $ is $z^{-1}$, as it pertains to the $U(1)$ group and $z= e^{i\phi}$. Therefore, the $PE$ can be expressed as
   \begin{equation}
        PE = \exp \left[ \sum_{r=1} \frac{1}{r} \left( (m_1 z)^r + \left(\frac{m_1^*}{z} \right)^r \right)\right].
    \end{equation}
    The Haar measure for $U(1)$ was given in \eqref{haar:u1}.
    %  \begin{equation}
    %   \int_{U(1)} d\mu_{U(1)} = \frac{1}{2\pi i} \oint_{|z|=1} \frac{dz}{z}. 
    % \end{equation}
   Consequently, the Molien-Weyl formula becomes
   \begin{equation}
       H = \frac{1}{2\pi i} \oint_{|z|=1}  \frac{dz}{z} \cdot \exp \left[ \sum_{r=1} \frac{1}{r} \left( (m_1 z)^r + \left(\frac{m_1^*}{z} \right)^r \right) \right]\,.
   \end{equation}
    Using the simplification in equation \eqref{log1} the \textit{refined} Hilbert series takes the following form
    \begin{equation}
       H = \frac{1}{2\pi i} \oint_{|z|=1}  \frac{dz}{z} \frac{z}{(1-m_1 z)(z- m_1^*)}. \label{Hm}
   \end{equation}
    Considering that $|m_1|<1$, the only pole within the unit circle $|z|=1$ is $z= m_1^*$. Upon evaluating the residue at $z= m_1^*$  the Hilbert series simplifies to
   \begin{equation}
       H(m_1,m_1^*) = \frac{1}{(1-m_1 m_1^*)}.
   \end{equation}
    By expressing in terms of a single variable, $m_1 \to q$ and $m_1^* \to q$, the \textit{unrefined} series is
   \begin{equation}
       H(q)= \frac{1}{(1-q^2)},
   \end{equation}
    while the Plethystic logarithm is
    \begin{equation}
         PL[H(q)]= \sum_{r=1}^\infty \frac{\mu(r) \log H(q^r)}{r} =  1\cdot q^2\,.
         \label{PL:m1}
    \end{equation}
    We can highlight a few aspects:
    \begin{itemize}
       \item $N(q)=1$ in the Hilbert series indicates that there are no interrelations or dependencies among the invariants;
       
       \item $D(q)=(1-q^2)$ indicates there is one basic invariant of degree 2 and all other invariants can be written as a polynomial of it.
       This is confirmed by the Plethystic logarithm in \eqref{PL:m1}.
       \end{itemize}

% \red{Tem que manter a notação já introduzida. Primary ainda não foi definido.
% Melhor se ater à noção de basic invariant.
% Também, com 1 invariante só, não faz sentido falar de independencia.
% }

\subsection{A Toy Model with Two Couplings Under $U(1)$ Symmetry}

     We now consider a theoretical model featuring two couplings, $m_1$ and $m_2$, both possessing equal charges and transforming under a $U(1)$ symmetry group. For the purpose of comparison and discussion, we will use the same notation as in \cite{jenkins_algebraic_2009}:
    \begin{equation}
        m_1 \to e^{i \phi} m_1, \quad m_2 \to e^{i \phi} m_2.
    \end{equation}
    Note that here we have four \textit{spurions}: $m_1,m_1^*,m_2,m_2^*$.

    This example is still simple enough, 
    % straightforward, 
    allowing us to list all the basic invariants and express the relations among them directly. Subsequently, we can confirm all these properties by calculting the Hilbert series through the Molien-Weyl formula.

\textbf{Identifying Basic Invariants}

    Basic invariants are polynomials of the spurions $m_1,m_1^*,m_2,m_2^*$ that remain invariant under the $U(1)$ symmetry transformation. They constitute a set, enabling any invariant within the system to be represented as a polynomial. None of the basic invariants can be expressed as a polynomial of other basic invariants on the set. We can directly list the basic invariants for our system:
    \begin{itemize}
        \item $I_1=m_1 m_1^*$,
        \item $I_2=m_2 m_2^*$,
        \item $I_3=m_1 m_2^*$,
        \item $I_4=m_2 m_1^*$.
    \end{itemize}
    It is important to note that within the set of basic invariants, not all the basic invariants are necessarily algebraically independent, and there may exist polynomial functions of the basic invariants that are identically equal to zero \cite{trautner_systematic_2019}. Such polynomial relationships are known as syzygies. We can choose among the basic invariants, the \emph{primary} invariants, which are not related by syzygies. So it is evident that the number of basic invariants is not less than that of primary invariants. The information about all the basic invariants can be obtained by calculating the Plethystic logarithm given by eq.\,\eqref{PL:log}.

\textbf{Identifying Primary Invariants}
    
    Primary invariants are those basic invariants that are algebraically independent. An important result is that the number of primary invariants equals the number of physical parameters in the theory. The following relation demonstrates that the basic invariants mentioned above are not algebraically independent:
    \begin{equation}
     I_1 I_2=I_3 I_4.   \label{syzygy}
    \end{equation}
    So the four invariant can not be primary. The equation \eqref{syzygy} is an example of a syzygy: a polynomial relation among the basic invariants.
    
    Thus, we may choose $I_1,I_2,I_3$ as our set of primary invariants, because $I_4$ can be related through the \textit{syzygy}. 

    The denominator of the Hilbert series encodes information about the primary invariants.

\textbf{Constructing the Hilbert Series:}

    To construct the Hilbert series for our theoretical model, the initial step is to build the $PE$:
    \begin{align}
        PE = &\exp \left[\sum_{r=1} \frac{1}{r} (m_1 e^{i \phi})^r \right] \exp \left[\sum_{r=1} \frac{1}{r} (m_1^* e^{-i \phi})^r \right] \nonumber \\
        &\exp \left[\sum_{r=1} \frac{1}{r} (m_2 e^{i \phi})^r \right] \exp \left[\sum_{r=1} \frac{1}{r} (m_2^* e^{-i \phi})^r \right].
    \end{align}    
    Taking $z=e^{i\phi}$, the $PE$ simplifies to
    \begin{align}
    PE = &\exp \left[\sum_{r=1} \frac{1}{r} (m_1 z)^r \right] \exp \left[\sum_{r=1} \frac{1}{r} \left(\frac{m_1^*}{z}\right)^r \right] \nonumber \\
    &\exp \left[\sum_{r=1} \frac{1}{r} (m_2 z)^r \right] \exp \left[\sum_{r=1} \frac{1}{r} \left(\frac{m_2^*}{z}\right)^r \right]\,.
    \end{align}
    Using \eqref{log1}, 
    % \begin{equation}
    %     \sum \frac{x^k}{k} = -\ln(1-x)\,,
    % \end{equation}
    the $PE$ becomes
     \begin{align}
    PE = &\exp \left[(-\ln{(1- m_1 z)})\left(-\ln{\left(1- \frac{m_1^*}{z} \right)}\right) (-\ln{(1- m_2 z)})\left(-\ln{\left(1- \frac{m_2^*}{z} \right)}\right) \right]\,.
    \end{align}
    Using the property that $e^{\ln{x}}=x$, we obtain
    \begin{equation}
        PE= \frac{1}{(1-m_1 z) \left( 1-\frac{m_1^*}{z} \right)(1-m_2 z) \left( 1-\frac{m_2^*}{z} \right)}\,.
    \end{equation}
    Writing $q_1, q_2,q_3,q_4$ instead of $m_1, m_1^*,m_2, m_2^*$, we have
    \begin{equation}
        PE= \frac{z^2}{(1-q_1 z) (z-q_2)(1-q_3 z) ( z-q_4)}. 
    \end{equation}
    Now, considering \eqref{H(q)} and the Haar measure for $U(1)$ group given in \eqref{haar:u1}, we can calculate the Hilbert series as
    \begin{equation}
         H(q_1,q_2,q_3,q_4)= \frac{1}{2 \pi i} \int_{|z|=1} \frac{dz}{z} \frac{z^2}{(1-q_1 z)(z-q_2) (1-q_3 z)(z-q_4)}.
    \end{equation}
    Considering the residues of the poles at $z=q_2, z=q_4$, we derive the Hilbert series
    \begin{equation}
        H(q_1,q_2,q_3,q_4)=  \frac{1-q_1 q_2 q_3 q_4}{(1-q_1 q_2) (1-q_2 q_3)(1-q_1 q_4)(1-q_3 q_4)}\,. \label{HSnr}
    \end{equation}
    The denominator of the Hilbert series is particularly insightful as it indicates the invariants $I_1$ through $I_4$. Meanwhile, the numerator compensates for the syzygy $I_1 I_2 = I_3 I_4=|m_1|^2|m_2|^2$: considering only the denominator, we would obtain in the expanded series a coefficient 2 for $q_1q_2q_3q_4$ coming from $q_1q_2 \times q_3q_4$ ($I_1 \times I_2$) and from $q_1q_4 \times q_2q_3$ ($I_3 \times I_4$) but the same factor in the numerator subtracts the coefficient to 1 which is the correct number.

    The \textit{unrefined} Hilbert series is
    \begin{equation}
        H(q)=\frac{1+q^2}{(1-q^2)^3} \label{hq}. 
    \end{equation}
    This series features a palindromic numerator with $d_N= 2$ and a denominator with $d_D=6$ and $p=3$, where $p$ equals both  the number of denominator factors and the number of parameters: $|m_1|,|m_2|$ and the relative phase between $m_1$ and $m_2$. The space of spurions $[m_1, m_1^*, m_2, m_2^*]$ comprise a linear space of $\dim V = 4$. Knop's theorem \eqref{knop} is then satisfied as $4 \geq 6 - 2 \geq 3$. Specifically, the denominator of the series \eqref{hq} indicates the existence of three primary invariants of degree 2. These primary invariants are algebraically independent and form the foundational elements of our invariant structure.  Additionally, the numerator in the equation reveals the presence of a basic invariant of degree 2. This basic invariant, represented by the term $q^2$ in the numerator, complements the primary invariants. This can be confirmed by  expanding the equation \eqref{hq} into a series in $q$: 
    \begin{equation}
        H(q)= 1 + 4 q^2+\cdots.
    \end{equation}
    This expansion indicates that there are four invariants of degree two, but only three of them are algebraically independent. The relation between these four basic invariants is given by the syzygy which is highlighted by the factor in the numerator.
    
    We can reach the same conclusion from the $PL$ function of the Hilbert series: 
    \begin{equation}
    PL(q) = 4q^2 - q^4\,.
    \end{equation}
    It indicates that there are four basic invariants of degree 2, though they are not all algebraically independent. The negative sign in the $PL$ corresponds to the fourth degree syzygy \eqref{syzygy}.
   
    To make an analogy to the quark invariants, it is interesting to analyze the CP properties of these invariants. The primary invariants can also be chosen as $I_1,I_2, I_3+I_4$, which are CP even. The fourth invariant can be chosen as $I_3-I_4$ which is CP odd, yet higher powers of this invariant can be related to the CP even invariants as shown in:
    \begin{equation}
        (I_3 - I_4)^2 = (I_3 + I_4)^2 - 4 I_3 I_4 = (I_3 + I_4)^2 - 4 I_1 I_2\,. \label{i3i4}
    \end{equation}  
    
    This syzygy illustrates the intricate relationship between these invariants. Note that $(I_3-I_4)^2$ is CP even and then it is expected that it can be expanded using the other CP even invariants.

% \red{O quadrado de um CP ímpar é par.}

% In summary, this theoretical model with two couplings under $U(1)$ symmetry provides insights into the structure of basic and primary invariants, their algebraic interdependencies, and the implications for CP symmetry within the framework.
    In summary, this theoretical model with two couplings under $U(1)$ symmetry illustrates the role of basic and primary invariants, their algebraic interdependencies, and how the CP properties of the invariants help us understand their relations.

    % These are invariants that remain invariant under CP transformation (CP-even),
    % \begin{itemize}
    %     \item $I_1=m_1 m_1^*$,
    %     \item $I_2=m_2 m_2^*$,
    %     \item $I_3=m_1 m_2^*$,
    %     \item $I_4=m_2 m_1^*$.
    % \end{itemize}
    %  Each of these invariants is the product of a term with its complex conjugate, rendering them real and unchanged under CP transformation. This real nature implies their conservation of CP symmetry.
     
    % Among the identified invariants, one exhibits CP-odd behavior, specifically $I_3 - I_4$. This invariant differs from the CP-even invariants as it changes sign under CP transformation, indicating a violation of CP symmetry.

\section{Quarks Invariants and Hilbert Series Analysis}
\label{sec:invs:quarks}

In this section, we delve into the formal construction of invariants within the quark sector. Our primary tool for this analysis will be the Hilbert series, which will aid in identifying both primary and basic invariants. By utilizing the Hilbert series, we can systematically establish the structure of these invariants and subsequently verify the quantity and degree of these invariants through the Plethystic logarithm.

We seek invariants of group
\begin{equation}
SU(3)_q\otimes SU(3)_u\otimes SU(3)_d\,, \label{group:SM,0,0}
\end{equation}
where the left-handed quark doublets $q_L$ transform as a triplet under $SU(3)_q$ and the right-handed quark singlets of up-type ($u_R$) and of down-type ($d_R$) transform as triplets under $SU(3)_u$ and $SU(3)_d$, respectively, which represent independent unitary transformations among the three generations of quarks. The explicit transformations are
\begin{equation}
        q_{iL} \to (U^q_{L})_{ij} q_{jL}; \quad 
        d_{iR} \to  (U^d_{R})_{ij} d_{jR}; \quad 
        u_{iR} \to (U^u_{R})_{ij} u_{jR};  
    \label{WB:->}
\end{equation}
where $U_L^q\in SU(3)_q, U^u_{R}\in SU(3)_u, U^d_{R}\in SU(3)_d$ and we have rewritten \eqref{WB}. In the SM, without Yukawa interactions, the group \eqref{group:SM,0,0} of transformations is a global symmetry of the quark sector.
The introduction of Yukawa couplings $Y^u$ and $Y^d$ breaks this symmetry explicitly. However, by treating the Yukawa matrices as spurions that transform under these symmetries, we can effectively restore these symmetries in the Lagrangian and track how this group of flavor symmetries is broken by the Yukawa interactions. WBT of the Yukawa couplings were given in \eqref{WBT}. Here, in contrast, the Yukawa Lagrangian \eqref{yukawa} is invariant by the transformation of fields \eqref{WB:->} if the spurions $Y^u,Y^d$ transform as
    \begin{equation}
        \begin{aligned}
            Y^u \to U^q_L Y^u U^{u\dagger}_R, \\
            Y^d \to U^q_L Y^u U^{d\dagger}_R.
        \label{Yu.Yd:U}
        \end{aligned}
    \end{equation}
So, under \eqref{group:SM,0,0},

\begin{align}
    Y^u\sim (3,\bar{3},1), \nonumber\\
    Y^d\sim (3,1,\bar{3}). \label{Yu:Yd}
\end{align}
It is easy to cancel out the transformations under $SU(3)_u$ and $SU(3)_d$ by considering $X_u,X_d$ in \eqref{Xu.Xd}. They are nontrivial only under $SU(3)_q$ and transform as
\begin{equation}
         \begin{aligned}
             X_u &\to U^q_L X_u U^{q\dagger}_L\,, \\
             X_d &\to U^q_L X_d U^{q\dagger}_L\,.
        \end{aligned}
    \label{sm:X.transf}
    \end{equation} 
Their representation is
    \begin{equation}
        X_u\sim X_d\sim (3 \otimes \Bar{3},1,1)\,,
    \end{equation}
where $3 \otimes \Bar{3} = 1 \oplus \text{adj}$.

\subsection{Two Quark Families}
\label{sec:SM:n=2}

  Exploring the quark sector for two quark families $(u,c)$ and $(d,s)$, we focus on constructing invariants through the Hilbert series, a powerful tool that allows the identification of both primary and basic invariants. These invariants are traces of matrices $X_u$ and $X_d$:
        \begin{equation}            
        \aver{X_u}, \aver{X_d}, \aver{X_u^2}, \aver{X_d^2}, \aver{X_u X_d}, \cdots \label{trace:xuxd}
        \end{equation}
    We will denote the trace as $Tr[A]=\aver{A}$\,\cite{jenkins_algebraic_2009}. The problem is to establish what are the basic invariants from which all invariants can be constructed.
    
    The number and degree of the invariants is given by the Hilbert series which can be calculated through the Molien-Weyl formula:
    \begin{equation}
        H = \int_{SU(2)} d\mu \cdot PE[X_{u,d}, R].
    \end{equation}  
    Using the Haar measure for $SU(2)$ in \eqref{su2} and the characters from Table \ref{tab:char}, the formula can be written as
    \begin{equation}
        H =  \frac{1}{2\pi i} \int \frac{dz}{z} (1-z^2) \cdot PE[X_{u,d}, R].
    \end{equation}
    The $PE$ is given by
    \begin{equation}
        PE[X_{u,d}, R] = \left[ \exp{\sum_r \frac{X_u^r}{r}} \left( z^{2r} + \frac{1}{z^{2r}} + 2 \right) \right] \cdot \left[ \exp{\sum_r \frac{X_d^r}{r}} \left( z^{2r} + \frac{1}{z^{2r}} + 2 \right) \right]. 
    \end{equation}   
    Using again the relation \eqref{log1}, the Plethystic exponential can be written as
    \begin{equation}
        PE[X_{u,d}, R] = \frac{1}{(1- X_u z^2) \left(1-\frac{X_u}{z^2}\right) (1-X_u)^2 (1- X_d z^2) \left(1-\frac{X_d}{z^2}\right) (1-X_d)^2}.
    \end{equation}
    So the formula becomes
    \begin{equation}
        H =  \frac{1}{2\pi i} \int_{|z|=1} dz (1-z^2) \frac{z^3}{(1- X_u z^2) (z^2-X_u) (1-X_u)^2 (1- X_d z^2) (z^2-X_d) (1-X_d)^2}
    \end{equation}
    where the poles within the unit circle are $z= \pm \sqrt{X_u}, z= \pm \sqrt{X_d}$. We derive the refined Hilbert series:
     \begin{equation}
        H(u,d) = \frac{1}{(1-u^2) (1-u^4) (1-d^2) (1-d^4) (1-u^2 d^2)\,,} \label{H(u,d)}
    \end{equation}
    which agrees with Ref.\,\cite{jenkins_algebraic_2009}. Here $X_u \equiv u^2$ and $X_d \equiv d^2$. So that the degree refers to the degree of $Y^{u,d}$.

    To find the \textit{unrefined} series, simply perform the substitution, $u,d \to q$: 
     \begin{equation}
         H(q) = \frac{1}{(1-q^2)^2 (1-q^4)^3}\,.\label{2quarks}
     \end{equation}
     Equation \eqref{2quarks} tells us that there are 2 primary invariants of degree 2 and 3 primary invariants of degree 4. From the numerator, we also conclude that there are no syzygies and all basic invariants are primary and algebraically independent.

    We can further confirm this through the $PL$, which yields the result:
    \begin{equation}
    PL(q) = 2q^2 + 3q^4.
    % + O[q]^7,
    \end{equation}
    This result corroborates the presence of two basic invariants of degree 2 and three basic invariants of degree 4. Moreover, the absence of negative terms in the series also indicates that there are no syzygies present.
    
    With the Hilbert series \eqref{2quarks}, we now know the number and degree of the basic and primary invariants. We can now search for the actual invariants, which are the traces \eqref{trace:xuxd}. It is easy to list the invariants shown in Table \ref{tab:inv:n=2}. We also show them in terms of physical parameters calculated in the $u$-diagonal basis \eqref{diag1} and \eqref{diag2}. The physical parameters are the Yukawas $y_u,y_c,y_d,y_s$ and the Cabibbo angle $\theta$. The calculation is explained in Appendix \ref{xuxd}. From these functions, it is clear that these invariants are independent. It is also clear that these five basic invariants are in one-to-one correspondence to the physical parameters.
    
\begin{table}[H]
\centering
\caption{\label{tab:inv:n=2}
Basic flavor Invariants for 2 quark families with their degrees and CP parity. 
}
\begin{tabular}{|c|c|c|}
\hline
Flavor Invariant & Degree & CP \\
\hline
$I_{2,0} = \aver{X_u} = y_u^2 + y_c^2$ & 2 & + \\
$I_{0,2} = \aver{X_d} = y_d^2 + y_s^2$ & 2 & + \\
$I_{4,0} = \aver{X_u^2} = y_u^4 + y_c^4$ & 4 & + \\
$I_{0,4} = \aver{X_d^2} = y_d^4 + y_s^4$ & 4 & + \\
$I_{2,2} = \aver{X_u X_d} = y_u^2 y_s^2 + y_c^2 y_d^2 + (y_s^2 - y_d^2)(y_c^2 - y_u^2) \cos^2(\theta)$ & 4 & + \\
\hline
\end{tabular}
\label{tab:flavor_invariants}
\end{table}    
    The space of the original variables $Y^u,Y^d$ has dimension $\dim V=16$ because each of them is a generic complex $2 \times 2$ matrix. From \eqref{2quarks} observe that $p=5$, corresponding to the five physical parameters: four Yukawas $y_u,y_c,y_d,y_s$ and one angle $\theta$ in the CKM. Still, from \eqref{2quarks} the numerator has degree $d_N=0$ and the denominator has $d_D=\sum_r d_r= 16$. Knop's theorem is satisfied as $16 \geq 16-0 \geq 5$.
    If we consider $X_u$ and $X_d$ as our spurions,
    $\dim V$ is reduced to 8 because they are hermitian matrices, and we should replace $q^2 \to q$ in the Hilbert series such that $d_D=8$ and $d_N=0$ is unmodified. So the Knop's theorem is also satisfied as $8 \geq 8-0 \geq 5$.
    
    As all invariants from Table \ref{tab:inv:n=2} are real, it implies the absence of CP odd invariants. This absence of CP odd invariants directly implies that our theoretical model does not accommodate CP violation.
   
    Given the basic invariants in Table\,\ref{tab:inv:n=2}, we need to analyze how the invariants of higher degree in \eqref{trace:xuxd} can be written as polynomials of these basic invariants\,\cite{jenkins_algebraic_2009}. The key relation is the Cayley-Hamilton theorem\,\footnote{Detailed in Appendix \ref{Cayley-H}.} which states for a $2\times 2$ matrix that
\begin{equation}
            A^2 = \aver{A} A + \frac{1}{2}[\aver{A^2} - \aver{A}^2] \mathbb{I}.
        \label{CH:n=2}
        \end{equation}
Taking the trace of this equation yields trivially $\aver{A^2} = \aver{A^2}$.
Further, when we multiply by $A$ and take the trace again, we find:
    \begin{align}
        \aver{A^3} &=  \frac{3}{2}\aver{A} \aver{A^2} - \frac{1}{2} \aver{A}^2 \aver{A}. 
    \label{CH:n=2:A3}
    \end{align}

    Thus, for $2 \times 2$ matrices, $\aver{A^n}$ for $n \geq 3$ , can be expressed in terms of $\aver{A},\aver{A^2}$. Therefore, the independent invariants involving only $X_u$ are $\aver{X_u}, \aver{X_u^2}$, and similarly for $X_d$. 
    
    Invariants involving both $X_u$ and $X_d$ can be formulated as \cite{jenkins_algebraic_2009}
    \begin{equation}
        \aver{X_u^{r_1} X_d^{s_1} X_u^{r_2} X_d^{s_2} \cdots}, \quad \text{for} \quad r_i, s_i
        < 2,
        \label{invar}
        % = \text{integers}. 
    \end{equation}
    where powers of $X_u$ or $X_d$ with exponents $r_i,s_i\ge 2$ can be reduced to lower powers according to the Cayley-Hamilton theorem \eqref{CH:n=2}.
    We are left with powers $\aver{(X_uX_d)^r}$ because any other combination will lead to $X_u^2$ or $X_d^2$.
    Due to \eqref{CH:n=2:A3}, we can exclude $r\ge 3$.
    
    In appendix \ref{identity} we show that the invariant with $r=2$ can be also written in terms of lower degree invariants. This reduction process highlights $\aver{X_u X_d}$ as the sole independent invariant involving both $X_u,X_d$. We conclude that all invariants can be written as a polynomial of the basic invariant in Table~\ref{tab:flavor_invariants}.
    
\subsection{Three quarks families}

    Similarly to the case with two families, the invariants for three families are the traces of the matrices $X_u$ and $X_d$ as given by equation \eqref{trace:xuxd}.

    Constructing the Hilbert series from Molien-Weyl formula and using the Haar measure for $SU(3)$ and the characters from Table \ref{tab:char}, we have
    \begin{equation}
        H = \int_{SU(3)} d\mu \cdot PE[X_{u,d},R].
    \end{equation}
   The construction of $PE$ from the characters leads to
    \begin{align}
    \small
    &PE[X_{u,d},R] =& \nonumber \\    
    &\exp{\left[ \sum_r \frac{X_u^r}{r} (z_1 z_2)^r \left( \frac{z_2^2}{z_1} \right)^r \left( \frac{z_1^2}{z_2} \right)^r + 3+ \left( \frac{z_2}{z_1^2} \right)^r \left( \frac{z_1}{z_2^2} \right)^r \left( \frac{1}{z_1 z_2}\right )^r \right]} \nonumber \\
    &\cdot \exp{\left[ \sum_r \frac{X_d^r}{r} (z_1 z_2)^r \left( \frac{z_2^2}{z_1} \right)^r \left( \frac{z_1^2}{z_2} \right)^r + 3 + \left( \frac{z_2}{z_1^2} \right)^r \left( \frac{z_1}{z_2^2} \right)^r \left( \frac{1}{z_1 z_2}\right )^r \right]}.
    \end{align}    
    Using \eqref{log1}, the $PE$ becomes:
    \begin{align}
    \small
      &PE[X_{u,d},R]=& \nonumber\\
      &\frac{z_1 z_2 z^2_1 z^2_2 z_1 z_2}{(1 - X_u z_1 z_2)(z_1 - X_u z^2_2)(z_2 - X_u z^2_1)(1-X_u)^3 (z^2_1 - X_u z_2)(z^2_2 - X_u z_1)(z_1 z_2 - X_u)} \times \nonumber \\
        \quad & \nonumber \\
        & \frac{z_1 z_2 z^2_1 z^2_2 z_1 z_2}{(1 - X_d z_1 z_2)(z_1 - X_d z^2_2)(z_2 - X_d z^2_1)(1-X_d)^3 (z^2_1 - X_d z_2)(z^2_2 - X_d z_1)(z_1 z_2 - X_d)} 
    \end{align}
    Then the formula can be simplified to
    \begin{equation}
    \begin{split}
        H =& \frac{1}{(2 \pi i)^2} \oint_{|z=1|} dz_1 \oint_{|z=1|} dz_2 (1-z_1 z_2) (z_2- z^2_1) (z_1 -z^2_2) z^6_1 z^6_1 \times \\
        \quad & \\
        &\frac{1}{(1 - X_u z_1 z_2)(z_1 - X_u z^2_2)(z_2 - X_u z^2_1)(1-X_u)^3 (z^2_1 - X_u z_2)(z^2_2 - X_u z_1)(z_1 z_2 - X_u)} \times \\
        \quad & \\
        & \frac{1}{(1 - X_d z_1 z_2)(z_1 - X_d z^2_2)(z_2 - X_d z^2_1)(1-X_d)^3 (z^2_1 - X_d z_2)(z^2_2 - X_d z_1)(z_1 z_2 - X_d)}.     
    \end{split}
    \end{equation}
    Replacing $X_u\to u^2$, $X_d\to d^2$, the integral leads to the Hilbert series multi-graded :
    \begin{equation}
    \footnotesize
         H(u,d)=\frac{1+u^6 d^6}{(1-u^2)(1-u^4)(1-u^6)(1-d^2)(1-d^4)(1-d^6)(1-u^2 d^2)(1-u^4 d^2)(1-u^2 d^4)(1-u^4 d^4)}
    \end{equation}
    Taking $u,d \rightarrow q$, we obtain the \textit{unrefined} Hilbert series:    \begin{equation}
         H(q)=\frac{1+q^{12}}{(1-q^2)^2(1-q^4)^3(1-q^6)^4(1-q^8)}. \label{H3family}
    \end{equation}
    The denominator reveals  the presence of two primary invariants of degree two, three of degree four, four of degree six, and one of degree eight\,\cite{jenkins_algebraic_2009}. The $p=10$ factors in the denominator should match the number of physical parameters. The latter encompasses six quark masses originating from the Yukawa matrices $Y^u$ and $Y^d$ (specifically $y_u, y_c, y_t, y_d, y_s, y_b$), three angles from the CKM matrix ($\theta_{12}, \theta_{13}, \theta_{23}$), and one CP-violating phase ($\delta$). In the space of the original variables $Y^u,Y^d$, the dimension $dim V=36$, accounts for the two $3\times 3$ matrices and their complex conjugates.
    Considering the degrees of the numerator and denominator, $d_N=12$ and $d_D=48$, Knop's theorem is satisfied as $36 \geq 48 - 12 \geq 10$. We could also consider $X_u$ and $X_d$ as the basic spurions. In this case, we should make the replacement $q^2 \to q$ in the Hilbert series, in which case $d_N = 6$ and $d_D = 24$. The dimension $dim V$ is reduced to 18 because $X_u,X_d$ are both $3\times 3$ and Hermitian. Again, Knop's theorem is satisfied as $18 \geq 24 - 6 \geq 10$. 
    
    % Basic Invariants:  
    We can determine the number and degree of the basic invariants through the examination of the $PL$ \cite{ Bento:2023owf, Bento:2021hyo}:
\begin{equation}
        PL(q)=2 q^2+3 q^4+4 q^6+q^8+q^{12}-q^{24}\,.
        \label{PL 3quarks}
    \end{equation}
We can count 11 basic invariants of the corresponding degrees. It is notable that the syzygy arises only at $q^{24}$, because is first term negative of the series. The Hilbert series numerator solely presents an entry at $q^{12}$, lacking any $q^{24}$ term. The term $(1-q^{12})$ is not present in the denominator since it represents a dependent invariant, as we will discuss below. The 11 basic invariants are known\,\cite{jenkins_algebraic_2009} and we list them in Table\,\ref{tab:basic:n=3}, together with their degree and CP parities. For the invariants that depend solely on either $X_u$ or $X_d$, we also show the expressions in terms of physical parameters. It is easy to see that the first 6 invariants correspond to the 6 Yukawas. We use the notation $I_{2n,2m}$ for an invariant of degree $X_u^n$ and $X_d^m$. The last five invariants depend also on the CKM matrix $V$. In Appendix \ref{Invariants: i22} we show that dependence utilizing the standard parametrization for $V$. From these expressions in terms of physical parameters, it is evident that the first 10 invariants mentioned correspond to the 10 physical parameters of the Standard Model in the Yukawa sector: the 6 Yukawas and four parameters in the CKM matrix.

We now discuss how the Cayley-Hamilton theorem can be used to write any invariant in terms of the basic invariants in Table\,\ref{tab:basic:n=3}. The Cayley-Hamilton theorem for a $3\times 3$ matrix $A$ is expressed as
     \begin{equation}
        A^3 = \aver{A}A^2 - \frac{1}{2}[\aver{A}^2 - \aver{A^2}]A + \frac{1}{6}[\aver{A}^3 - 3 \aver{A}\aver{A^2} + 2 \aver{A^3}]\mathbb{I}_{3\times 3}.
    \label{n=3:CH}
    \end{equation} 
    Taking the trace leads to the trivial result $\aver{A^3} = \aver{A^3}$. Multiplying by $A$ and taking the trace again, we derive
      \begin{equation}
        \aver{A^4} =  \frac{1}{6}\aver{A}^4 -\aver{A}^2 \aver{A^2} + \frac{4}{3} \aver{A^3}\aver{A} + \frac{1}{2}\aver{A^2}^4. 
    \label{A4}
    \end{equation}
    This implies any $\aver{A^n}$, with $n\geq 4$, can be rewritten in terms of  $\aver{A}, \aver{A^2}, \aver{A^3}$. Hence, for three families, for invariants involving only $X_u$, we only need to consider $\aver{X_u^n}$ up to degree 3, i.e., $\aver{X_u}, \aver{X_u^2}, \aver{X_u^3}$. The case of invariants involving only $X_d$ is similar.
% \celso{
% For instance,
% **{}
%     , for $3\times 3$ matrix $X$, the theorem facilitates the following decomposition:
% \begin{equation}
%         X^3 = \aver{X}X^2 - \frac{1}{2}[\aver{X}^2 - \aver{X^2}]X + \frac{1}{6}[\aver{X}^3 - 3 \aver{X}\aver{X^2} + 2 \aver{X^3}]\mathbb{I}_{3\times 3}.
%     \label{n=3: xu}
%     \end{equation} 
% Taking $X\to X_u$ and multiply \eqref{n=3: xu} by $X_d$ and taking the trace simplifies the expression to involve invariants of lower orders
% \begin{equation}
%         \aver{X_u^3 X_d} = \aver{X_u}\aver{X_u^2 X_d} - \frac{1}{2}[\aver{X_u}^2 - \aver{X_u^2}]\aver{X_u X_d} + \frac{1}{6}[\aver{X_u}^3 - 3 \aver{X_u}\aver{X_u^2} + 2 \aver{X_u^3}]\aver{X_d}.
%     \label{n=3: xu:xd}
%     \end{equation} 
% In \eqref{n=3: xu:xd} we can find the following invariants:
% \begin{equation}
%     \aver{X_u}, \quad \aver{X_u^2}, \quad \aver{X_u^3}, \quad \aver{X_u X_d},  \quad \aver{X_u^2 X_d}, \quad \aver{X_d}.  
% \end{equation}
% To obtain the other irreducible invariants, simply multiply \eqref{n=3: xu} by $X_d^2, X_d^3$ and $YY^\dagger$.
% \red{Isso aqui seria no meio do parágrafo seguinte. Mas exige um esforço colocar.}

    Invariants involving both $X_u$ and $X_d$ follow the form of equation \eqref{invar}, with $r_i = 1, 2$ and $s_i = 1, 2$. This leads to traces of products of $X_u$, $X^2_u$, $X_d$, $X^2_d$. While this suggests an infinite number of invariants, many are not independent. For arbitrary $3 \times 3$ matrices $A$, $B$ and $C$, the Cayley-Hamilton theorem also implies a relation between the trace $\aver{ABAC}$ where the matrix $A$ is repeated with $\aver{A^2BC}$ and $\aver{A^2CB}$ where $A$ is not repeated. This identity allows us to reduce traces where the same matrix appears more than once in terms of invariants where $X_u$,$X_u^2$,$X_d$,$X_d^2$ each occurs at most once. The identity is\,\cite{jenkins_algebraic_2009} 
    \begin{align}
        0 = &\aver{A}^2 \aver{B} \aver{C} - \aver{BC} \aver{A}^2 - 2 \aver{AB} \aver{A}\aver{C} \nonumber \\
        &- 2 \aver{AC} \aver{A} \aver{B} + 2 \aver{ABC} \aver{A} + 2 \aver{ACB}\aver{A} \nonumber\\
        &- \aver{A^2} \aver{B} \aver{C} + 2 \aver{AB} \aver{AC} + \aver{A}^2 \aver{BC}\nonumber \\
        &+ 2 \aver{C} \aver{A^2 B} + 2 \aver{B} \aver{A^2 C} - 2 \aver{A^2 BC} \nonumber \\ 
        &- 2 \aver{A^2 CB} - 2 \aver{ABAC}\,.
    \label{ABAC}
    \end{align}
See Appendix \ref{identity} for details. We then obtain the invariants of Table\,\ref{tab:basic:n=3} where the uniqueness of the last degree 12 invariant needs further discussion.     
\begin{table}[H]
\centering
\caption{\label{tab:basic:n=3}
The 11 basic flavor Invariants for 3 quark families with their degrees and CP parity.}
\begin{tabular}{|c|c|c|}
\hline
Flavor Invariants & Degree & CP \\
\hline
\( I^+_{20} = \aver{X_u} =  y_u^2 + y_c^2 + y_t^2\) & 2 & + \\
\( I^+_{40} = \aver{X_u^2} = y_u^4 + y_c^4 + y_t^4\) & 4 & + \\
\( I^+_{60} = \aver{X_u^3} = y_u^6 + y_c^6 + y_t^6\) & 6  & + \\
\( I^+_{02} = \aver{X_d}= y_d^2 + y_s^2 + y_b^2\) & 2 & +  \\
\( I^+_{04} = \aver{X_d^2} =y_d^4 + y_s^4 + y_b^4\) & 4 & + \\
\( I^+_{06} = \aver{X_d^3} = y_d^6 + y_s^6 + y_b^6\) & 6 & + \\
\( I^+_{22} = \aver{X_u X_d}\) & 4 & +  \\
\( I^+_{24} = \aver{X_u X_d^2}  \) & 6 & + \\
\( I^+_{42} = \aver{X_u^2 X_d} \) & 6 & + \\
\( I^+_{44} = \aver{X_u^2 X_d^2} \) & 8 & + \\
\(I^{(-)}_{66}= \aver{X^2_U X^2_D X_U X_D} - \aver{X^2_D X^2_U X_D X_U}\) & 12&-\\
\hline
\end{tabular}
\label{tab:invariantes3damily}
\end{table}
    We can now discuss the CP properties of the basic invariants listed in Table~\ref{tab:basic:n=3}. The first ten invariants from this table are CP even. This contrasts with the final invariant in the list, which is CP odd. Looking at $I_{66}^{(-)}$, it is clear that we can write two degree 12 invariants:
     \begin{align}
        I^{(-)}_{66}= \aver{X^2_U X^2_D X_U X_D} - \aver{X^2_D X^2_U X_D X_U} \label{I-},\\
        I^{(+)}_{66}= \aver{X^2_U X^2_D X_U X_D} + \aver{X^2_D X^2_U X_D X_U},  \label{I+}
    \end{align}
which are CP odd and CP even, respectively. The term of degree 12 in the numerator of the Hilbert series \eqref{H3family} indicates that there is one invariant of degree 12 other than those given by products of denominator factors.
This corresponds to $I^{(-)}_{66}$, which is a basic invariant and cannot be reduced with \eqref{ABAC}.
Then the Hilbert series implies that the other degree-twelve invariant, $I^{(+)}_{66}$, is not a basic invariant and, indeed, it can be written as a polynomial in the other CP even invariants.
This identity is\,\cite{jenkins_algebraic_2009}
    \begin{align}
        3I^{(+)}_{6,6} &= I^{3}_{2,0} I^{3}_{0,2} - I_{2,0}I_{4,0}I^{3}_{0,2} - 3I_{2,2}I^{2}_{2,0} I^{2}_{0,2} + 3I_{4,2}I_{2,0}I^{2}_{0,2} - I_{0,4}I^{3}_{2,0}I_{0,2} \nonumber\\
        &+ 3I_{2,4}I^{2}_{2,0}I^{0}_{0,2} -3I_{4,4}I_{2,0}I_{0,2} + I_{0,4}I_{6,0}I_{0,2} + 3I_{2,4}I_{4,2} + 3I_{2,2}I_{4,4} \nonumber \\
        &+ I_{0,6}I_{2,0}I_{4,0} - I_{0,6}I_{6,0}.
    \end{align}    

Although the invariant $I^{(-)}_{66}$ cannot be written in terms of lower degree invariants, it is not algebraically independent of the other 10 basic invariants.  This is due to the fact that the square 
$(I^{(-)}_{66})^2$, which is CP even, can be written in terms of lower degree CP even invariants. 
So we have a polynomial relation involving the 11 basis invariants, including $I^{(-)}_{66}$ known as 
a syzygy. 
This syzygy involving $I^{(-)}_{66}$ is also discussed in the work \cite{Bento:2023owf}, where it is presented using a different set of invariants.

The previous discussion can be also understood in terms of physical parameters.
As shown in Appendix \ref{Invariants: i22}, the CP even invariants $I_{22},I_{24},I_{42},I_{44}$ depend only on the cosine of the CKM phase $\delta$. The sign of $\delta$ cannot be determined from these invariants.
Nonetheless, this crucial information is contained within the CP odd invariant $I^{(-)}_{66}$ which is proportional to $\sin\delta$. Indeed, 
    \begin{equation}
        I^{(-)}_{6,6}  = -\frac{1}{3} \aver{[X_u, X_d]^3}  = \det \mathcal{C} ,
        \label{I66}
    \end{equation}
is the CP odd invariant already discussed in \eqref{Invariant3} which is proportional to the Jarlskog invariant \eqref{CKM.J}.
The fact that the square $(I^{(-)}_{6,6})^2$ can be expressed in terms of CP even invariants is well-known \cite{Bento:2023owf, jenkins_algebraic_2009}: the unitarity triangle can be determined by measuring CP-preserving sides rather than angles; once the triangle's sides are known, any ambiguity is resolved by the sign of the Jarlskog invariant.

%% file: capitulos/4-VLQS.tex
\chapter{Vector-Like Quarks}

   % The Standard Model and Its Limitations: 
   % While the 
   The SM of particle physics, described by the gauge group $SU(3)_c \times SU(2)_L \times U(1)_Y$,  has been immensely successful in corroborating experimental findings. However, the plethora of free parameters and several unresolved questions suggest that the SM is not the ultimate theory. This realization underscores the necessity to explore realms beyond the SM.
   
   From a bottom-up perspective, one of the simplest yet profound extensions of the SM involves the introduction of Vector-Like Quarks (VLQs). This extension has garnered significant interest, particularly due to the potential contributions of VLQs to new physics phenomena, which are expected to be observable in experiments conducted at the Large Hadron Collider (LHC). VLQs are theorized to play a crucial role in elucidating physics beyond the established SM framework. Their unique properties and interactions could unveil new phenomena, especially in high-energy particle collisions. For a review, see Refs.\,\cite{alves_vector-like_2023,aguilar-saavedra_handbook_2013}.

% \red{
% Movi a citação do paragrafo 1 para o 2.
% Na citação acima, incluí o review to Saavedra.}

% \red{
% O livro do Branco 
% \cite{Branco:1999fs},
% voce cita em algum parágrafo falando de CP.
% }

% Integration of VLQs into the SM:
    % One notable extension of the SM involves the addition of VLQ. The integration of VLQs into the SM heralds a promising avenue for explicating the mass hierarchies of particles. Their presence could significantly influence the mass spectrum of other particles, offering insights into why certain particles manifest greater mass than others. Furthermore, VLQs might affect decay amplitudes and contribute to Charge-Parity (CP) violation. These models have been explored through the study,  including those by Branco and Alves,  have delved into how heavy isosinglet quarks might affect CP-violating asymmetries in neutral-B decays, considering scenarios dominated by Z-exchange in $B^0 - \overline{B^0}$ mixing and standard box-diagram contributions.

% \red{Evite esse negócio de colocar subtítulos.}

    % Gauge Theories and VLQs: 
    % In this context of gauge theories,
    VLQs are envisaged as hypothetical particles of spin $1/2$, with a distinct characteristic: both left- and right-handed components posses identical gauge quantum numbers and undergo similar transformations under the SM gauge group $SU(3)_c \times SU(2)_L \times U(1)_Y$. 
    Particularly, we can highlight the following properties for isosinglet VLQs:
    \begin{enumerate}
        \item  Isosinglet VLQs, which share the same quantum numbers as the SM right-handed quarks, allows for the natural mixing between VLQs and SM quarks, suppressed by the electroweak scale over the mass scale of the VLQs. This mixing alters the VLQs interactions with the SM gauge bosons.
        \item VLQs, unlike Standard Model quarks, do not obtain their masses via Yukawa couplings to a Higgs doublet.  Instead, their mass generation mechanism is fundamentally different, with their mass scale not tied to the weak scale.        
        \item The potential mixing of VLQs with SM quarks can modify their interactions with the Z, W, and Higgs bosons. Notably, singlet VLQs primarily contribute corrections to the left-handed currents coupled to the Z and W bosons, akin to triplet VLQs but with an opposite sign.
    \end{enumerate}

  VLQs  play a pivotal role in models featuring spontaneous CP violation (SCPV) where the Lagrangian maintains CP symmetry while the vacuum state breaks CP spontaneously. By expanding the scalar sector to include a complex singlet, SCPV can be realized, allowing for the generation of a complex CKM matrix. This is crucial for matching experimental evidence which points to a non-zero and distinct phase $\gamma$ in the CKM matrix \cite{branco_geometrical_1984, branco_vector-like_2022, botella_new_2005, collaboration_utfit_2006}. 
  
   Furthermore, VLQs offer solutions to longstanding puzzles in particle physics, such as the Strong CP problem, providing an intriguing alternative to the axion model. The proposal by Barr and Nelson\,\cite{nelson_naturally_1984, barr_solving_1984}, and later, Bento, Branco, and Parada\,\cite{bento_minimal_1991}, shows that the Strong CP problem can be solved within a minimal model framework, underscoring the versatility of VLQs in resolving fundamental issues in theoretical physics.
   
   VLQs are also compatible with the concept of a “desert” between the electroweak scale and a higher mass scale such as the Grand Unified Theory (GUT) scale, contributing to the unification of coupling constants without necessitating Supersymmetry\,\cite{seng_reduced_2018}. 
   
   VLQs might also offer a potential explanation for the observed deviations from $V_{CKM}$ unitarity. This suggests a broader framework for understanding particle interactions beyond the Standard Model, as discussed in references \cite{dermisek_unification_2013, seng_reduced_2018}.
 
  In experimental contexts, VLQs can be pair-produced at hadron colliders due to their interactions with gluons, leading to distinctive decay signatures into Standard Model particles. These decays, governed by the gauge quantum numbers of the VLQs, offer insights into the properties and potential interactions of VLQs with SM particles, providing a unique window into physics beyond the Standard Model \cite{aguilar-saavedra_handbook_2013}.

   A significant challenge within the SM is its inability to account for the observed baryon asymmetry of the Universe (BAU) through CP-violating effects. In contrast, the extended SM with down-type VLQs introduces CP-odd weak basis invariants (WBIs) at a lower mass order, suggesting that VLQ models could play a crucial role in generating the necessary CP violation for BAU, highlighting the potential of VLQs to address one of the major unsolved problems in cosmology \cite{albergaria_cp-odd_2023}.

% \red{
% A discussão dentro do MP já foi feita em outro capítulo. Cite a seção.
% }

\section{Vector-like quarks that couple to the SM}

    If the scalar sector only includes $SU(2)_L$ doublets, as in the SM, new VLQs coupling to SM quarks. With renormalizable couplings can appear in seven gauge multiplets of the gauge group $SU(3)_C\otimes SU(2)_L\otimes U(1)_Y$. These multiplets are categorized as singlets, doublets, and triplets of $SU(2)_L$ \cite{aguilar-saavedra_handbook_2013, del_aguila_effective_2000, del_aguila_observable_2000}:

   \begin{table}[h]
\centering
\caption{ Representation of VLQ Multiplets: This table categorizes the VLQ multiplets that interact with SM quarks via Yukawa couplings, detailing 2 singlets, 3 doublets, and 2 triplets.}
\begin{tabular}{|c|c|c|c|c|c|c|c|}
\hline
Multiplet & $U_{L,R}$ & $D_{L,R}$ & $\begin{pmatrix}
    U\\
    D
\end{pmatrix}_{L,R}$ & $\begin{pmatrix}
    X\\
    U
\end{pmatrix}_{L,R}$ & $\begin{pmatrix}
    D\\
    Y
\end{pmatrix}_{L,R}$ & $\begin{pmatrix}
    X\\
    U\\
    D
\end{pmatrix}_{L,R}$ & $\begin{pmatrix}
    U\\
    D\\
    Y
\end{pmatrix}_{L,R}$\\ \hline
 $SU(2)_L$ & 1 & 1 & 2 & 2 & 2 & 3 & 3 \\ \hline
\end{tabular}
\label{Rep:multiplets}
\end{table}
\noindent The new fields $U$, $D$ have electric charges $2/3$ and $-1/3$, respectively. Note that some multiplets might include quarks $X$ with an electric charge of $5/3$, and $Y$ with a charge of $-4/3$.  In scenarios with only one such multiplet, the weak and mass eigenstates of these quarks $X$ or $Y$ coincide. In this dissertation, we will focus on extensions of the SM  with only one singlet  down-type VLQ $D_L,D_R$. We will often denote them as $B_L,B_R$.

    In general, as the field $D_R$ has the same quantum numbers as the SM $d_R$ fields, the new fields will mix with the SM fields and the physical states of down-type quarks will be mixtures of the SM quarks and the new VLQ. This will give rise to four physical down-type quark mass eigenstates $(d, s, b, D)$ of charge $-1/3$.

\section{One singlet VLQ of down-type}

    In the rest of the dissertation, we will focus on the SM extended by the addition of an isosinglet quark $B_L,B_R$ 
    % ($n_d=1$) 
    of down type. Then the colored fields of the model are
     \begin{equation}
         q_{iL}=
        \begin{pmatrix}
            u_{iL} \\
            d_{iL}
        \end{pmatrix} ;       
        \quad u_{iR} ; \quad d_{iR} ;  \quad B_{L}; \quad B_{R}.
    \label{fields:vlq}
    \end{equation}
    % where $q_{iL}$ represents the left-handed quark doublets for the three generations (indexed by $i$). The right-handed counterparts are singlets: $u_{iR}$ for up-type and $d_{iR}$ for down-type quarks. 
    The quantum numbers under $SU(3)_c\otimes SU(2)_L\otimes U(1)_Y$ of these fields are,
    \begin{equation}
        q_L \sim (3, 2, 1/6); \quad u_r \sim (3,1,2/3); \quad d_R \sim (3, 1,-1/3) \sim B_R \sim B_L. 
    \end{equation}
    This notation means that the left-handed quarks ($q_{iL}$) are doublets under $SU(2)_L$ with a hypercharge of $1/6$, while the Right-handed down-type quarks ($d_{iR}$) and the VLQs ($B_R$ and $B_L$) are singlets with hypercharges of $-1/3$.

% \red{Não precisa de $n_d=1$. Não foi definido. Se precisar depois, introduza depois. O $n$ já é 3.}
    
    % In the specific scenario of introducing a single VLQ ($n_d=1)$ into the SM with three families ($n=3$), 
    The general Yukawa Lagrangian can be expressed as:
    \begin{equation}
        -\mathcal{L} = \overline{q_L} Y^u \Tilde{H}u_R + \overline{q_L}\Tilde{Y}^d H \begin{pmatrix}
            d_R\\
            B_R
        \end{pmatrix} + \overline{B_L}\Tilde{M}^B \begin{pmatrix}
            d_R\\
            B_R
        \end{pmatrix}. 
        \label{lagragian 1vlq}
    \end{equation}
        % The components of this Lagrangian for the case of a single VLQ ($n_d = 1$) and three SM families are detailed as follows:
    The Yukawa couplings and mass matrices of this Lagrangian are as follows:
    \begin{itemize}
        \item $Y^u$ is the Yukawa coupling matrix for up-type quarks of size $3\times 3$.
        \item $\Tilde{Y}^d$ represents the extended Yukawa coupling matrix for down-type quarks, including interactions with the VLQ. This matrix is a $3 \times (3+1)$ matrix.
        \item $\Tilde{M}^B$ is the bare mass matrix for the VLQs
        of size $1 \times (3+1)$.
        % matri
    \end{itemize}
    % Such that, in the case of adding just one VLQ to the SM with three families, the matrices take the form:    
    % \begin{align*}
    % Y^u &\sim 3\times3, \\
    % \Tilde{Y}^d &\sim 3 \times 4, \\
    % \Tilde{M}^B &\sim 1 \times 4. 
    % \end{align*}
   % This approach can also be applied to introducing an up-type VLQ (See \cite{alves_vector-like_2023}).
Similarly, we can consider more than one VLQs or consider VLQs of up type\,\cite{alves_vector-like_2023}.
    
We can also make the block structure in $\tilde{Y}^B$ and $\tilde{M}^B$ apparent as
    \begin{equation}
    \begin{aligned}
    \tilde{Y}^d &= 
        \begin{pmatrix}
        Y^d & Y^B
        \end{pmatrix},
    \cr
    \tilde{M}^B &= 
        \begin{pmatrix}
        \overline{M^B} & M^B
        \end{pmatrix}, 
    \end{aligned} \label{base completa}
    \end{equation}
where $Y^d$ is of size $3\times 3$, $Y^B$ is $3\times 1$, $\overline{M^B}$ is $1 \times 3$ and $M^B$ is $1\times 1$. Using the block structure above, the Lagrangian \eqref{lagragian 1vlq} is rewritten as
    \begin{equation}
        -\mathcal{L} = \overline{q_{L}}Y^u \Tilde{H} u_{R} + \overline{q_{L}} Y^d H d_{R} + \overline{q_{L}}H Y^B_{r} B_{R} + \overline{B_{L}} \, \overline{M^B} B_{R} +\overline{B_{L}} M^B B_{R}.
    \label{Lagragian-used} 
    \end{equation}
    % Here, $\overline{M^B}$ and $M^B$ denote general complex matrices, being $n_d \times 3$ and $n_d \times n_d$, respectively. 

%     Upon examining the Lagrangian framework involving the inclusion of VLQs, a comparison with the model proposed by Branco et al. \cite{branco_addition_1986} reveals notable parallels, particularly in the singlet states and mass term structures. The Lagrangian as per Branco et al. is expressed as:
%     \begin{align}
%         -\mathcal{L} =  \overline{q}_{L} H Y_d d_{R} + \overline{q}_{L}\Tilde{H} Y_u u_{R} + \overline{q}_{L} H \overline{Y_d} D_{R} + \overline{D}_{L}\,(\overline{M_d}) d_{R} + \overline{D}_{L}(M_d) D_{R} \label{Lagragian-branco}
%     \end{align}

% \red{A notação barra do review não vai ser usada. Reduzir. ??}
    
%    When comparing this to our current setup, the following correspondences are observed:
%     \begin{itemize}
%     \item $\left.\begin{array}{c}
%         \begin{matrix}
%             D_R \sim B_R, \\
%             D_L \sim B_L
%         \end{matrix}
%         \end{array}\right\} \text{The VLQs in both models.}$
%     \item $\left.\begin{array}{c}
%         \begin{matrix}
%             \overline{M_d} \sim \overline{M}^B, \\
%             M_d \sim M^B
%         \end{matrix}
%         \end{array}\right\} \text{ Mass terms in both models.}$
%      \item $\left.\begin{array}{c}
%         \begin{matrix}
%             \overline{Y_d} \sim Y^B.
%         \end{matrix}
%         \end{array}\right\} \text{New coupling to the extra right-handed field.}$
% \end{itemize}

In general, without any symmetry, $d_{iR}$ and $B_R$ have the same gauge quantum numbers, and they are indistinguishable. So it is always possible to perform a weak basis transformation on the space of four fields $(d_{iR},B_R)$. Therefore, we can utilize WBTs to simplify some couplings.
Specifically, it is possible to select a basis where certain components of the mass matrix vanish\,\cite{alves_vector-like_2023}. For example, it is always possible to adopt a basis where
    \begin{equation}
    \begin{aligned}
    \tilde{Y}^d &= 
        \begin{pmatrix}
        Y^d & Y^B
        \end{pmatrix},
    \cr
    \tilde{M}^B &= 
        \begin{pmatrix}
        \mathbf{0} & M^B
        \end{pmatrix}.
    \end{aligned} \label{base yd mb}
    \end{equation} 
In this basis, the matrix $\tilde{M}^B$ includes a zero block $\mathbf{0}$, indicating the absence of direct mass mixing between the SM down-type quarks and the VLQs, and $M^B$, representing the VLQ mass term. In this chosen basis, the Lagrangian for the extended SM simplifies to:
\begin{equation}
        -\mathcal{L}=   \overline{q}_{L} H Y^d d_{R} + \overline{q}_{L}\Tilde{H} Y^u  u_{R} +
        \overline{q}_{L} H Y^B B_{R} + \overline{B}_{L} M^B B_{R}+ h.c.  \label{Lfvlq}
    \end{equation}
Prior to electroweak symmetry breaking, this is the mass basis for the VLQ $B_L,B_R$. The rest of the SM fields are massless.

Therefore, in this basis, the $3 \times 1$ Yukawa coupling $Y^B$ has physical meaning as the coupling between the VLQ and the SM quarks:
    \begin{equation}
\label{YB:MB.diag}
    Y^B = \begin{pmatrix}
    |Y_1^B| e^{i\alpha_1} \\
    |Y_2^B| e^{i\alpha_2}\\
    |Y_3^B| e^{i\alpha_3}
    \end{pmatrix}.
    \end{equation}

\section{Mass diagonalization} 

     After Electroweak Symmetry Breaking (EWSB), the Higgs doublet in the unitary gauge is represented as 
\
\begin{equation}
H = \begin{pmatrix}
    0\cr \displaystyle\frac{v+h}{\sqrt{2}} 
\end{pmatrix}
\,,
\end{equation}    
where $h$ is the Higgs field and $v \approx 246$ GeV  is the vacuum expectation value of the Higgs field. This process results in additional mass terms in the Lagrangian \eqref{Lfvlq} for quarks, including the VLQ.
% , which are crucial for generating the masses of these particles.
    %  From eq.\, \eqref{lagragian 1vlq} we can be written only mass terms for down-type as
    % \begin{equation}
    %     \mathcal{L}_{mass} = \overline{q}_L \Tilde{Y}^d \begin{pmatrix}
    %         d_R\\
    %         B_R
    %     \end{pmatrix} + B_L^\dagger \Tilde{M}^B \begin{pmatrix}
    %         d_R\\
    %         B_R
    %     \end{pmatrix}
    % \end{equation}
    
    The down-type quark mass matrix, including the VLQ contribution, is a $4\times 4$ matrix by blocks as in equation \eqref{base completa}, denoted as $\mathcal{M}_d$
    % \begin{equation}
    % -\mathcal{L}_{\text{mass}} = \left( \begin{array}{cc}
    % \overline{d}_{L} & \overline{B}_{L} \end{array} \right)
    % \mathcal{M}_d \left( \begin{array}{c}
    % d_{R} \\
    % B_R\end{array} \right)    
    % + h.c.
    % \end{equation}
    \begin{equation}
    \mathcal{M}_d = \left( \begin{array}{c}
    \frac{v}{\sqrt{2}}\Tilde{Y^d} \\[1ex] \hdashline
    \Tilde{M^B} 
    \end{array} \right) =  \left( \begin{array}{c;{2pt/2pt}c}
    \frac{v}{\sqrt{2}}Y^d &  \frac{v}{\sqrt{2}} Y^B  \\ \hdashline[2pt/2pt]
    \overline{M}^B & M^B
\end{array} \right)
    \label{MassYd}
    \end{equation}
    Generally, this matrix is not symmetric nor Hermitian. It is  typically expected that $\tilde{M}^B \gg \frac{v}{\sqrt{2}}\tilde{Y}^d $. Here, we have already considered that $\mathcal{M}_u$ is diagonal, i.e., $\mathcal{M}_u = \mathcal{D}_u =\diag(m_u,m_c,m_t)$. 
    
    To diagonalize $\mathcal{M}_d$, biunitary transformations are employed. This involves finding unitary matrices  $\mathcal{W}$ such that,
    \begin{equation}
    \mathcal{W}_{L}^\dagger \mathcal{M}_{d} \mathcal{W}_{d_R} = \mathcal{D}_d, \quad \mathcal{D}_d 
%     =\left( \begin{array}{c;{2pt/2pt}c}
%     \hat{Y}^d \frac{v}{\sqrt{2}} & 0 \\ \hdashline[2pt/2pt]
%     0 & M_B
% \end{array} \right)
=\diag(m_d,m_s,m_b,M^B).
    % \text{ with } \mathcal{D}= \left( \begin{array}{c}
    % \hat{Y}^d  \\
    % \hdashline
    % \hat{M}^B
    % \end{array} \right) 
    \label{svd}
    \end{equation}
    Here, $\mathcal{W}_{L},\mathcal{W}_{d_R}$  are $4\times 4$ unitary matrices. $\mathcal{D}_d$ is a diagonal matrix containing the physical masses of the quarks, including the VLQ mass $M^B$. The diagonalization is performed by the change of basis
\begin{equation}
\begin{aligned}
\begin{pmatrix}
d_{iL} \cr B_L
\end{pmatrix}&\to \mathcal{W}_L \begin{pmatrix}
d_{iL} \cr B_L
\end{pmatrix}
\cr
\begin{pmatrix}
d_{iR} \cr B_R
\end{pmatrix}&\to \mathcal{W}_{d_R} \begin{pmatrix}
d_{iR} \cr B_R
\end{pmatrix}
\end{aligned}
\label{diag.d.fields}
\end{equation}
The diagonalization matrices $\mathcal{W}_{d_R}$ and $\mathcal{W}_L$ can be divided into two blocks
    \begin{equation}
    \mathcal{W}_{L,{d_R}} = \begin{pmatrix}
    A_{L,{d_R}}\\
    \hdashline
    B_{L,{d_R}} \\
    \end{pmatrix}.
\label{A.block}
    \end{equation}
    With the introduction of a single down-type VLQ in the SM, $A_{L,{d_R}}$ are $3 \times (3 + 1)$ matrices and $B_{L,d_R}$ is a $1 \times (3 + 1)$ matrix. In terms of these blocks and the physical masses, the expressions for $Y^d$, $Y^B$ and $M^B$ are
    \begin{equation}
    \begin{aligned}
        \frac{v}{\sqrt{2}}Y^d = A_L \mathcal{D}_d A_{d_R}^{\dagger}, \quad 
        \frac{v}{\sqrt{2}}Y^B= A_L \mathcal{D} B_{d_R}^{\dagger}, \\
        \overline{M}^B = B_L \mathcal{D}_d A_{d_R}^{\dagger}, \quad M^B= B_L\mathcal{D} B_{d_R}^{\dagger}. \label{Physis:mass} 
    \end{aligned}
    \end{equation}

    % Here,  $Y^d$ is a $3\times 3$ matrix, $M^B$ is a scalar (since we consider one VLQ) and $Y^B$ is $3\times 1$ a generic matrix. $B_L$ is only a phase that can be described as $1\times 1$ matrix.
    
    The unitarity of the diagonalization matrices is equivalent to
    \begin{align}
    \begin{pmatrix}
    A_{L,{d_R}} \\
    B_{L,{d_R}} \\
    \end{pmatrix} \begin{pmatrix}
    A_{L,{d_R}}^{\dagger} & B_{L,{d_R}}^{\dagger} \\
    \end{pmatrix} & = \begin{pmatrix}
        A_{L,{d_R}} A_{L,{d_R}}^\dagger & A_{L,{d_R}} B_{L,{d_R}}^\dagger\\
        B_{L,{d_R}} A_{L,{d_R}}^\dagger & B_{L,{d_R}} B_{L,{d_R}}^\dagger
    \end{pmatrix}=\begin{pmatrix}
    \mathbb{I}_3 & 0 \\
    0 & 1 \\
    \end{pmatrix}, \\
    \begin{pmatrix}
    A_{L,{d_R}}^{\dagger} & B_{L,{d_R}}^{\dagger} \\
    \end{pmatrix} \begin{pmatrix}
    A_{L,{d_R}}\\
    B_{L,{d_R}} \\
    \end{pmatrix} &= A_{L,{d_R}}^{\dagger}A_{L,{d_R}} + B_{L,{d_R}}^{\dagger}B_{L,{d_R}} = \mathbb{I}_{3+1}. \label{identitys}
    \end{align}

\section{Gauge Interactions in the SM Extended with a Single Down-Type VLQ}

    In the SM extended to include a single down-type VLQ, the gauge interactions are enriched by the presence of the VLQ. The charged current interactions mediated by the  $W$ boson are described by the Lagrangian:
    \begin{align}
        \mathcal{L}_W = -\frac{g}{\sqrt{2}} J_\mu^- W^+_\mu + h.c., 
    \end{align}
    where, $W^+_\mu$ denotes the W boson field.  
    This part of the Lagrangian describes transitions between up and down-type quarks.
    
    Electromagnetic interactions for quarks are governed by:
    \begin{equation}
        \mathcal{L}_A = -e J_\mu^{\text{em}} A_\mu
    \end{equation}
    where the Lagrangian involves the electromagnetic current $J^\mu_{\text{em}}$  and the photon field $A_\mu$. 

    Weak neutral current interactions, mediated by the $Z$ boson are denoted by
    \begin{equation}
        \mathcal{L}_Z = -\frac{g}{2\cos{\theta_W}} J^Z_\mu Z_\mu,
    \end{equation}
% \red{Aqui não tem um menos na frente?} \eduardo{sim}
highlighting the quark interactions with the $Z$ boson, with $\theta_W$ representing the Weinberg angle. 
 
    In the Standard Model extended to include Vector-Like Quarks (VLQs), these new quarks interact with both photons and the $Z$ boson, despite being $SU(2)_L$ singlets, due to their non-zero weak hypercharge which allows them to have electric charges equal to SM quarks and mix with them. 
    % Interestingly, in the flavor basis of this extended model, there are no Flavor-Changing Neutral Currents (FCNCs) mediated by the $Z$ boson, consistent with the SM's behavior where such transitions are rare or forbidden. 
In the flavor basis of this extended model, all the currents above are diagonal in flavor:
\begin{equation}
     \begin{aligned}
        J^-_{\mu} &= \sum_{i=1}^3\overline{u}_{Li} \gamma^\mu d_{Li}
        % + h.c. 
        \\
        J^Z_{\mu} &= J^3_{\mu} - \sin^2{\theta_W}\cdot J^\mu_{\text{em}},\\          
        J^3_{\mu} &= \sum_{i=1}^3\overline{u}_{Li} \gamma^\mu  u_{Li} - \overline{d}_{Li} \gamma^\mu  d_{Li},\\
        J^\mu_{\text{em}} &= \frac{2}{3}\sum_{i=1}^3 \overline{u}_i \gamma^\mu  u_i -\frac{1}{3} \bigg(\sum_{i=1}^3\overline{d}_i \gamma^\mu  d_i + \overline{B}\gamma^\mu B \bigg), 
        \label{currents}
    \end{aligned}
\end{equation}
where $\psi=\psi_L+\psi_R$ for all $\psi=u_i,d_i$ and $B$. However, in the mass eigenstate basis, particularly for down-type quarks, the mixing between VLQs and SM quarks lead to FCNCs, altering the quarks' couplings to the $Z$ boson. After the change to the mass basis \eqref{diag.d.fields}, the currents become
    \begin{equation}    
    \begin{aligned}
        J^-_{\mu} &=  \overline{u}_{L} \gamma^\mu 
        % W_L^\dagger 
        A_{L} \begin{pmatrix}
d_{L} \cr B_L
\end{pmatrix}
    + h.c.,  \\
        J^3_{\mu} &=  \overline{u}_L \gamma^\mu  u_L - \begin{pmatrix}
\overline{d_{L}} & \overline{B_L}
\end{pmatrix} \gamma^\mu A_{L}^\dagger A_{L}  \begin{pmatrix}
d_{L} \cr B_L
\end{pmatrix}\\ 
        J^\mu_{\text{em}} &= \frac{2}{3} (\overline{u} \gamma^\mu  u) -\frac{1}{3} (\overline{d} \gamma^\mu  d + \overline{B}\gamma^\mu B). \label{currents}
    \end{aligned}
    \end{equation}
where we use a matricial notation in flavor space
and $A_L$ is the subblock in \eqref{A.block}. Here, we can note that the electromagnetic current, $J^\mu_{\text{em}}$, remains unchanged when transitioning from the flavor to the mass basis. This model also introduces deviations from the unitarity of the CKM matrix due to the inclusion of VLQs, thereby expanding the SM's quark sector with new interactions and phenomena \cite{alves_vector-like_2023}.
    % With $\Tilde{V}$ being a $3 \times (3+1)$ non-unitary mixing matrix, defined as
    % \begin{equation}
    %     \Tilde{V}=
    %     % W_L^\dagger 
    %     A_{L}.
    % \end{equation}
    % It is clear that the CKM matrix is not more unitary, and is represent as a $3\times 4$ matrix, denoted by $\Tilde{V}$.
The new CKM matrix appearing in the interaction with the $W$ boson is now $3\times 4$ and we denote it as
    \begin{equation}
        \Tilde{V}=
        % W_L^\dagger 
        A_{L}.
    \label{ckm:3x4}
    \end{equation}
It is clear that even the $3\times 3$ subblock of $\tilde{V}$ is no longer unitary. 
    
    The equation for $J^3_{\mu}$ in \eqref{currents} highlights the potential for FCNCs in the down-quark sector at the tree level, a phenomenon not present within the Standard Model. The effect of FCNC is given by the off-diagonal entries of the $4\times 4$ Hermitean matrix
    \begin{equation}
        F^d= A_{L}^\dagger A_{L}= \tilde{V}^\dagger \tilde{V}.
    \end{equation}
Using the identity \eqref{identitys}, we can also write
    \begin{equation}
        F^d
        % = A_{L}^\dagger A_{L} 
        = \mathbb{I}- B_{L}^{\dagger} B_{L}\,.
    \end{equation}
    It is clear that $F^d$ deviates from the identity matrix, unlike in the SM, indicating the presence of FCNCs.

\section{Weak Basis Transformations} \label{WB:VLQ}

    % In the SM extended to include VLQ, WBT  serve a crucial role in streamlining the Lagrangian without altering the inherent physical properties of the model, such as quark masses, mixing angles, and couplings. These transformations essentially constitute a change of basis within the flavor space, thereby preserving the observable physical quantities.
    
    % \subsection{Transformations in the Extended Model}
    
    The general form of WBT for the fields \eqref{fields:vlq} can be expressed as
    \begin{align}
        &q_L \to U_L^q q_{L}; \quad
        u_R\to U_R^u u_R;\quad
        B_L \to W_L^B B_L = e^{i\theta} B_L; \quad \begin{pmatrix}
            d_R\\
            B_R
        \end{pmatrix} \to \mathcal{W}_{d_R} \begin{pmatrix}
            d_R\\
            B_R
            \end{pmatrix}. \label{wbtransformation}
    \end{align}   
    The left-handed quark doublets, $q_L$, transform under a $3 \times 3$ unitary matrix $U^q_L$, whereas the left-handed VLQs, $B_L$, undergo a transformation that can be represented by a phase. The right-handed down-type quarks, including the VLQ, are collectively
    transformed by a $4 \times 4$ unitary matrix $\mathcal{W}_{d_R}$.
    
    The field transformations \eqref{wbtransformation}, induces the following transformations on the couplings in the Lagrangian \eqref{lagragian 1vlq}:
\begin{equation}
\begin{aligned}
\tilde{Y}^d&\to U_L^{q\dagger} \Tilde{Y}^d \mathcal{W}_{d_R}
\cr
\tilde{M}^B&\to e^{-i\theta} \Tilde{M}^B \mathcal{W}_{d_R}\,.
\end{aligned}
\label{trans: vlq}
\end{equation}
On the effective mass matrix \eqref{MassYd}, the transformations above act as 
 \begin{equation}
    \mathcal{M}_d = \left( \begin{array}{c;{2pt/2pt}c}
    \frac{v}{\sqrt{2}}Y^d &  \frac{v}{\sqrt{2}} Y^B  \\ \hdashline[2pt/2pt]
    \overline{M}^B & M^B
\end{array} \right) \to \begin{pmatrix}
    \frac{v}{\sqrt{2}} U_L^{q\dagger} Y^d &  \frac{v}{\sqrt{2}} U_L^{q\dagger} Y^B  \\
   W_L^{B\dagger} \overline{M}^B &  W_L^{B\dagger} M^B
\end{pmatrix} \mathcal{W}_{d_R}
    \label{MassYd: transformed}
    \end{equation}

 % From these transformations we can write

With the basic transformation law \eqref{trans: vlq}, we can write how some combinations transform.
For example,
 \begin{equation}
 \small
\begin{aligned}
     & Y^u Y^{u\dagger} \to U_L^{q\dagger} Y^u Y^{u\dagger}U_L^{q}, \\
     & Y^{u\dagger} Y^u \to W_R^{u\dagger} Y^{u\dagger} Y^u W_R^u, \\
     &\Tilde{Y}^d \Tilde{Y}^{d\dagger}  \to U_L^{q\dagger} \Tilde{Y}^d \Tilde{Y}^{d\dagger} U_L^{q},\\
     &\Tilde{M^B} \Tilde{M}^{B\dagger} \to W_L^{B\dagger} \Tilde{M^B} \Tilde{M}^{B\dagger} W_L^B,\\
     &\Tilde{Y}^{d\dagger} \Tilde{Y}^d \to \mathcal{W}_{d_R}^\dagger   \Tilde{Y}^{d\dagger} \Tilde{Y}^d \mathcal{W}_{d_R},\\
     &\Tilde{M}^{B\dagger} \Tilde{M^B} \to \mathcal{W}_{d_R}^\dagger \Tilde{M}^{B\dagger} \Tilde{M^B}  \mathcal{W}_{d_R},\\
     &\Tilde{Y}^d \Tilde{M}^{B\dagger} \Tilde{M^B} \Tilde{Y}^{d\dagger}\to U_L^{q\dagger} \Tilde{Y}^d  \Tilde{M^B}^\dagger \Tilde{M^B} \Tilde{Y}^{d\dagger}  U_L^{q},
     \label{WBT:VLQ}
 \end{aligned}
 \end{equation}
where $W^B_L=e^{i\theta}$ is just a phase.

The transformations \eqref{MassYd: transformed} also allows us to define some special basis where some couplings are simpler. For example,
the unitary rotations $\mathcal{W}_{d_R}$ always allow the vanishing of one of the following subblocks \cite{alves_vector-like_2023}:
\begin{itemize}
\item A WB exists where $Y^B=0$ 
% for all VLQ singlets 
(vanishing $Y^B$).
\item A WB exists where $\overline{M^B}=0$ (vanishing $\overline{M^B}$).
\end{itemize}
We will often use the second basis, which is possible by choosing $\mathcal{W}_{d_R}$ to be such that it cancels $\overline{M^B}$. This is the basis \eqref{base yd mb}.
% Using these condition, we have that \eqref{MassYd: transformed} becomes,

% \begin{equation}
%     \mathcal{M}_d = \begin{pmatrix}
%     \frac{v}{\sqrt{2}} U_L^{q\dagger} Y^d &  \frac{v}{\sqrt{2}} U_L^{q\dagger} Y^B  \\
%   0 &  W_L^{B\dagger} M^B
% \end{pmatrix} \mathcal{W}_{d_R} 
%     \label{MassYd: MB0}
%     \end{equation}
Further residual transformations of the form \eqref{MassYd: transformed}, keeping vanishing $\overline{M^B}$, allows 
\begin{equation}
Y^u \to U_L^{q\dagger} Y^u W_R^{u},\quad
     Y^d \to U_L^{q\dagger} Y W^d_R, \quad 
     Y^B\to U_L^{q\dagger}Y^B e^{i\theta}. \label{trans: vlq:mass}
 \end{equation}
% \eqref{MassYd: MB0} becomes, 
% Under transformations the matrix \eqref{MassYd: transformed}
Choosing these transformations appropriately allows further simplification of couplings in \eqref{MassYd: transformed} to 
\begin{equation}
\mathcal{M}_d =  \begin{pmatrix}
    V' \hat{Y}^d & Y^B \\
    0 & M^B 
    \end{pmatrix}\,,
\label{phys.basis}
     \end{equation}
where here $M^B$ is real positive and $V'$ is a unitary matrix close to the SM CKM matrix.
 % Hence, we have:
 % \begin{equation}
 %     \Tilde{Y}^d \to U_L^{q\dagger} \Tilde{Y}^d \mathcal{W}_{d_R}, \quad \Tilde{M^B} \to W_L^{B\dagger} \Tilde{M}^B \mathcal{W}_{d_R}, \quad Y^u \to U_L^{q\dagger} Y^u W_R^{u}. \label{trans: vlq}
 % \end{equation}
After rephasing the fields $B_L,B_R$, one can still eliminate one of the complex parameters, resulting in
    \begin{equation}
        Y^B = \begin{pmatrix}
            |Y_1^B| \\
            |Y_2^B| e^{i\alpha_2}\\
            |Y_2^B| e^{i\alpha_3}
        \end{pmatrix}.
    \label{phys.basis:YB}
    \end{equation}
After rephasing of $q_{Li}, d_{Ri}$,
we can take $V'$ as in the standard parametrization with 4 parameters.
Then no further transformation is possible.

% \subsection{Parameter Counting in the Extended Model}
\section{Parameter Counting in the Extended Model}
\label{vlq:param.count}

In the basis of \eqref{phys.basis} and \eqref{phys.basis:YB}, we can count the number of physical parameters in the Lagrangian \eqref{Lfvlq} of the SM extended with the inclusion of a singlet down type VLQ. We can count 16 parameters, of which 10 are related to the original SM. The breakdown of these parameters is as follows \cite{cherchiglia_consequences_2020} :
    \begin{itemize}
        \item From $Y^u=\hat{Y}^u$: 3 real parameters.
        \item From $Y^d=V'\hat{Y}^d$: 6 real parameters and 1 phase.
        \item From $Y^B$ in \eqref{phys.basis:YB}: 3 real parameters and 2 phases.
        \item The term $M^B$ is a real mass term, counted as one real parameter.
    \end{itemize}
    % The parameter count for $Y^B$ can be summarized as $2 n_d - 1$ (where $n_d$ denotes the number of VLQs). For a single VLQ ($n_d = 1$), this results in:
    % \begin{equation}
    % n^2_F + 1 + 2n_F n_d = 3^2 +1 +2\cdot3\cdot1 =16
    % \end{equation}
    % In this formula, $F$ represents the number of quark families, and the number 1 accounts for the phase in the Standard Model. Thus, 16 parameters, the model achieves a comprehensive description of the quark sector, encompassing both the SM quarks and the added VLQ. 

    In the basis above, one can also make the counting of CP violating phases. 
% This is equal to the number of phases in $C_L$  minus the number of independent phase field redefinitions, 
%     \begin{align}
%         n_{CP} &= n(n+n_d) - \frac{n(n-1)}{2} - (2n + n_d -1) \\
%         &= (n-1)n_d + \frac{1}{2}(n-1)(n-2)= 3.
%     \end{align}
%     CP is conserved if CKM can be made real. This is the case if all $n_{CP}$ phases vanish. Leading to $n=3$ and $n_d=1$ we have 3 CP violating phase, the same as for an extra family $n=4$ and $n_d=0$. 
That number is three compared to one in the SM.
See other equivalent forms to count parameters in 
Ref.\,\cite{alves_vector-like_2023}. In general, the number of physical parameters, and in particular of CP violating phases, grows rapidly with the addition of more VLQs. 
% We will stick to these cases which incorporate many of the new features of the addition of new quark fields. For an extra down quark isosinglet, $V_CKM$ consists of the first 3 rows of a $4 \times 4$ unitary matrix.

% \section{CP violation}
\section{CP violation with one VLQ}

    To determine the restrictions on quark mass matrices implied by CP invariance, one has to consider the most general CP transformation which leaves the gauge currents invariant \cite{branco_addition_1986}
\begin{equation}
    \begin{aligned}
          q_L &\to W_L \gamma^0 C \overline{q_L}^T, 
          \qquad
          B_L \to  W^B_L \gamma^0 C
            \overline{B_L}^T, 
          \\
          \begin{pmatrix}
              d_R\\
              B_R
          \end{pmatrix} &\to \mathcal{W}_{d_R} \gamma^0 C \begin{pmatrix}
              \overline{d_R}^T\\
              \overline{B_R}^T
          \end{pmatrix}\,,
    \end{aligned}
\end{equation}
where $C$ represents the charge conjugation matrix. The matrices $W$ are the same as the diagonalizing matrices \eqref{diag.d.fields}.
    % The quark mass matrices satisfy the conditions delineated in equation \eqref{WBT:VLQ}. These matrices, as previously mentioned, comprise $W_L$ a $3\times 3$ matrix, $W_L^B$ which is an $n_d \times n_d$ matrix, and $\mathcal{W}_{d_R}$, a $(3+n_d) \times (3+n_d)$ matrix. All these matrices are unitary and operate within the flavor space. Additionally, $C$ represents the charge conjugation matrix.
% \red{Em principio as matrizes não são as mesmas de \eqref{trans: vlq}.}

    The conditions necessary for CP-invariance can be written in terms of quantities which are invariant under a change of WB.
    % , i.e., in order for the quark mass matrices to be CP invariant, the condition to be satisfied, independently of the mechanism responsible for generating quark masses\cite{branco_addition_1986}.  
    % For instance, if under a change of WB and we
Let us introduce the following hermitian matrices:
    \begin{align}
        X_{dB} &\equiv 
        \tilde{Y}^d \tilde{Y}^{d\dagger}
        =
        Y^d Y^{d\dagger}+ Y^B Y^{B\dagger} = X_d + Y^B Y^{B\dagger} \nonumber\\
        X'_{dB} &\equiv 
        \tilde{Y}^{d\dagger} \tilde{Y}^d=
        \begin{pmatrix}
            Y^{d\dagger} Y^d & Y^{d\dagger} Y^B \\
            Y^{B\dagger} Y^d & Y^{B\dagger} Y^B 
        \end{pmatrix} \nonumber \\
        X_{M^B} &\equiv 
        \tilde{M}^{B\dag} \tilde{M}^B= \begin{pmatrix}
            0 & 0 \\
            0 & M^{B\dagger} M^B 
        \end{pmatrix}
       \label{X:s}
     \end{align}
    % This set of relations represents the necessary and sufficient conditions for having CP invariance. 
    % Multiplying equations \eqref{WBT:VLQ} enables the derivation of necessary conditions for CP invariance in gauge interactions, as expressed in terms of WB invariants.
The process of building WB invariants, as detailed in \cite{branco_addition_1986}, involves multiplying 
% \eqref{WBT:VLQ} 
\eqref{X:s} by themselves, evaluating appropriate commutators and taking the trace.
Through this method, one can readily derive a set of invariants
    \begin{align}
        \aver{[X_{dB}^a, X_u^b]^r}=0, \nonumber\\
        \aver{[X_{dB}^{'a} X_{M^B}^b]^r} =0  \label{conditions}
    \end{align}    
    where $r$ is odd and $a,b$ are arbitrary positive integers. It is obvious that the conditions above are WB independent. 

    In the SM limit, the first family of invariants in eq. \eqref{conditions} simply yields the CP-odd $I_{SM}$ of eq. \eqref{Invariant3} at the lowest non-trivial order. The remaining families of CP-odd invariants arise only in the presence of VQLs.
    % down- and up-type 
    % VLQs, respectively.

    In the given context, additional independent CP-odd Weak Basis WB invariants can be constructed, which are instrumental in analyzing CP violation phenomena. One of these invariants, shown in \cite{alves_vector-like_2023, albergaria_cp-odd_2023,del_aguila_cp_1998}, can be written as
    \begin{equation}
        \operatorname{Im}\aver{X_{M^B} \Tilde{Y}^{d\dagger} X_{u} \Tilde{Y}^d X'_{dB}} \label{imtraces}
    \end{equation}
% \red{Escreva esse invariante na notação de \eqref{X:s}.} \eduardo{Já fiz}
In general, finding a set of necessary and sufficient conditions for CP invariance in terms of WB invariants is complicated. 
    
One can also construct CP-even WB invariants, such as
    \begin{equation}
        \aver{(X_{dB})^a (X_u)^b}, \quad \aver{(X'_{dB})^a (X_{M^B})^b},
    \end{equation}
with $a$ and $b$ positive integers. A sufficiently large number of independent CP-even and CP-odd WB invariants may be used to reconstruct the quark mixing matrix $V$ \cite{albergaria_cp-odd_2023}.

    The introduction of VLQs into the SM framework inherently generates new avenues for CP violation. 
    To elucidate this, let us assume an extreme chiral limit where the masses of the up, charm, down, and strange quarks are zero ($m_u = m_c = m_d = m_s = 0$). In this specific limit, the CP violation in the SM, represented by $I_{SM}$, vanishes due to the degeneracy in the masses of quarks with identical charges. Nonetheless, the interaction dynamics involving VLQs introduce CP-violating phases that remain physically significant, even within this chiral limit, as highlighted in \cite{del_aguila_cp_1998}.

%% file: capitulos/5-Resultados.tex
\chapter{Flavor invariants for one VLQ}\label{cap:resultados}

In this chapter, we study the flavor invariants of the SM enriched with a single VLQ. 
We will treat the Lagrangian \eqref{Lagragian-used} in the mass basis of the VLQ and the Lagrangian \eqref{lagragian 1vlq} in the general form. Our focus is on dissecting the basic invariants that emerge from the Hilbert series analysis, categorizing them into CP-odd and CP-even types.

% \red{Tratar o $B_R$ diferente não é um cenário diferente. É só uma escolha de base que é mais física v.s. tratar todos os $d_R,B_R$ como indinstinguíveis.}

\section{Hilbert series in VLQ mass basis} \label{Sec:5.1}

We first consider the SM with one down type singlet VLQ in the mass basis of the VLQ as described in \eqref{Lfvlq}. For a theory with a single down-type VLQ, the flavor symmetry group in the absence of $Y^u,Y^d,Y^B$ is characterized by:
\begin{equation}
        G_F = SU(3)_q\otimes SU(3)_u\otimes SU(3)_d \otimes U(1)_{\text{VLQ}},
    \label{GF:vlq:MB}
\end{equation}
where the group $SU(3)_q\otimes SU(3)_u\otimes SU(3)_d$ is the same as in the SM quark fields as described in \eqref{group:SM,0,0}. The factor $U(1)_{\text{VLQ}}$ corresponds to a global phase rotation associated with the VLQ fields:
    \begin{align}
    \label{vlq:number}
    B_L &\to e^{i\theta} B_L, \\
    B_R &\to e^{i\theta} B_R,
    \end{align}
    with $\theta$ being a real phase parameter. 
We can see that the VLQ mass term involving $M^B$ in \eqref{Lfvlq} remains invariant under these phase transformations.

When the Yukawa couplings $Y^u,Y^d,Y^B$ are turned on, the group \eqref{GF:vlq:MB} is explicitly broken.
Invariance is restored if the Yukawa couplings $Y^u,Y^d$ are spurions transforming as \eqref{Yu.Yd:U} and $Y^B$ transforms as
\begin{equation}
Y\to U_L^q Y e^{-i\theta}\,,
\end{equation}
This is the inverse of the WBT in \eqref{trans: vlq:mass}.
To simplify the notation, we will adopt $Y^B\to Y$ in this chapter.
Similarly to the SM, we can disregard $SU(3)_u$ and $SU(3)_d$ by considering $X_u,X_d$ which transform only under $SU(3)_q$ as in \eqref{sm:X.transf}.

% Given the transformation laws for both $Y^{u,d}$ and 
% $X_{u,d}$ as specified in equations \eqref{trans: vlq} and \eqref{WBT:VLQ} respectively, 
% the group is defined as follows. 
So we can consider the group
\begin{equation}
\label{su3.u1:VLQ}
    SU(3)_q\otimes U(1)_{\text{VLQ}}\,,
\end{equation}
under which the spurions transform as
% \red{
% Escreva a lei de transformação para $Y^B$.
% Então cite a lei de transformação para $Y^u,Y^d$ do Cap4.
% Cite a lei de transformação para $X_u,X_d$ do Cap4.
% Daí o grupo relevante vira $SU(3)_q\otimes U(1)_{\text{VLQ}}$,
%  etc.
% }

    % Furthermore, the condition highlighted in equation \eqref{conditions} ensures that the matrix $M^B$ is always “excluded” from the transformations that involve the SM quarks and their Yukawa couplings. This exclusion is a result of the chosen basis where $\overline{M^B}$, representing a component of the VLQ mass matrix, is set to zero. Consequently, the $M^B$ matrix, which encapsulates the mass terms specific to the VLQ, maintains its form and does not participate in the transformations that affect the SM quark sector. 
% The spurions $X_u,X_d$  transform as $3_q\otimes \bar{3}_q$ and $Y^B\sim (3_q)_{-1}$  carries a $U(1)$ charge of $-1$. Under the group \eqref{su3.u1:VLQ} they are defined as:
    \begin{equation}
    \begin{aligned}
    &X_u \sim X_d \sim (3 \otimes\Bar{3},0), \\ 
    &Y \sim (3,-1),\\
    & Y^{\dagger} \sim (\overline{3},+1).
    \end{aligned}
    \end{equation}
Similarly to $X_u,X_d$, the combination $YY^\dag$transforms as 
\begin{equation}
    YY^\dagger \to U_L^q YY^\dagger U_L^{q\dagger},
\end{equation}
such that
\begin{equation}
    YY^\dagger \sim (3\otimes\Bar{3},0)
\end{equation}
Note that we do not consider $YY^\dag$ as a basic spurion because it should have rank one due to $Y$ being a column vector.
   
    From these spurions $X_u, X_d, Y, Y^\dag$ we can construct the Hilbert series. The Hilbert series $H(x)$ enumerates the invariants of the theory. So, taking $X_u,X_d,Y,{Y}^\dag\to x$, we obtain the \emph{unrefined} Hilbert series:
   \begin{equation}
    H(x) = \frac{{1 + 2x^4 + 4x^5 + 5x^6 + 2x^7 + 2x^8 + 2x^9 + 5x^{10} + 4x^{11} + 2x^{12} + x^{16}}}{{(1 - x^4)^3 (1 - x^3)^6 (1 - x^2)^4 (1 - x)^2}}, \label{HS:resultados}
    \end{equation}
    where $x$ is a complex variable representing all the spurions. See Appendix \ref{case2} for the details of the calculation. The denominator of the $H(x)$ carries the information about the primary invariants, i.e., those invariants that are algebraically independent. There are fifteen factors in the denominator of the $H(x)$ in \eqref{HS:resultados}, which means there are a total of 15 primary flavor invariants, three invariants of degree four, six invariants of degree three, four invariants of degree two and two invariants of degree one. The e Plethystic logarithm of the Hilbert series, which helps us identify the  basic invariants, is
    \begin{equation}
    PL[H(q)]= \sum_{r=1}^\infty \frac{\mu(r) \log H(q^r)}{r},    
   \end{equation}
   \begin{equation}
   PL= 2x + 4x^2 + 6x^3 + 5x^4 + 4x^5 + 5x^6 + 2x^7 - x^8 - 6x^9 + O(x^{10}). \label{PL:basics}
    \end{equation}
The first negative coefficients indicate the presence of one syzygy of degree 8 and six syzygies of degree 9. From the positive terms of the $PL$ in \eqref{PL:basics}, there are 28 basic invariants in the generating set, which we list in Table \ref{tab:flavor_invariants}, with 9 CP-odd and 19 CP-even invariants. The 9 CP-odd invariants are not primary because their square can be written in terms of others CP-even invariants. CP violation is manifested in the CP-odd invariants. Changing the signs from minus to plus in the listed CP odd invariants renders them reducible, i.e., expressible in terms of previously defined invariants. Further discussion on how we obtained the list of basic invariants in Table \ref{tab:flavor_invariants} is given in the next subsection.

    % [Table \ref{tab:flavor_invariants} would typically be presented here in a formal paper, listing all the basic flavor invariants along with their properties.]
    % \newpage

In Table~\ref{tab:flavor_invariants}, the notation $I_{(2n)(2m)(2l)}$ denotes an invariant of degree $2n$ in $Y^u$, degree $2m$ in $Y^d$ and degree $2l$ in $Y$.  We denote this degree as the physical degree in comparison with the degree in the Hilbert series \eqref{HS:resultados} and the Plethystic logarithm \eqref{PL:basics} where $X_u,X_d,Y,Y^\dag$ all count as degree one.
When ambiguity arises, we also include the label $\pm$ depending on if it is a sum or subtraction of two terms. The sign also coincides with the CP parity. The invariants that contain only $X_u$ and $X_d$ are the same as in SM, listed in Table \ref{tab:invariantes3damily}.
Here, the only CP odd invariant of the SM \eqref{I-} is denoted as $I^-_{660}$.

To test the validity of Knop's theorem, we note that the dimension $\dim V=24$ comes from the number of 9 parameters from $X_u$, 9  from $X_d$ and 6 from $Y$. The theorem is satisfied with $p=15$, $d_D=40$, $d_N=16$. In the Hilbert series \eqref{HS:resultados}, $p=15$ refers to the number of denominator factors. This number should match the number of physical parameters contained in $X_u,X_d$ and $Y$. As in the SM, cf.\,Table \ref{tab:basic:n=3}, $X_u$ and $X_d$ contains 10 physical parameters, in which we exclude $I^{(-)}_{66}$. There is 3 up-type masses, 3 down-type masses, 1 CP odd and 3 CP even mixing parameters. $Y$ in \eqref{YB:MB.diag} contains five physical parameters because one phase can be removed by rephasing using \eqref{vlq:number}. The total is correctly 15.

   \begin{table}[h]
\centering
\caption{Summary of the 28 basic flavor invariants. Along with their physical degree, the degree in the Plethystic logarithm \eqref{PL:basics}, and CP parity. 
We use the commutator $[X_u, X_d] \equiv X_u X_d - X_d X_u$ and the anticomutator $\{X_u, X_d\} \equiv X_u X_d + X_d X_u$.}
\begin{tabular}{|c|c|c|c|}
\hline
Flavor Invariant & Phys. degree & degree in \eqref{PL:basics} & CP \\
\hline
\( I_{200} = \aver{X_u} \) & 2 & 1 &+ \\
\( I_{020} = \aver{X_d} \) & 2 & 1 &+ \\
\( I_{400} = \aver{X^2_u} \) & 4 & 2 & + \\
\( I_{040} = \aver{X^2_d} \) & 4 & 2 & + \\
\( I_{220} = \aver{X_u X_d} \) & 4 & 2 & + \\
\( I_{002} = \aver{YY^\dagger} \) & 2 & 2 & + \\
\( I_{600} = \aver{X^3_u} \) & 6 & 3 & + \\
\( I_{060} = \aver{X^3_d} \) & 6 & 3 & + \\
\( I_{420} = \aver{X^2_u X_d} \) & 6 & 3 & + \\
\( I_{240} = \aver{X_u X^2_d} \) & 6 & 3 & + \\
\( I_{202} = \aver{X_u YY^\dagger} \) & 4 & 3 & + \\
\( I_{022} = \aver{X_d YY^\dagger} \) & 4 & 3 & + \\
\( I_{440} = \aver{X_u^2X_d^2} \) & 8 & 4 & + \\
\( I_{402} = \aver{X_u^2YY^\dagger} \) & 6 & 4 & + \\
\( I_{042} = \aver{X_d^2YY^\dagger} \) & 6 & 4 & + \\
\( I^+_{222} = \aver{\{X_u,X_d\} YY^\dagger }\) & 6 & 4 & + \\
\( I^-_{222} = \aver{[X_d,X_u]YY^\dag} \) & 6 & 4 & - \\
\( I^-_{422} = \aver{[X^2_u, X_d]YY^\dag} \) & 8 & 5 & - \\
\( I^+_{422} = \aver{\{X_u^2,X_d\}YY^\dag} \) & 8 & 5 & + \\
\( I^-_{242} = \aver{[X_d^2,X_u]YY^\dag} \) & 8 & 5 & - \\
\( I^+_{242} = \aver{\{X^2_d,X_u\}YY^\dag} \) & 8 & 5 & + \\
\( I_{660}^- = \aver{X_u^2X_d^2X_uX_d}-\aver{X_d^2X_u^2X_dX_u} \) & 12 & 6 & - \\
\( I^-_{442} = \aver{[X^2_u,X^2_d] YY^\dag} \) & 10 & 6 &- \\
\( I^+_{442} = \aver{\{X^2_u,X^2_d\} YY^\dag} \) & 10 & 6 &+ \\
\( I_{622}^- = \aver{X_u^2X_dX_uYY^\dag} - \aver{X_uX_dX_u^2YY^\dag} \) & 10 & 6 &- \\
\( I^-_{262} = \aver{X_d^2X_uX_dYY^\dag}-\aver{X_dX_uX_d^2YY^\dag} \) & 10 & 6 &- \\
\( I^-_{642} = \aver{X_u^2X_d^2X_uYY^\dag}-\aver{X_uX_d^2X_u^2YY^\dag} \) & 12 &7 &- \\
\( I^-_{462} = \aver{X_d^2X_u^2X_dYY^\dag}-\aver{X_dX_u^2X_d^2YY^\dag} \) & 12 & 7&- \\
\hline
\end{tabular}
\label{tab:flavor_invariants}
\end{table}

% \red{Eduardo: acrescente o no. de eq. acima referindo-se aos 10 parametros contidos em Xu, Xd no MP.}
\newpage
\section{Reduction of invariants of higher degree}

% The Cayley-Hamilton theorem given in \eqref{n=3:CH} allows for the reduction of powers of the matrices
% $X_u,X_d,YY^\dag$
% , such as $X_u$ and $X_d$, in terms of their lower powers. The equation \eqref{A4} shows that for higher order, we can be written in terms of lower order. The fixed degree of $YY^\dagger$ representing the VLQ mass terms, when combined with the Cayley-Hamilton theorem, elucidates the structure of invariants involving $YY^\dagger$. 

The invariants involving only $X_u,X_d$ are the same as in the SM, and they were listed in Table~\ref{tab:invariantes3damily}.
For invariants involving $YY^\dag$,
we only need the first power because
\begin{equation}
    (YY^\dagger)^2 = (Y^\dagger Y)YY^\dagger. 
\end{equation}
Considering that $Y$ is a $3 \times 1$ generic matrix, we have that $YY^\dagger$ is a $3\times 3$ matrix, and $Y^\dagger Y$ is a $1\times 1$ matrix and real.
For a similar reason, the matrix $YY^\dag$ needs to appear only once inside the trace.
% How is it a $1\times 1$ matrix, this term stays out of trace. This can be seen explicitly, as
% \begin{equation}
% \begin{aligned}
%      \aver{X^3 (YY^\dagger)^2} &= \aver{X} \aver{X^2 (YY^\dagger)^2} - \frac{1}{2}[\aver{X}^2 - \aver{X^2}] \aver{X (YY^\dagger)^2} \\
%      & + \frac{1}{6}[\aver{X}^3 - 3 \aver{X}\aver{X^2} + 2 \aver{X^3}]\aver{(YY^\dagger)^2}\\
%      &= (Y^\dagger Y) \aver{X} \aver{X^2 YY^\dagger} - \frac{1}{2}[\aver{X}^2 - \aver{X^2}] (Y^\dagger Y)\aver{X YY^\dagger} \\
%      & + \frac{1}{6}[\aver{X}^3 - 3 \aver{X}\aver{X^2} + 2 \aver{X^3}](Y^\dagger Y) \aver{YY^\dagger}.
% \end{aligned}
% \end{equation}
% From this we conclude that the greatest physical degree is 2 for trace of $YY^\dagger$.

% The same way that, 
% \begin{equation}
%     \aver{(YY^\dagger)^3}= \aver{Y^\dagger Y(YY^\dagger)^2} = \aver{(Y^\dagger Y) (Y^\dagger Y)(YY^\dagger)} =(Y^\dagger Y) (Y^\dagger Y) \aver{YY^\dagger}. 
% \end{equation}
% Given this, we have to
% \begin{equation}
%     \aver{(YY^\dagger)^n}= (Y^\dagger Y)^{n-1}\aver{YY^\dagger}.
% \end{equation}
% where $(Y^\dagger Y)^{n-1}$ is real. Therefore, the highest degree of trace of $YY^\dagger$ is 2. 

So we have at our disposal, the 5 matrices 
\begin{equation}
\label{powers}
\{X_u,X_u^2,X_d,X_d^2,YY^\dag\},
\end{equation}
forming a chain inside the trace, as in \eqref{invar}.
Powers greater than 3 can be reduced by the Cayley-Hamilton theorem \eqref{n=3:CH}.
Repetition of any of these matrices can be discarded because the identity \eqref{ABAC} allows the reduction.
Table\,\ref{tab:flavor_invariants} lists up to $I^+_{242}$ the traces with at most three matrices of the set \eqref{powers}
in the chain.
The invariants $I^\pm_{442}$ are of the same type.
For some cases, we also separate between the invariant with labels $\pm$ indicating the sum and subtraction between complex conjugates, because
\begin{equation}
\aver{AB}^*=\aver{A^\dag B}\,,
\end{equation}
for Hermitian $B$ and generic $A$.
So the invariants with $+$ are real, while the invariants with $-$ are purely imaginary.
One example is  
\begin{equation}
    I^+_{442}= \aver{X_u^2 X_d YY^\dagger + X_d X_u^2 YY^\dagger},
\end{equation}
which is real and not reducible.
% This invariant stands out for its non-reducibility, encapsulating interactions at a fundamental level without being expressible in terms of simpler invariants. It arises naturally when applying the Cayley-Hamilton theorem to decompose higher powers of $X_u$ and multiplying by $X_d YY^\dagger$, highlighting its foundational role in the model's flavor structure.

The invariants $I^-_{660},I^-_{622},I^-_{262},I^-_{642},I^-_{462}$ have four matrices in the chain. The first is the same as in the SM and we note that only the invariants with minus sign are listed.
That is because the same invariants with plus sign are reducible.

Let us analyze how we can reduce them. For example, let us consider
\begin{equation}
    I_{622}^+ = \aver{X_u^2X_dX_uYY^\dag} + \aver{X_uX_dX_u^2YY^\dag}\label{I622},
\end{equation}
which is not in the list of Table~\ref{tab:flavor_invariants} and should be reducible. This invariant can be simplified through the application of the mathematical identity \eqref{ABAC} involving the product of matrices $A,B,C$. By considering the second term of \eqref{I622} as $\aver{\underbrace{X_u}_{\text{A}} \underbrace{X_d X_u}_{\text{B}} \underbrace{X_u}_\text{A}  \underbrace{YY^\dagger}_{\text{C}}}$ and applying
the identity, one can express it in terms of lower degree invariants as 
\begin{equation}
\begin{aligned}
    I^+_{622} &= \aver{X_u^2 X_d X_u YY^\dagger} +\frac{1}{2}\aver{X_u}^2\aver{X_d X_u}\aver{YY^\dagger} - \frac{1}{2}\aver{X_d X_u YY^\dagger}\aver{X_u}^2  \\
    &- \aver{X_u^2 X_d}\aver{X_u}\aver{YY^\dagger}- \aver{X_u YY^\dagger} \aver{X_u} \aver{X_d X_u} + \aver{X_u X_d X_u YY^\dagger}\aver{X_u}  \\
    &+ \aver{X_d X_u^2 YY^\dagger}\aver{X_u} -\frac{1}{2} \aver{X_u^2}\aver{X_d X_u}\aver{YY^\dagger} + \aver{X_u^2 X_d}\aver{X_u YY^\dagger}  \\
    &+ \frac{1}{2}\aver{X_u^2}\aver{X_d X_u YY^\dagger} +\aver{ YY^\dagger}  \aver{X_u^3 X_d} +  \aver{X_d X_u} \aver{X_u^2 YY^\dagger} \\
    &- \aver{X_u^2 X_d X_u YY^\dagger} -\aver{X_d X_u^3 YY^\dagger}
\,.
\end{aligned}
\end{equation}
The first term is cancelled, and the term $\aver{X_u X_d X_u YY^\dagger}$ can be reduced again by the identity \eqref{ABAC}. The
terms $\aver{X_u^3 X_d},  \aver{X_d X_u^3 YY^\dagger}$ can be reduced by the Cayley-Hamilton theorem such that the invariant becomes
\begin{equation}
\begin{aligned}
    I^+_{622} &= \frac{2}{3} \aver{X_u}^3 \aver{X_d} \aver{YY^\dag}  - \frac{2}{3} \aver{X_u}^3 \aver{X_d YY^\dag} - \aver{X_u}^2 \aver{X_d}  \aver{X_u YY^\dag} \\
    & - \aver{X_u}^2 \aver{X_u  X_d} \aver{YY^\dag} - \aver{X_u} \aver{X_u^2} \aver{X_d} \aver{YY^\dag} + \aver{X_u^2} \aver{X_u} \aver{X_d  YY^\dag} \\
    & + \aver{X_u}^2 \aver{(X_d X_u + X_u X_d)  YY^\dag} - \aver{X_u} \aver{(X_d X_u^2 + X_u^2 X_d) YY^\dag}\\
    & +  \aver{X_u} \aver{X_d} \aver{X_u X_u YY^\dag} + \aver{X_u X_d} \aver{X_u^2 YY^\dag} +  \aver{X_u} \aver{X_u^2 X_d} \aver{YY^\dag}\\
    & +  \aver{X_u^2 X_d} \aver{X_u YY^\dag} + \frac{1}{3}  \aver{X_u^3} \aver{X_d} \aver{YY^\dag}  - \frac{1}{3} \aver{X_u^3} \aver{X_d YY^\dag} \,.
\end{aligned}
\end{equation}
The reduced form involves only the invariants of Table~\ref{tab:flavor_invariants}.

\section{Comparison with Ref.\,\cite{albergaria_cp-odd_2023}}

In Section \ref{Sec:5.1}, we derived results that allow for comparison with the findings presented in Ref.\,\cite{albergaria_cp-odd_2023}, which studies invariants of the WBT \eqref{wbtransformation} in a generic basis. One of the key  quantities used to formulate conditions for the invariants is $H_d^{(r)}$, which can be written in our notation as
\begin{equation}
    H_d^{(r)}= \frac{v^2}{2}\Tilde{Y}^d \Big(\frac{v^2}{2}\Tilde{Y}^{d\dagger} \Tilde{Y}^d + \Tilde{M}^{B\dagger} \Tilde{M}^B\Big)^{r-1} \Tilde{Y}^{d\dagger}, \label{Hdr}
\end{equation}
or as defined in \eqref{X:s}
\begin{equation}
     H_d^{(r)}= \frac{v^2}{2}\Tilde{Y}^d \Big(\frac{v^2}{2} X'_{dB} + X_{M^B}\Big)^{r-1} \Tilde{Y}^{d\dagger}.
\end{equation}
For $r=1$ it  simplifies to
\begin{equation}
    H_d^{(1)}= \frac{v^2}{2}\Tilde{Y}^d \Tilde{Y}^{d\dagger}= \frac{v^2}{2}X_{dB},
\end{equation}
where $X_{dB}$ was defined in \eqref{X:s}. Since they are proportional, we consider $X_{dB}$ instead of $H_d^{(1)}$.

Considering $r=2$, which is the case we will often use, we have that
\begin{equation}
    H_d^{(2)}= \frac{v^2}{2}(X_d^2 +YY^\dagger X_d + X_d YY^\dagger + (YY^\dagger)^2) + YY^\dagger |M|^2. 
\end{equation}
From here on we simplify the notation of $M^B\to M$. We can write the quantity \eqref{Hdr} in terms of physical masses using \eqref{Physis:mass}: 
\begin{equation}
    H_d^{(r)} = A_L \mathcal{D} \mathcal{W}_{d_R}^{\dagger} ({W}_{d_R} \mathcal{D}^\dagger A_L^\dagger A_L \mathcal{D} \mathcal{W}_{d_R}^\dagger + {W}_{d_R} \mathcal{D}^\dagger B_L^\dagger B_L \mathcal{D} \mathcal{W}_{d_R}^\dagger)^{r-1} \mathcal{W}_{d_R}  \mathcal{D} A_L^\dagger.  
\end{equation}
Using the identity \eqref{identitys}, we get
\begin{equation}
    H_d^{(r)} = A_L \mathcal{D} \mathcal{W}_{d_R}^{\dagger} ({W}_{d_R} \mathcal{D}^\dagger \mathcal{D} \mathcal{W}_{d_R}^\dagger)^{r-1} \mathcal{W}_{d_R}  \mathcal{D} A_L^\dagger = A(\mathcal{D}^2)^r A_L^\dagger, \label{Hd}
\end{equation}
such that
\begin{equation}
     H_d^{(2)} =A_L(\mathcal{D}^2)^2 A_L^\dagger
\end{equation}

Ref.~\cite{albergaria_cp-odd_2023} lists in its eq.~(25) a set of 7 CP odd invariants which, if vanishing, would correspond to the necessary and sufficient conditions for CP invariance. They consider a model with one singlet VLQ of up-type, so we adapt their invariants to our case. In the following, we will write these invariants  in terms of our basic invariants of Table \ref{tab:flavor_invariants} by specializing to the vanishing $\overline{M^B}$ basis in \eqref{base yd mb}. We write six of these invariants in the following:
    \begin{equation}
        \begin{aligned}
            \aver{[X_{dB},X_u]H_d^{(2)}} &= I_{222}^- |M|^2 , \\
            \aver{[X_{dB},X_u^2]H_d^{(2)}} &= -I_{422}^- |M|^2 ,\\
            \aver{[X_{dB}^2,X_u]H_d^{(2)}} &= (I_{242}^-  + I_{222}^- Y^\dagger Y) |M|^2,\\
            \aver{[X_{dB}^2,X_u^2]H_d^{(2)}} &= (I_{422}^- Y^\dagger Y - I_{442}^- ) |M|^2, \\
              \aver{\{X_{dB},X_u\} [X_u^2,H_d^{(2)}]} &= \big[(I_{660}^- - I_{622}^- Y^\dagger Y + I_{642}^- Y^\dagger Y  + I_{222}^- I_{402} + I_{422}^- I_{202} \\
              &  - I_{200} I_{020} I^-_{422} + I_{020}  I^-_{622} + I_{220} I^-_{422} + I_{200} I^-_{442} - I^-_{642} \big] \frac{v^2}{2} \\
               &+\big[-I_{200} I^-_{442} - I_{622}^- + \frac{1}{2} (I_{200})^2 I^-_{222} + \frac{1}{2} I_{200} I^-_{222}\big]|M|^2 ,\\
             \aver{\{X_{dB}^2,X_u^2 \} [X_u,H_d^{(2)}]} &= \big[I_{200} I_{442}^-  - I_{200} I_{422}^- + \frac{1}{2} I_{242}^- I_{400} - \frac{1}{2} I_{242}^- (I_{200})^2 \\
             & - \frac{1}{2} I_{222}^-(I_{200})^2 Y^\dag Y + \frac{1}{2} I_{222}^- I_{400} Y^\dag Y \\
             & +I_{642}^- + I_{402} I_{222}^- + I_{202} I_{422}^-\big] |M|^2        
        \end{aligned}
    \end{equation}
The right-hand side is written in terms of our basic invariants. Here, it is already possible to observe that they are written in terms of only 7 of our invariants, where the invariants  $I^-_{262}$ and $I^-_{462}$ do not appear.
% are not taken into account, 
% these two basic invariants vanishing with the conditions $\aver{[X_{dB},X_u^2]H_d^{(2)}}$ and $\aver{[X_{dB}^2,X_u]H_d^{(2)}}$, respectively.  This indicates an inconsistency with the Hilbert series and its basic invariants. 

The seventh invariant of Ref.\,\cite{albergaria_cp-odd_2023} is,
    \begin{equation}
     \begin{aligned}
        \aver{[X_{dB}, X_u]^3} &= \aver{[(X_d + YY^\dagger), X_u]^3},  \\
       &=  3 [I^-_{660} - I_{200} I_{020} I^-_{422} + I_{020}  I^-_{622} + I_{220} I^-_{422} + I_{200} I^-_{442} - 2 I^-_{642} \\
       & - I^-_{222}I_{402} + I^-_{622} Y^\dagger Y - I^-_{422}I_{202}]\frac{v^2}{2},
       \label{minimal:set}
    \end{aligned}
    \end{equation}
    which is also written in terms of the same 7 invariants already mentioned. It is worth mentioning that in the fifth line of equation (25) of \cite{albergaria_cp-odd_2023} there is a typo, since $\aver{\{H^{(2)}_u, h_d^2\}[ h_d, H^{(2)}_u]}=0$. To be consistent with the right-hand side, this invariant
    would be  $\aver{\{h_u^2, h_d^2\}[ h_d, H^{(2)}_u]}$. In the notation adopted in this dissertation, it becomes  $\aver{\{X_{dB}^2, X_u^2\}[X_u, H^{(2)}_d]}$.
    
    % Using the equation \eqref{Physis:mass} we have
    % \begin{equation}
    %     \begin{aligned}
    %         \aver{[X_{dB},X_u] H_d^{(2)}} &= 2i m^2_{d\alpha} m^2_{ui} m^4_{d\beta} \Im[F_{\alpha \beta} V^*_{\alpha i} V_{\beta i}]\\
    %         & \vdots
    %     \end{aligned}
    % \end{equation}

   % Some questions may arise regarding the existence of additional conditions beyond the minimal conditions. However, within this scope, when manipulating the degree within these conditions, we find, for example, that
   %  \begin{equation}
   %      \aver{[X_{dB}^2,X_u] \{X_u^2, H_d^{(2)}\}}. 
   %  \end{equation}
   %  This condition cannot be reduced and written in terms of the basics. Therefore, it is not a condition that can be adopted. For higher degrees in these conditions, the result remains unchanged, and only 7 CP-violating invariants continue to appear.
% \red{Não entendi o que quis dizer e omiti.}
    
\section{Invariants considering $(d_R,B_R)$} \label{Sec: 5.4}

Here we consider the Lagrangian \eqref{lagragian 1vlq}, where $(d_{iR},B_R)$ can be transformed by a $SU(4)$ rotation, implying that the relevant flavor group when $\tilde{Y}^d,\tilde{M}^B$ are switched off is 
    \begin{equation}
\label{group:dR.BR}
        SU(3)_q \otimes SU(4)_{d_R} \otimes U(1)_{B_L}.
    \end{equation}
As usual, we already disregard $SU(3)_u$ by considering $X_u$ instead of $Y^u$. Differently from \eqref{vlq:number}, here the $U(1)_{B_L}$ charge is only for $B_L$. The transformations related to spurions are the inverse of \eqref{trans: vlq}, namely,
    \begin{equation}
        \begin{aligned}
\tilde{Y}^d&\to U_L^{q} \Tilde{Y}^d \mathcal{W}_{d_R}^\dagger,
\cr
\tilde{M}^B&\to e^{i\theta} \Tilde{M}^B \mathcal{W}_{d_R}^\dagger 
\cr
X_u&\to U_L^{q} X_u U_L^{q\dagger} \,.
\end{aligned}
    \end{equation}
Therefore, with respect to the group \eqref{group:dR.BR}, the spurions $X_u$, $\tilde{Y}^B$ and $\tilde{M}^B$ transform as follows:
    \begin{equation}
        \begin{aligned}
            X_u &\sim  (3_q \otimes \bar{3}_q,0,0),\\
            \tilde{Y}^d  &\sim (3_q,\bar{4}_D,0),\\
            \tilde{Y}^{d\dagger}  &\sim (\bar{3}_q, 4_D,0),\\  
            \tilde{M}^B  &\sim (0,\bar{4}_D,1),\\
            \tilde{M}^{B\dagger} &\sim (0,4_D,-1), 
            \label{tranformations:5.4}
        \end{aligned}
    \end{equation}
Again, we consider both $\tilde{M}^B,\tilde{M}^{B\dagger}$ instead of $\tilde{M}^{B\dagger}\tilde{M}^B$ because it is rank one.

Taking all spurions in \eqref{tranformations:5.4} equal to $x$, the expression for the
unrefined Hilbert series is
    \begin{align}
    H(x) &= \frac{1}{(2\pi i)^6} \oint_{|z_1|=1} dz_1 \oint_{|z_2|=1} dz_2 \oint_{|z_3|=1} dz_3 \oint_{|z_4|=4} dz_1 \oint_{|z_5|=1} dz_5 \oint_{|z_6|=1} dz_6  \times \nonumber \\
    &\times (1-z_2 z_3)(z_3-z^2_2)(z_2-z^2_3)(1-z_4 z_6)(z_5-z^2_4)(z_6-z_4 z_5)(z_4 z_6-z^2_5) \times \nonumber \\
    &\times (z_4 - z_5 z_6)(z_5-z^2_6) z_1^4 z_2^{12} z_3^{12} z_4^8 z_5^8 z_6^8  PE,\label{Hilbert:su4} 
    \end{align}
where the $PE$ is
    \begin{align}
        PE&= \frac{1}{(1-x)^3 (1-x z_2z_3)(z_3- xz_2^2)(z_2z_3- x)(z_3^2- xz_2)(z_2- xz_3^2)(z_2^2- xz_3)} \nonumber\\
       &\times \frac{1}{(z_4-x z_2)(z_5- xz_2z_4)(z_6- xz_2z_5)(1- xz_2z_6)(z_3z_4- x)(z_3z_5- xz_4)(z_3z_6-xz_5)}  \nonumber\\
       &\times \frac{1}{(z_3-x z_6)(z_2z_4-x z_3)(z_2 z_5-x z_3 z_4)(z_2 z_6-x z_3z_5)(z_2- xz_3z_6)} \nonumber \\
       &\times \frac{1}{(1-xz_4z_3)(z_2-xz_4)(z_3-xz_4z_2)(z_4-xz_5z_3)(z_4z_2-xz_5)(z_4z_3-xz_5z_2)(z_5-xz_6z_3)} \nonumber\\
       &\times \frac{1}{(z_5z_2-xz_6)(z_5 z_3-xz_6z_2)(z_6-xz_3)(z_2z_6-xz_2)} \nonumber\\
       &\times \frac{1}{(z_4-xz_1)(z_5-xz_1z_4)(z_6-xz_1z_5)(1-xz_1z_6)} \nonumber\\
       &\times \frac{1}{(1-xz_4)(z_4z_1-xz_5)(z_5z_1-xz_6)(z_1z_6-x)}.
    \end{align}
See Appendix \ref{Integrais: dr Br} for details.

Since the integral is highly complex and demands significant computational power, it was not feasible to solve the residue integrals directly  using the software Mathematica. Therefore, to tackle the $SU(4)$ case and obtain the few first terms of the Hilbert series, the method of Lehman and Martin in \cite{lehman_hilbert_2015} is employed. This method involves expanding Equation \eqref{Hilbert:su4} into series in $x$ and performing residue integration for each variable $z_i$. Then each pole in $z_i$ will be at zero, and the result should match the exact residue integral over all variables, followed by series expansion in $x$. After this procedure, we obtain the following result:
    \begin{equation}
        H(x)= 1 + x + 4x^2 + 6x^3 + \cdots \label{Hilb:Su(4)}
    \end{equation}
This series tells us that the number of linearly independent invariants is one for degree one, four for degree two and 6 for degree three. Note that this is the Hilbert series in  \eqref{Hilbert} and not the e Plethystic logarithm in \eqref{PL:log}. 
% \red{Faltou especificar que a expansão era em $x$ e os polos são em $z_i$! Isso é muito importante para entender.}

% Using the relation $\aver{H_d^{(r)} X_u^s}$ \cite{albergaria_cp-odd_2023}, it is possible to obtain the even invariants referring to equation \eqref{Hilb:Su(4)}, in which they are listed in Table \ref{tab:flavor:SU(4)}.

In Table~\ref{tab:flavor:SU(4)} we list the invariants of low degree that corresponds to the Hilbert series \eqref{Hilb:Su(4)}. The table contains one invariant of degree one, four of degree two and two of degree three. For degree one, the number is the same as in the Hilbert series. For degree two, corresponding to $4x^2$, the missing invariant is $(I_{200})^2$. A similar reasoning applies to $6x^3$ and the number 6 includes all the products of lower degree invariants.
\begin{table}[H]
\centering
\caption{Summary of flavor invariants with their degrees and CP parity.
The notation is similar to Table~\ref{tab:flavor_invariants}.
}
\begin{tabular}{|c|c|c|c|}
\hline
Flavor Invariant & Phys. degree & Degree in \eqref{Hilb:Su(4)} & CP \\
\hline
\( I_{200} = \aver{X_u} \) & 2 & 1 &+ \\
\( I_{400} = \aver{X_u^2} \) & 4 & 2&+ \\
\( I_{020} = \aver{\Tilde{Y}^d \Tilde{Y}^{d\dagger}} \) & 4 & 2 &+ \\
\( I_{002} = \aver{\Tilde{M}^B \Tilde{M}^{B\dagger}} \) & 2 & 2 & + \\
\( I_{220} = \aver{\Tilde{Y}^d \Tilde{Y}^{d\dagger}X_u} \) & 4 & 3 &+ \\
\( I_{300} = \aver{X_u^3} \) & 6 & 3&+ \\
% \( I_{040} = \aver{(\Tilde{Y}^d \Tilde{Y}^{d\dagger})^2} \) & 4 & 4 &+ \\
% \( I_{030} = \aver{(\Tilde{Y}^d \Tilde{Y}^{d\dagger})^3} \) & 6 & + \\
% \( I_{240} = \aver{(\Tilde{Y}^d \Tilde{Y}^{d\dagger})^2 X_u } \) & 6 & + \\
% \( I_{420} = \aver{(\Tilde{Y}^d \Tilde{Y}^{d\dagger}) X_u^2 } \) & 6 & + \\
% \( I_{402} = \aver{(\Tilde{M}^B \Tilde{M}^{B\dagger})X_u^2} \) & 6 & + \\
% \( I_{222} = \aver{(\Tilde{Y}^d \Tilde{Y}^{d\dagger})(\Tilde{M}^B \Tilde{M}^{B\dagger})X_u} \) & 6 & + \\
\hline
\end{tabular}
\label{tab:flavor:SU(4)}
\end{table}

% The physical parameter of the single down-quark consistis in 7 masses, in which three masses are given by up-quark $\mathcal{M}_u= \diag(m_u,m_c, m_t)$ and the three standard down-type quarks plus the mass of the VLQ given by $\mathcal{D}_d= (m_d, m_s, m_b, M)$, 3 CP-violating phases, and 6 angles.
The physical parameters of the model consist in 7 masses, the three masses of up quarks $\mathcal{M}_u= \diag(m_u,m_c, m_t)$, and the three masses of standard down-type quarks plus the mass of the VLQ given by $\mathcal{D}_d= (m_d, m_s, m_b, M)$.
The extended $3\times 4$ CKM matrix 
\eqref{ckm:3x4} contains three CP-violating phases and 6 angles.
The total of parameters is then 16 parameters.
This should be the number of denominator factors \eqref{HS:deno} of the Hilbert series had we managed to calculate it.

The computation of the unrefined Hilbert series for a model incorporating a fourth quark family, as detailed in \cite{hanany_hilbert_2011}, reveals its complexity, featuring a numerator with over 6000 terms. This complexity is mirrored in the case of Vector-Like Quarks (VLQs), where the transition to an SU(4) symmetry complicates the computation by increasing the number of integrations to be considered. Specifically, for our VLQ model, the Hilbert series calculation requires the evaluation of 6 residue integrals across a function comprising 46 terms, highlighting the significant computational challenges inherent in these analyses.

% \red{
% A gente tem 9 básicos CP odd, 
% Branco-Albergaria escrevem 7 e dizem que são necessário e suficientes.
% Essas duas coisas não são compatíveis.
% }

%% file: capitulos/6-Conclusion.tex
\chapter{Conclusion}

% \red{Não precisa falar só do caso de VLQ.
% Faça um parágrafo antes sobre o MP, sobre invariantes, sobre a serie de Hilbert.
% São coisas que você estudou na dissertação.
% Do jeito que está, só tem coisa do cap.5.
% }

In this dissertation, a study of the SM was conducted, focusing on CP violation through the CKM mechanism. We reviewed CP violation via weak basis invariants, which in the SM is described by the Jarlskog invariant. This example illustrates the usefulness of formulating CP violation in terms of weak basis invariant and also of characterizing the theory in terms of flavor invariants. So, flavor invariants might be important for exploring the parameter space of extended models in a basis-independent manner.

To explore flavor invariants, we have reviewed the use of the Hilbert series. The rigorous application of the Hilbert series for the identification and classification of flavor invariants introduces a novel approach to probing the structure of invariants. The investigation reveals that all flavor invariants can be generated from a finite set of basic invariants, constructible as polynomials of these elementary entities. These invariants are systematically enumerated using the Hilbert series, with the quantity of basic invariants further detailed through the plethystic logarithm. Both tools, alongside the Molien-Weyl formula, facilitate a deep dive into the algebraic structure of the invariants, enabling the determination of their numbers, degrees, and their interrelations (syzygies). This methodological framework greatly aid the explicit construction of all basic invariants.

After reviewing the application of the Hilbert series for the SM quark sector, a systematic investigation of flavor invariants within the quark sector of the SM extended by a VLQ was undertaken. Emphasizing that physical observables should remain invariant under flavor basis transformations, the study thoroughly explores CP-even and CP-odd weak basis invariants. We managed to compute the Hilbert series in the mass basis of the VLQ and flavor invariants comprise 28 basic flavor invariants in the generating set—9 of which are CP odd and the remaining CP even. The computational demands have so far hindered the calculation of integrals for the full $SU(4)$ group considering $(d_{iR},B_R)$. However, as shown in Sec.\ref{Sec: 5.4},
employing the Lehman and Martin method, we derived the first few terms of the expanded Hilbert series representing linearly independent, rather than basic, invariants. We have also compared the invariants of Ref.\,\cite{albergaria_cp-odd_2023} with ours and have shown that all the former can be written in terms of the latter.

% The application of invariant theory in flavor physics has proven exceptionally insightful.
% Future investigations into flavor invariants within comprehensive models, as well as establishing connections between low-energy and high-energy scale invariants, are anticipated to yield significant advancements. 

% \bigskip
% \hrule
% Contrary to initial assessments from the plethystic logarithm, a refined analysis identifies 7, instead of 9, basic generators as the minimal conditions for CP violation.

% \red{Essa é a conclusão do paper do Branco. Não do resultado da análise da serie de Hilbert.}

%% file: postextual/apendices.tex
% ----------------------------------------------------------
% Apêndices
% ----------------------------------------------------------

% ---
% Inicia os apêndices
% ---
\begin{apendicesenv}

% Imprime uma página indicando o início dos apêndices
\partapendices
% ----------------------------------------------------------
\chapter{Spontaneous Symmetry Breaking} \label{SSB:A}

    Let us first examine the scenario where a discrete symmetry is spontaneously broken. The Lagrangian of a real scalar field is
    \begin{equation}
        \mathcal{L}= \frac{1}{2} \partial_\mu \phi \partial^\mu \phi - V(\phi). \label{LSSB}
    \end{equation}
    It is observed that the potential is a function of the scalar field $\phi$, which remains invariant under transformation.
    \begin{equation}
        V(\phi) = V(-\phi).
    \end{equation}
    hence, it is evident that Lagrangian \eqref{LSSB} exhibits symmetry related to the parity transformation of the scalar field $\phi$
    \begin{equation}
        \phi \mapsto -\phi,
    \end{equation}
    and explicit equation for the potential is
    \begin{equation}
        V(\phi)=\frac{1}{2}\mu^2 \phi^2 + \frac{1}{4}|\lambda|\phi^4
    \end{equation}
   in the case where the parameter $\mu^2$ is positive, the minimum field value for the scalar field is
    \begin{equation}
        \frac{\partial V(\phi)}{\partial \phi} = 0 \mapsto \phi_0 = 0,
    \end{equation}
    this outcome is related to the theory's quantum ground state. So, in the case where $\mu^2 > 0$, the fundamental state is
    \begin{equation}
        \langle \phi \rangle = 0.
    \end{equation}
    For small oscillations around the equation above, the Lagrangian is described as
    \begin{equation}
        \mathcal{L}=\frac{1}{2} \partial_\mu \phi \partial^\mu \phi + \frac{1}{2}\mu^2 \phi^2,
    \end{equation}
    where the parameter $\mu$ describes the mass of a free particle. 

    In the case $\mu^2 <0$, the potential is
    \begin{equation}
        V(\phi)=\frac{1}{2}\mu^2 \phi^2 + \frac{1}{4}|\lambda|\phi^4
    \end{equation}
    determining the ground state, we find that 
    \begin{equation}
        \frac{\partial V(\phi)}{\partial \phi} = \mu^2 \phi + |\lambda|\phi^3
    \end{equation}
    and fundamental state is 
    \begin{equation}
        \phi_0^2=\frac{-\mu^2}{\lambda} \rightarrow \phi_0=\pm \sqrt{\frac{\mu^2}{\lambda}}=\pm v.
    \end{equation}
    in the equation above, it is apparent that the ground state is degenerate, as there is more than one configuration. Defining the field around the minimum, we have that 
    \begin{equation}
        \phi' = \phi- v,   
    \end{equation}
    such that the Lagrangian becomes
    \begin{equation}
        \mathcal{L}= \frac{1}{2} \partial_\mu \phi' \partial^\mu \phi' - \mu^2 \left(\frac{\phi'^4}{4v^2} + \frac{\phi'^3}{v} + \phi'^2 - \frac{v^2}{4} \right).
    \end{equation}
    The Lagrangian loses its invariance under the parity transformation of the scalar field, indicating spontaneous symmetry breaking, where the Lagrangian around the vacuum field lacks the original symmetry \cite{quigg1985gauge}.

    \section*{Spontaneous breaking of continuous global symmetry}

    Now, we will address the case where spontaneously broken symmetry is continuous and global. The Lagrangian for two scalar fields $\phi_1$ and $\phi_2$ is
    \begin{equation}
    \mathcal{L} = \frac{1}{2} \left[ \partial_\mu \phi_1 \partial^\mu \phi_1 + \partial_\mu \phi_2 \partial^\mu \phi_2 \right] - V (\phi_1^2 + \phi_2^2)
    \end{equation}

    The Lagrangian is invariant under transformations of the rotation group $SO(2)$,
    \begin{equation}
    \Phi = \begin{pmatrix} \phi_1 \\ \phi_1 \end{pmatrix} \rightarrow \begin{pmatrix} \cos\theta & \sin\theta \\ -\sin\theta & \cos\theta \end{pmatrix} \begin{pmatrix} \phi_1 \\ \phi_2 \end{pmatrix}.
    \end{equation}
    Written the potential in function of $\Phi^2$ is 
    \begin{equation}
    V(\Phi^2) = \frac{1}{2}\mu^2\Phi^2 + \frac{1}{4}| \lambda| (\Phi^2)^2, 
    \end{equation}
    where $\Phi^2 = \phi_1^2 + \phi_2^2$.
    
    In the case where $\mu^2 > 0$, the theory has a unique minimum, 
    \begin{equation}
    \langle\Phi\rangle_0 = \begin{pmatrix} 0 & 0 \end{pmatrix}.
    \end{equation}
    
    The Lagrangian for small oscillations of the fields around the vacuum is
    \begin{equation}
    \mathcal{L} = \frac{1}{2} \left( \partial_\mu \phi_1 \partial^\mu \phi_1 - \mu^2 \phi_1^2 \right) + \frac{1}{2} \left( \partial_\mu \phi_2 \partial^\mu \phi_2 - \mu^2 \phi_2^2 \right).         
    \end{equation}
    This Lagrangian describes two free particles with a mass $\mu$. When $\mu^2 < 0$, the theory's minimum is degenerate and satisfies the relation 
    \begin{equation}
        \langle\Phi^2\rangle_0 = -\frac{\mu^2}{|\lambda|} = v^2, 
    \end{equation} 
    where one of the possible vacuum configurations is 
    \begin{equation}
        \langle\Phi\rangle_0 = \begin{pmatrix} v \\ 0 \end{pmatrix},
    \end{equation} 
    defining the field $\Phi'$ around the vacuum configuration $\langle\Phi\rangle_0$ as 
    \begin{equation}
        \Phi' = \Phi - \langle\Phi\rangle_0 = \begin{pmatrix} \eta \\ \xi \end{pmatrix}
    \end{equation} 
    Therefore, the Lagrangian for $\Phi'$ is
    \begin{equation}
        \mathcal{L} = \frac{1}{2} \left( \partial_\mu \eta \partial^\mu \eta + 2\mu^2 \eta^2 \right) + \frac{1}{2} \left( \partial_\mu \xi \partial^\mu \xi \right). 
    \end{equation}

    The Lagrangian no longer maintains its invariance under transformation. This Lagrangian depicts two particles: the $\eta$ particle, whose mass squared equals $-2\mu^2$, and the $\xi$ particle, which remains massless.

    The emergence of massless bosons aligns with Goldstone's theorem, which postulates that the spontaneous breakdown of continuous global symmetry correlates with the existence of massless bosons in the spectrum of the theory, as discussed in reference. Furthermore, the count of massless bosons, often referred to as Nambu-goldstone bosons, matches the count of broken generators of the symmetry \cite{quigg1985gauge, thomson2013modern}.
    
% ----------------------------------------------------------
\chapter{CP violation via Trace} \label{Appendix trace}

Starting from the commutativity relation, we have that
    \begin{equation}
        \begin{aligned}
    \text{Tr}[H_u,H_d]^3 &= \text{Tr}[(X_u X_d - X_d X_u)^3] \\
    &= \text{Tr}\begin{aligned}[t]
        &(X_u X_d X_u X_d X_u X_d - X_u X_d X_d X_u X_u X_d - \\
        & X_d X_u X_u X_d X_u X_d + X_d X_u X_d X_u X_u X_d - \\
        & X_u X_d X_u X_d X_d X_u + X_u X_d X_d X_u X_d X_u + \\
        & X_d X_u X_u X_d X_d X_u - X_d X_u X_d X_u X_d X_u)
    \end{aligned}
\end{aligned}
\end{equation}
Considering the linearity of the trace operation and the cyclic property of the trace \cite{anton2013elementary}, we can implement significant simplifications in expressions involving matrix products. The cyclic property of the trace states that for any matrices A and B, the trace of their product $Tr(AB)$ is equal to the trace of $Tr(BA)$. This property allows us to cancel or rearrange terms in a given expression    
    \begin{align}
    \text{Tr}[X_u,X_d]^3 = \text{Tr}\begin{aligned}[t]
        &(- X_u X_d^2 X_u^2 X_d - X_d X_u^2 X_d X_u X_d +\\
        & + X_d X_u X_d X_u^2 X_d - X_u X_d X_u X_d^2 X_u + \\
        & + X_u X_d^2 X_u X_d X_u + X_d X_u^2 X_d^2 X_u )
    \end{aligned}
    \end{align}
    Rearranging the terms, we have
    \begin{align}
    \text{Tr}[X_u,X_d]^3 = \text{Tr}\begin{aligned}[t]
        &(- X_u X_d^2 X_u^2 X_d - X_u^2 X_d X_u X_d^2 - X_d^2 X_u^2 X_d X_u +  \\
        & + X_d X_u^2 X_d^2 X_u + X_d X_u^2 X_d^2 X_u +X_d X_u^2 X_d^2 X_u)
    \end{aligned}
    \end{align}

    \begin{align}
    \text{Tr}[X_u,X_d]^3 &= 3\text{Tr}(- X_u X_d^2 X_u^2 X_d + X_d X_u^2 X_d^2 X_u) \\
    &= 6 i \text{Tr}(X_u^2 X_d^2 X_u X_d). \hspace{4cm} \Box   \end{align}

% ----------------------------------------------------------

% \chapter{Group $SO(2)$} \label{Product}

% Verifying that $R(\theta)R(\phi) = R(\theta + \phi)$
%     \begin{equation}
%         R(\theta)R(\phi) =  \begin{pmatrix}
%                 \cos{\theta} \cos{\phi} - \sin{\theta} \sin{\phi} & \cos{\theta} \sin{\phi} + \sin{\theta} \cos{\phi} \\
%               -\sin{\theta} \cos{\phi} - \cos{\theta} \sin{\phi}  & -\sin{\theta} \sin{\phi} + \cos{\theta} \cos{\phi}            
%         \end{pmatrix},
%     \end{equation}
%     using the trigonometry relations, 
%     \begin{align}
%         \cos{a \pm b} = \cos{a} \cos{b} \mp \sin{a} \sin{b}, \\
%         \sin{a \pm b} = \sin{a} \cos{b} \pm \cos{a} \sin{b},
%     \end{align}
%     we conclude that $R(\theta)R(\phi) = R(\theta + \phi)$. 

%-----------------------------------------------------------

\chapter{Geometric Series} \label{Appendix geometric series}

   A geometric series is a sequence of terms obtained by repeatedly multiplying an initial term $a$ by a constant ratio $r$ \cite{weber2003essential}. Mathematically, this can be expressed as:
    \begin{equation}
        \sum_{k=0}^\infty a r^k = ar^0 + ar^1 + ar^2 +...
    \end{equation}
    Here, $r$ is equivalent to the product $m_1 m_1^*$, where $m_1$ and $m_1^*$ are specific constants related to our series. This expression lays the groundwork for our exploration.

   We can demonstrate it as follows: consider the partial sum $S_n$ of the first $n$ terms of our geometric series,
    \begin{eqnarray}
        S_n &=& ar^0 + ar^1 + ar^2 +... + a r^{n-1} \nonumber \\
        &=& ar^0 + ar^1 + ar^2 +... + a r^n r^{-1}.
    \end{eqnarray}
    Note that $ar^0=a$,  which is our initial term. Now, let us multiply the entire series by $r$, yielding: 
    \begin{equation}
        rS_n = ar + ar^2 + ar^3 +... + a r^n
    \end{equation}
    Subtracting $rS_n$ from $S_n$ given us:
    \begin{eqnarray}
        S_n - rS_n &=&  (ar^0 + ar^1 + ar^2 +... + a r^{n-1})-(ar + ar^2 + ar^3 +... + a r^n), \nonumber\\
        &=& ar^0 - ar^n, \nonumber\\
        &=& a(1-r^n).
    \end{eqnarray}
    This simplifies to:
    \begin{equation}
       S_n(1-r)= a(1-r^n) \nonumber 
    \end{equation}
    Rearranging, we find the formula for the partial sum $S_n$ of a geometrical series
    \begin{equation}
        S_n= \frac{a(1-r^n)}{1-r}; \quad r\neq1. 
    \end{equation}
    Series Convergence: For $|r|<1$ and $n\to \infty$, $r^n$ approaches 0, causing the series to converge to
    \begin{equation}
        S = \frac{1}{1-r}. 
    \end{equation}
    in terms of $m_1 m_1^*$, 
    \begin{equation}
        S = \frac{1}{1-(m_1 m_1^*)}. 
    \end{equation}

%-----------------------------------------------------------

% \chapter{Wyel Chamber and Simple roots}

%-----------------------------------------------

\chapter{Cayley-Hamilton theorem} \label{Cayley-H}
The Cayley-Hamilton theorem is a statement about square matrices and their characteristic polynomials. The characteristic polynomial of a $n\times n$  matrix is defined by \cite{anton2013elementary}:
    \begin{equation}
        f(\lambda) \equiv \text{Det}(\lambda \mathbb{I}_n -A)= \lambda^n + a_{n-1} \lambda^{n-1} + \cdots a_1 \lambda +a_0,
    \end{equation}
    where $\mathbb{I}_n$ is the n-dimensional identity matrix and $a_0, a_1, \cdots a_{n-1}$ are coefficients determined by the matrix A itself. The theorem asserts that every square matrix $A$ satisfies its own characteristic equation:
    \begin{equation}
        f(A)= A^n + a_{n-1} A^{n-1} + \cdots a_1 A + a_0 =0. 
    \end{equation}
   This powerful statement allows us to express $A^n$  as a linear combination of lower powers of $A$. 
    
    Special cases for $n=2$ and $n=3$:    
    
    For n=2:When applying the Cayley-Hamilton theorem to a $2 \times 2$ matrix, we find
    \begin{align} 
        A^2 = \aver{A} A - \frac{1}{2}[\aver{A}^2 - \aver{A^2}]\mathbb{I}_{2\times2}.
    \end{align}
    Taking the trace of this equation, we arrive at:
    \begin{align}
        \aver{A^2} &= \aver{A}\aver{A} + \frac{1}{2}[\aver{A^2} - \aver{A}^2] \aver{\mathbb{I}}, \nonumber \\
        \aver{A^2} &= \aver{A}^2 + \frac{1}{2}[\aver{A^2} - \aver{A}^2] 2,  \nonumber \\
        \aver{A^2} &= \aver{A}^2 + \aver{A^2} - \aver{A}^2, \nonumber \\
        \aver{A^2} &= \aver{A^2}.
    \end{align}
   
    Further manipulation by multiplying $A$ and taking the trace again yields:
    \begin{align}
        A^3 &= \aver{A} A^2 + \frac{1}{2}[\aver{A^2} - \aver{A}^2] \mathbb{I}\cdot A \nonumber \\
        \aver{A^3} &= \aver{A} \aver{A^2} + \frac{1}{2}[\aver{A^2} - \aver{A}^2] \aver{A} \nonumber \\
        \aver{A^3} &= \aver{A} \aver{A^2} + \frac{1}{2} \aver{A^2} \aver{A} - \frac{1}{2} \aver{A}^2 \aver{A}  \nonumber \\
        \aver{A^3} &=  \frac{3}{2}\aver{A} \aver{A^2} - \frac{1}{2} \aver{A^2} \aver{A}. 
    \end{align}
    \newpage
    For n=3: For a $3 \times 3$ matrix, the equation becomes a more complex,
    \begin{align}
        A^3 = \aver{A}A^2 - \frac{1}{2}[\aver{A}^2 - \aver{A^2}]A + \frac{1}{6}[\aver{A}^3 - 3 \aver{A}\aver{A^2} + 2 \aver{A^3}]\mathbb{I}_{3\times 3}. \label{A3}
    \end{align}
    Taking the trace and manipulating further, we find:
    \begin{align}
        \aver{A^3} &= \aver{A}\aver{A^2} - \frac{1}{2}[\aver{A}^2 - \aver{A^2}]\aver{A} + \frac{1}{6}[\aver{A}^3 - 3 \aver{A}\aver{A^2} + 2 \aver{A^3}]\aver{\mathbb{I}_{3\times 3}},\\
        \aver{A^3} &= \aver{A}\aver{A^2} - \frac{1}{2}\aver{A}^2 \aver{A} + \frac{1}{2} \aver{A^2}\aver{A} + \frac{1}{6}[\aver{A}^3 - 3 \aver{A}\aver{A^2} + 2 \aver{A^3}]\aver{\mathbb{I}_{3\times 3}},\\
        \aver{A^3} &= \aver{A}\aver{A^2} - \frac{1}{2}\aver{A}^2 \aver{A} + \frac{1}{2} \aver{A^2}\aver{A} + \frac{1}{2}\aver{A}^3 - \frac{3}{2} \aver{A}\aver{A^2} +  \aver{A^3},\\
         \aver{A^3} &=  \aver{A^3}. 
    \end{align}
    Multiplying the equation \eqref{A3} by $A$, we get:
     \begin{align}
        A^4 = \aver{A}A^3 - \frac{1}{2}[\aver{A}^2 - \aver{A^2}]A^2 + \frac{1}{6}[\aver{A}^3 - 3 \aver{A}\aver{A^2} + 2 \aver{A^3}]\mathbb{I}_{3\times 3} A.
    \end{align}    
    Taking the trace again, we find
     \begin{align}
        \aver{A^4} &= \aver{A}\aver{A^3} - \frac{1}{2}[\aver{A}^2 - \aver{A^2}]\aver{A^2} + \frac{1}{6}[\aver{A}^3 - 3 \aver{A}\aver{A^2} + 2 \aver{A^3}]\aver{A} ,\\
        \aver{A^4} &= \aver{A}\aver{A^3} - \frac{1}{2}\aver{A}^2 \aver{A^2} + \frac{1}{2} \aver{A^2}\aver{A^2} + \frac{1}{6} \aver{A}^3 \aver{A} - \frac{1}{2} \aver{A}^2\aver{A^2} + \frac{1}{3} \aver{A^3} \aver{A}\\
        \aver{A^4} &=  \frac{1}{6}\aver{A}^4 -\aver{A}^2 \aver{A^2} + \frac{4}{3} \aver{A^3}\aver{A} + \frac{1}{2}\aver{A^2}^4. 
    \end{align}

%----------------------------------------------------------

\chapter{Chosen Basis and properties of trace} \label{xuxd}

    The matrices $X_u$ and $X_d$ are defined in the chosen basis as follows:
    \begin{equation}
    X_u = \begin{bmatrix}
    y_u^2 &  &  \\
     & y_c^2 &  \\
     &  & y_t^2
    \end{bmatrix};
    \quad
   X_d = V \cdot
            \begin{bmatrix}
    y_d^2 &  &  \\
     & y_s^2 &  \\
     &  & y_b^2
    \end{bmatrix}
        \cdot V^\dagger \label{basis}
    \end{equation}
    Here, $X_u$ is a diagonal matrix with squared Yukawa couplings of the up-type quarks as its elements. $X_d$, on the other hand, involves the CKM matrix, $V$, indicating a basis transformation applied to the down-type quark Yukawa couplings.
    
    The trace of $X_u$ is is straightforward to calculate:
    \begin{equation}
        \aver{X_u}= y_u^2+y_c^2+y_d^2.
    \end{equation}
     Calculating the trace of $X_d$ we can utilize the cyclic property of trace, which asserts that the trace of the product of matrices remains invariant under cyclic permutations of those matrices:
    \begin{equation}
        \aver{ABC}= \aver{CAB}= \aver{BCA}.
    \end{equation}
   Applying this to $X_d$, we find:
    \begin{equation}
        \aver{X_d} = \aver{V \hat{X_d} V^\dagger} = \aver{V^\dagger V \hat{X_d}} = y_d^2 + y_s^2 + y_b^2 
    \end{equation}
    Here, $V$ is the CKM matrix, which is unitary, satisfying $V^\dagger V = \mathbb{I}$, and $X_d$. The same applies to higher orders of $X_d$. In this context, $\hat{X_d}$ is a diagonal matrix.

    For the trace of the product of both matrices, we explicitly calculate:
    \begin{equation}
        \aver{X_u X_d}= \aver{\begin{bmatrix}
    y_u^2 &  &  \\
     & y_c^2 &  \\
     &  & y_t^2
    \end{bmatrix} V \begin{bmatrix}
    y_d^2 &  &  \\
     & y_s^2 &  \\
     &  & y_b^2
    \end{bmatrix} V^\dagger}
    \end{equation}
So, for two quark families, the interaction can be written in terms of the Cabibbo angle, leading to a simplified expression for the trace of the product:
% Using the Cabibbo angle, we can express $y_{d'}$ and $y_{s'}$ as follows:   
%     \begin{align}
%         y_{d'}&= V_{ud} y_d + V_{us} y_s \\
%         y_{s'}&= V_{cd} y_d + V_{cs} y_s \\
%         y_{d'} &= \cos{\theta_C} y_d + \sin{\theta_C} y_s \\
%         y_{s'} &= -\sin{\theta_C} y_d + \cos{\theta_C} y_s
%     \end{align}
%     Using these relations and the fact that the generic CKM matrix is given by:
%     \begin{equation}
%       V=  \begin{bmatrix}
%     V_{ud} & V_{us} & V_{ub}  \\
%     V_{cd} & V_{cs} & V_{cb} \\
%     V_{td} & V_{ts} & V_{tb}
%     \end{bmatrix} 
%     \end{equation}
%    However, we make an approximation of the CKM matrix in the context of two flavors (up-down and charm-strange) with the Cabibbo angle$\theta$. In this approximation, the relevant elements of the CKM matrix are:
%     \begin{equation}
%         V_{ud} \approx \cos{\theta}, \quad V_{us} \approx \sin{\theta}, \quad V_{cd} \approx -\sin{\theta}, \quad V_{cs} \approx \cos{\theta}
%     \end{equation}
%    Other elements, such as $V_{ub},V_{cb}, V_{tb}, V_{td}, V_{ts}$ are considered small and can be neglected. Utilizing the trigonometric relation $\cos^2{\theta} = 1 - \sin^2{\theta}$we obtain the following relation:
    \begin{equation}
        \aver{X_u X_d} = y_u^2 y_s^2 + y_c^2 y_d^2 + (y_s^2 - y_d^2)(y_c^2 - y_u^2) \cos^2(\theta)
    \end{equation}

% ----------------------------------------------------------
\chapter{Reduction formulas}
% and syzigies} 
\label{identity}

Here we detail various reduction formulas that can be used to reduced higher degree invariants in terms of lower degree invariants in Sec.\,\ref{sec:invs:quarks}.
They rely on the Cayley-Hamilton theorem.

First, for two quark families in the SM, we show how to reduce $\aver{(X_uX_d)^2}$ in terms of $\aver{X_u^2X_d^2}$ and other lower degree invariants, as discussed in Sec.\,\ref{sec:SM:n=2}.

The Cayley-Hamilton theorem for a $2\times 2$ matrix $A$ was given in \eqref{CH:n=2}.
    % , we have
    % \begin{align} 
    %     A^2 = \aver{A} A - \frac{1}{2}[\aver{A}^2 - \aver{A^2}]\mathbb{I}_{2\times2}.
    % \end{align}
    Multiplying it by $A^2$ and taking the trace, it becomes
    \begin{equation}
        \aver{A^4} = \aver{A} \aver{A^3} - \frac{1}{2}[\aver{A}^2 - \aver{A^2}]\aver{A^2}.
    \end{equation}  
    Substituting $A\to A+B$ and taking the order $A^2B^2$,
\begin{align}
        4\aver{A^2B^2} + 2 \aver{ABAB}&= 3\aver{AB^2}\aver{A} + 3\aver{A^2B}\aver{B} - \frac{1}{2}\aver{A}^2\aver{B^2} -2\aver{A}\aver{B}\aver{AB} \cr
    &\quad
    -\frac{1}{2}\aver{A^2}\aver{B}^2 + 2 \aver{AB}^2 + \aver{A^2}\aver{B^2}.
    \end{align}
So we can always write $\aver{ABAB}$ in terms of $\aver{A^2B^2}$ and products of traces of lower degree.

Now, for three quark families in the SM, let us deduce the reduction formula \eqref{ABAC}.
We take the reduction formula for $\aver{A^4}$ in \eqref{A4}, which follows from the Cayley-Hamilton theorem \eqref{n=3:CH}, and substitute $A \rightarrow A + B + C$.
Then we obtain \eqref{ABAC} by collecting only the terms of order $A^2BC$\,\cite{jenkins_algebraic_2009}.
The identity should be valid separately for the different degrees in $A,B,C$ because the original identity is valid for generic matrices $A,B,C$.

    The identity \eqref{ABAC} relates the traces of different combinations of matrix products like $A^2BC$, $BCA^2$, $ABAC$, etc., representing different ways of multiplying the matrices $A$, $B$, and $C$. The order of multiplication is significant due to the non-commutativity of matrix multiplication up to the cyclic property of the trace.
    The identity is formed by a specific linear combination of these different matrix products such that, when summed together, the final result is zero. This indicates a specific relationship between these matrix products. The identity allows for the 
    elimination of traces $\langle ABAC \rangle$ where the same matrix is repeated.

% \textbf{Syzygies and Their Identification}
%    Now, let us write the identity for $\aver{A} = 0$:
%     \begin{equation}
%         2\aver{ABAC}=-2\aver{A^2\{B,C\}}
%     +2\aver{A^2B}\aver{C}+2\aver{A^2C}\aver{B} +2\aver{AB}\aver{AC}+\aver{A^2}[\aver{BC}-\aver{B}\aver{C}]    
%     \end{equation}
    
%     This identity assists in discovering the syzygies of a theory. Syzygies represent a way to express that certain linear combinations of generators of an ideal or module are zero.

%     The identity can be viewed as a specific form of syzygy, in that it expresses a zero relation among linear combinations of matrix products. In other words, it identifies a particular relationship (a syzygy) among the invariants constructed from these matrices. 

% Redução para $x^4$ \eduardo{aqui tenho que fazer para ordem 4 no caso?}

% ----------------------------------------------------------
\chapter{Hilbert Series for VLQ in mass basis
} 
\label{case2}

In this chapter, we delve into the derivation of the Hilbert series for VLQ, proving in more details to obtain the Hilber series \eqref{HS:resultados} in the mass basis of the VLQ $B_L,B_R$. We apply the Molien-Weyl formula in \eqref{H(q)}, combined with the $PE$ function in \eqref{PE}. The Hilbert series,$H$,  can be obtained through the integration over group measures and is given by:
    \begin{equation}
        \mathcal{H} = \int_{SU(3)} d\mu \int_{U(1)} d\mu \cdot PE[X_{u},R]\cdot PE[X_{d},R] \cdot PE[Y,R] \cdot PE[Y^{\dagger},R].
    \end{equation}
    where the four PE terms correspond to the four spurions: $X_u, X_d, Y, Y\dagger$. The spurions transform as follows:
    \begin{equation}
    \begin{aligned}
    &X_u \sim X_d \sim (3 \otimes\Bar{3},0), \\ 
    &Y \sim (3,-1),\\
    & Y^{\dagger} \sim (\overline{3},+1).
    \end{aligned}
    \end{equation}

    The characters for the respective transformations are:
    \begin{equation}
        \begin{aligned}
          \chi_3 &= z_1 + \frac{z_2}{z_1} + \frac{1}{z_2},\\
          \chi_{\Bar{3}} &= z_2 + \frac{z_1}{z_2} + \frac{1}{z_1}, \\
          \chi_1 &= \frac{1}{z_3},\\
          \chi_{\Bar{1}} &= z_3.
           \label{char-33}
    \end{aligned}
    \end{equation}
    Therefore, we can express the characters for each spurion $X_u, X_d, Y,Y^\dagger$, as: 
    \begin{equation}
        \begin{aligned}
            \chi_{3\times \Bar{3}} &= z_1 z_2 +\frac{z_1^2}{z_2} + \frac{z_2^2}{z_1} + 3 + \frac{z_2}{z_1^2} + \frac{z_1}{z_2^2} + \frac{1}{z_1 z_2},\\
             \chi_{3\times 1} &= \left( z_1 + \frac{1}{z_2} + \frac{z_2}{z_1}\right)\frac{1}{z_3} = \frac{z_1}{z_3} + \frac{1}{z_2 z_3}+ \frac{z_2}{z_1 z_3}, \\
        \chi_{\Bar{3}\times \Bar{1}} &= \left( z_2 + \frac{1}{z_1} + \frac{z_1}{z_2}\right)z_3 = z_2 z_3 + \frac{z_3}{z_1}+ \frac{z_3 z_1}{z_2}. 
        \end{aligned}
    \end{equation}
    The character $\chi_{3\times \overline{3}}$ is the same for both $X_u$ and $X_d$ since they both transform identically.
    \newpage
    Plethystic Exponential and Haar Measure: The PE and Haar measure for $X_u$ and $X_d$ from the characters involved are constructed as follows:
    \begin{align}
    &PE[X_{u,d},R] = \nonumber \\
    &\exp{\left\{ \sum_r \frac{X_u^r}{r} \left[ (z_1 z_2)^r+ \left( \frac{z_2^2}{z_1} \right)^r +\left( \frac{z_1^2}{z_2} \right)^r + 3+ \left( \frac{z_2}{z_1^2} \right)^r+ \left( \frac{z_1}{z_2^2} \right)^r +\left( \frac{1}{z_1 z_2}\right )^r \right]\right\}} \nonumber \\
    &\cdot \exp{\left\{ \sum_r \frac{X_d^r}{r} \left[(z_1 z_2)^r +\left( \frac{z_2^2}{z_1} \right)^r+ \left( \frac{z_1^2}{z_2} \right)^r + 3 + \left( \frac{z_2}{z_1^2} \right)^r + \left( \frac{z_1}{z_2^2} \right)^r + \left( \frac{1}{z_1 z_2}\right )^r \right] \right\}},
    \end{align}  
    using the logarithm identity: 
     \begin{equation}
        \sum \frac{x^k}{k} = -\ln(1-x), \label{log}
    \end{equation}
    to simplify the expressions, 
    \begin{align}
    \small
      &PE[X_{u,d},R]= \nonumber  \\
      &\frac{z_1 z_2 z^2_1 z^2_2 z_1 z_2}{(1 - X_u z_1 z_2)(z_1 - X_u z^2_2)(z_2 - X_u z^2_1)(1-X_u)^3 (z^2_1 - X_u z_2)(z^2_2 - X_u z_1)(z_1 z_2 - X_u)} \times \nonumber \\
        \quad & \nonumber \\
        & \frac{z_1 z_2 z^2_1 z^2_2 z_1 z_2}{(1 - X_d z_1 z_2)(z_1 - X_d z^2_2)(z_2 - X_d z^2_1)(1-X_d)^3 (z^2_1 - X_d z_2)(z^2_2 - X_d z_1)(z_1 z_2 - X_d)}. 
    \end{align}

    The Haar measure for $SU(3)$ is given by
    \begin{equation}
        \int_{SU(3)} d\mu = \frac{1}{(2 \pi i)^2} \oint_{|z=1|} \frac{dz_1}{z_1} \oint_{|z=1|} \frac{dz_2}{z_2} (1-z_1 z_2) \left(1-\frac{z^2_1}{z_2} \right) \left(1-\frac{z^2_2}{z_1} \right). \label{Haar}
    \end{equation}

    For $Y$ and $Y\dagger$, the PE is constructed using the characters and logarithm identity, resulting in:
    \begin{align}
    PE[Y,R] &= \exp{\left[ \sum_r \frac{Y^r}{r} \left(\frac{z_1^r}{z_3^r} + \frac{1}{z_2^r z_3^r}+ \frac{z_2^r}{z_1^r z_3^r}\right)\right]}, \\
    &= \frac{z_1 z_2 z_3^3}{(z_3 - Y z_1) (z_2 z_3 - Y) (z_1 z_3 - Y z_2)}.   \end{align}
    and 
     \begin{align}
    PE[Y^\dagger,R] &= \exp{\left[ \sum_r \frac{Y^{\dagger r}}{r} \left(z_3^r z_2^r + \frac{z_3^r}{z_1^r}+ \frac{z_3^r z_1^r}{z_2^r}\right)\right]}, \\
    &= \frac{z_1 z_2 }{(1 - Y^\dagger z_3 z_2) (z_1  - Y^\dagger z_3) (z_2 - Y^\dagger z_3 z_1)}.  
    \end{align}
     The Haar measure for $U(1)$ is
     \begin{equation}
      \int_{U(1)} d\mu_{U(1)} = \frac{1}{2\pi i} \oint_{|z_3|=1} \frac{dz_3}{z_3}.  
    \end{equation}

    Taking $X_u,X_d\to x$ and $Y,Y^\dag \to x$,and simplifying the terms, the Hilbert series is found to be:

   \begin{align}
    H &= \int\frac{dz_1}{z_1} \int\frac{dz_2}{z_2} \int\frac{dz_3}{z_3} \left[ \frac{(1 - z_1 z_2)(z_2 - z_1^2)(z_1 - z_2^2)z_1^8 z_2^8 z_3^2}{(1 - x z_1 z_2)(z_1 - x z_2^2)(z_2 - x z_1^2)(1 - x)^3(z_1^2 - x z_2)(z_2^2 - x z_1)} \right. \nonumber \\
    &+ \frac{(1 - z_1 z_2)(z_2 - z_1^2)(z_1 - z_2^2)z_1^8 z_2^8 z_3^2}{(z_1 z_2 - x)(1 - x z_1 z_2)(z_1 - x z_2^2)(z_2 - x z_1^2)(1 - x)^3(z_1^2 - x z_2)(z_2^2 - x z_1)} \nonumber \\
    &+ \left. \frac{(1 - z_1 z_2)(z_2 - z_1^2)(z_1 - z_2^2)z_1^8 z_2^8 z_3^2}{(z_1 z_2 - x)(z_3 - x z_1)(z_2 z_3 - x)(z_1 z_3 - x z_2)(1 - x z_2 z_3)(z_1 - x z_3)(z_2 - x z_1 z_3)} \right].
    \end{align}
    resulting in
    \begin{equation}
    H(x) = \frac{{1 + 2x^4 + 4x^5 + 5x^6 + 2x^7 + 2x^8 + 2x^9 + 5x^{10} + 4x^{11} + 2x^{12} + x^{16}}}{{(1 - x^4)^3 (1 - x^3)^6 (1 - x^2)^4 (1 - x)^2}}. \label{HSresultados}
        \end{equation}

%--------------------------------------------------

\chapter{Hilbert series with $d_R,B_R$} \label{Integrais: dr Br}

    Here we demonstrate the result obtained in Section \ref{Sec: 5.4}. We know that the employed group is $SU(3)q \otimes SU(4){d_R} \otimes U(1)_{VLQ}$ and it transforms under \eqref{tranformations:5.4}. Again, we apply the Molien-Weyl formula, combined with the plethystic exponential (PE), the Hilbert series becomes
     \begin{equation}
        \mathcal{H} = \int_{SU(4)} d\mu \int_{SU(3)} d\mu \int_{U(1)} d\mu \cdot PE[X_{u},R]\cdot  PE[Y,R] \cdot PE[Y^{\dagger},R] PE[Y,R] PE[M^{\dagger},R].
    \end{equation}
 We can express this in terms of the Haar measure given in \eqref{haar:u1}, \eqref{haar su3}, 
     \begin{align}
         H &= \frac{1}{(2\pi i)^6} \oint_{|z_1|=1} \frac{dz_1}{z_1} \oint_{|z_2|=1} \frac{dz_2}{z_2} \oint_{|z_3|=1} \frac{dz_3}{z_3} \oint_{|z_4|=1} \frac{dz_4}{z_4} \oint_{|z_5|=1} \frac{dz_5}{z_5} \oint_{|z_6|=1} \frac{dz_6}{z_6}\\
         &\times (1-z_2 z_3) \left(1- \frac{z^2_2}{z_3} \right) \left(1- \frac{z^2_3}{z_2}\right) (1-z_4 z_6) \left(1- \frac{z^2_4}{z_5} \right) \left(1- \frac{z_4 z_5}{z_6}\right) \left(1- \frac{z_5^2}{z_4 z_6} \right) \\
         &\times \left(1- \frac{z_5 z_6}{z_4}\right) \left(1- \frac{z_6^2}{z_5}\right) \\
         &\times PE[X_{u},R]\cdot  PE[Y,R] \cdot PE[Y^{\dagger},R] PE[M,R] PE[M^{\dagger},R]
     \end{align}

     Since these groups transform under \eqref{tranformations:5.4}, we can write the characters associated with each of the spurions as listed in Table \ref{tab:char}
     \begin{equation}
     \begin{aligned}
         \chi_{3\times \Bar{3}} &= z_1 z_2 + \frac{z^2_1}{z_2} + \frac{z^2_2}{z_1}+ \frac{z_1}{z^2_2}+\frac{z_2}{z^2_1} + \frac{1}{z_1 z_2} + 3,\\
         \chi_{3\times \Bar{4}} &= \frac{z_2}{z_4} + \frac{z_2 z_4}{z_5} + \frac{z_2 z_5}{z_6} + z_2 z_6+\frac{1}{z_3 z_4}+\frac{z_4}{z_3z_5} + \frac{z_5}{z_3 z_6} \frac{z_6}{z_3}+\frac{z_3}{z_2 z_4}+\frac{z_3 z_4}{z_2 z_5} + \frac{z_3 z_5}{z_2 z_6}+ \frac{z_3 z_6}{z_2}, \\
         \chi_{4\times \Bar{3}} &= \frac{z_4}{z_2} + \frac{z_2 z_4}{z_3} + \frac{z_3 z_5}{z_4} + z_4 z_3 +\frac{1}{z_2 z_6}+\frac{z_5}{z_4z_2} + \frac{z_5 z_2}{z_4 z_3} \frac{z_6 z_3}{z_5}+\frac{z_6}{z_2 z_5}+\frac{z_6 z_2}{z_3 z_5} + \frac{z_3}{z_6}+ \frac{z_2}{z_6 z_3}, \\
         \chi_{4\times \Bar{1}} &= \frac{z_1}{z_4} + \frac{z_1 z_4}{z_5}+ \frac{z_1 z_5}{z_6}+ z_1 z_6,\\
         \chi_{\Bar{4}\times \Bar{1}} &= \frac{z_4}{z_1} + \frac{z_5}{z_4 z_1}+ \frac{z_6}{z_1 z_5}+ \frac{1}{z_1 z_6}. 
         \end{aligned}
     \end{equation}
    Where each group is related to
   \begin{align}
        U(1) &\to z_1, \nonumber\\
        SU(3) &\to z_2, z_3, \nonumber \\
        SU(4) &\to z_4, z_5,z_6. \nonumber \\
    \end{align}
    
   Using the expressions \eqref{PE} and \eqref{log}, and making the same simplification as in the last appendix, we have that:
    \begin{equation}
    \begin{aligned}
        PE[X_u,R] &= \frac{z_2^4 z_3^4}{(1-X_u)^3 (1-X_u z_2z_3)(z_3- X_uz_2^2)(z_2z_3- X_u)}\\
        &\times \frac{1}{(z_3^2- X_uz_2)(z_2- X_uz_3^2)(z_2^2- X_uz_3)}
    \end{aligned}
    \end{equation}
    \begin{equation}
        \begin{aligned}
        PE[Y,R] &= \frac{z_2^4 z_3^4 z_4^3 z_5^3 z_6^3}{(z_4-Y z_2)(z_5- Y z_2z_4)(z_6- Y z_2z_5)(1- Y z_2z_6)}\\
        &\times\frac{1}{(z_3z_4- Y)(z_3z_5- Y z_4) (z_3z_6-Y z_5)(z_3-Y z_6)}\\
       &\times \frac{1}{(z_2 z_4-Y z_3)(z_2 z_5-Y z_3 z_4)(z_2 z_6-Y z_3z_5)(z_2- Y z_3z_6)}
    \end{aligned}
    \end{equation}
    \begin{equation}
    \begin{aligned}
        PE[Y^{\dagger},R] &= \frac{z_2^4 z_3^4 z_4^3 z_5^3 z_6^3}{(1-Y^{\dagger}z_4z_3)(z_2-Y^{\dagger}z_4)(z_3-Y^{\dagger}z_4z_2)(z_4-Y^{\dagger}z_5z_3)}\\
        &\times \frac{1}{(z_4z_2-Y^{\dagger}z_5)(z_4z_3-Y^{\dagger}z_5z_2) (z_5-Y^{\dagger}z_6z_3)(z_5z_2-Y^{\dagger}z_6)}\\
       &\times \frac{1}{(z_5 z_3-Y^{\dagger}z_6z_2)(z_6-Y^{\dagger}z_3)(z_2 z_6-Y^{\dagger})(z_6 z_3-Y^{\dagger}z_2)}
    \end{aligned}
    \end{equation}
       \begin{align}
           PE[M,R] &= \frac{z_4 z_5 z_6}{(z_4-M z_1)(z_5-M z_1z_4)(z_6-M z_1z_5)(1-M z_1z_6)}
       \end{align}
        \begin{align}
            PE[M^{\dagger},R]&= \frac{z_1^4 z_4 z_5 z_6}{(z_1-M^{\dagger}z_4)(z_4z_1-M^{\dagger}z_5)(z_5z_1-M^{\dagger}z_6)(z_1z_6-M^{\dagger})}
        \end{align}

        Simplifying the terms and setting all spurions equal to $x$, we obtain
         \begin{align}
    H(x) &= \frac{1}{(2\pi i)^6} \oint_{|z_1|=1} dz_1 \oint_{|z_2|=1} dz_2 \oint_{|z_3|=1} dz_3 \oint_{|z_4|=4} dz_1 \oint_{|z_5|=1} dz_5 \oint_{|z_6|=1} dz_6  \times \nonumber \\
    &\times (1-z_2 z_3)(z_3-z^2_2)(z_2-z^2_3)(1-z_4 z_6)(z_5-z^2_4)(z_6-z_4 z_5)(z_4 z_6-z^2_5) \times \nonumber \\
    &\times (z_4 - z_5 z_6)(z_5-z^2_6) z_1^3 z_2^{10} z_3^{10} z_4^5 z_5^5 z_6^5  PE.\label{Hilbert:su4} 
    \end{align}
    The PE, afer factoring out some terms, is
    \begin{align}
        PE&= \frac{1}{(1-x)^3 (1-x z_2z_3)(z_3- xz_2^2)(z_2z_3- x)(z_3^2- xz_2)(z_2- xz_3^2)(z_2^2- xz_3)} \nonumber\\
       &\times \frac{1}{(z_4-x z_2)(z_5- xz_2z_4)(z_6- xz_2z_5)(1- xz_2z_6)(z_3z_4- x)(z_3z_5- xz_4)(z_3z_6-xz_5)}  \nonumber\\
       &\times \frac{1}{(z_3-x z_6)(z_2z_4-x z_3)(z_2 z_5-x z_3 z_4)(z_2 z_6-x z_3z_5)(z_2- xz_3z_6)} \nonumber \\
       &\times \frac{1}{(1-xz_4z_3)(z_2-xz_4)(z_3-xz_4z_2)(z_4-xz_5z_3)(z_4z_2-xz_5)(z_4z_3-xz_5z_2)(z_5-xz_6z_3)} \nonumber\\
       &\times \frac{1}{(z_5z_2-xz_6)(z_5 z_3-xz_6z_2)(z_6-xz_3)(z_2z_6-x)(z_3 z_6 -x z_2)} \nonumber\\
       &\times \frac{1}{(z_4-xz_1)(z_5-xz_1z_4)(z_6-xz_1z_5)(1-xz_1z_6)} \nonumber\\
       &\times \frac{1}{(z_1-xz_4)(z_4z_1-xz_5)(z_5z_1-xz_6)(z_1z_6-x)}
    \end{align}

\chapter{Flavor Invariants Explicitly} \label{Invariants: i22}

Constructing the non-trivial invariants from Table \ref{tab:invariantes3damily}, we are writing them in the following basis:
\begin{equation}
\langle X_u V X_d V^\dagger \rangle
\end{equation}
where $V$ is the CKM matrix. The invariant $I_{22}$ is
\begin{align}
I_{22}= &yc^2 yd^2 \cos^2(\theta_{23}) \sin^2(\theta_{12}) + y_b^2 y_u^2 \sin^2(\theta_{13}) \nonumber \\
&+ ys^2 y_t^2 \cos^2(\delta) \cos^2(\theta_{23}) \sin^2(\theta_{12}) \sin^2(\theta_{13}) 
 \nonumber\\
&+ys^2 y_t^2 \cos^2(\theta_{23}) \sin^2(\delta) \sin^2(\theta_{12}) \sin^2(\theta_{13})  \nonumber\\
&+yd^2 y_t^2 \sin^2(\theta_{12}) \sin^2(\theta_{23})  \nonumber \\
&+yc^2 ys^2 \cos^2(\delta) \sin^2(\theta_{12}) \sin^2(\theta_{13}) \sin^2(\theta_{23})  \nonumber\\
&+yc^2 ys^2 \sin^2(\delta) \sin^2(\theta_{12}) \sin^2(\theta_{13}) \sin^2(\theta_{23})  \nonumber \\
&+\cos^2(\theta_{13}) (y_b^2 y_t^2 \cos^2(\theta_{23}) + ys^2 y_u^2 \sin^2(\theta_{12}) + y_b^2 yc^2 \sin^2(\theta_{23}))  \nonumber\\
&+\frac{1}{2} \cos^2(\theta_{12}) (2 yd^2 y_u^2 \cos^2(\theta_{13}) + (2 yc^2 ys^2 + yd^2 y_t^2 - yd^2 y_t^2 \cos(2 \theta_{13})) \cos^2(\theta_{23})  \nonumber \\
&+(yc^2 yd^2 + 2 ys^2 y_t^2 - yc^2 yd^2 \cos(2 \theta_{13})) \sin^2(\theta_{23}))  \nonumber \\
&+(yd^2 - ys^2) (yc^2 - y_t^2) \cos(\delta) \cos(\theta_{12}) \sin(\theta_{12}) \sin(\theta_{13}) \sin(2 \theta_{23}).
\end{align}
\newpage
\begin{equation}
\begin{aligned}
\small
    I_{24}= &\frac{1}{32} (8 y_b^4 y_c^2 + 12 y_c^2 y_d^4 + 12 y_c^2 y_s^4 + 8 y_b^4 y_t^2 + 12 y_d^4 y_t^2 + 12 y_s^4 y_t^2 + 16 y_b^4 y_u^2 \\
    & + 8 y_d^4 y_u^2 + 8 y_s^4 y_u^2 - 4 (y_d^4 - y_s^4) (y_c^2 + y_t^2 - 2 y_u^2) \cos(2 \theta_{12})\\
    &- 2 (y_d^4 - y_s^4) (y_c^2 + y_t^2 - 2 y_u^2) \cos(2 (\theta_{12} - \theta_{13}))\\
    &+ 8 y_b^4 y_c^2 \cos(2 \theta_{13}) - 4 y_c^2 y_d^4 \cos(2 \theta_{13}) - 4 y_c^2 y_s^4 \cos(2 \theta_{13}) + 8 y_b^4 y_t^2 \cos(2 \theta_{13})\\
    &- 4 y_d^4 y_t^2 \cos(2 \theta_{13}) - 4 y_s^4 y_t^2 \cos(2 \theta_{13}) - 16 y_b^4 y_u^2 \cos(2 \theta_{13}) + 8 y_d^4 y_u^2 \cos(2 \theta_{13})\\
    &+ 8 y_s^4 y_u^2 \cos(2 \theta_{13}) - 2 y_c^2 y_d^4 \cos(2 (\theta_{12} + \theta_{13})) + 2 y_c^2 y_s^4 \cos(2 (\theta_{12} + \theta_{13}))\\
    & - 2 y_d^4 y_t^2 \cos(2 (\theta_{12} + \theta_{13})) + 2 y_s^4 y_t^2 \cos(2 (\theta_{12} + \theta_{13})) + 4 y_d^4 y_u^2 \cos(2 (\theta_{12} + \theta_{13}))\\
    &- 4 y_s^4 y_u^2 \cos(2 (\theta_{12} + \theta_{13})) - 6 y_c^2 y_d^4 \cos(2 (\theta_{12} - \theta_{23})) + 6 y_c^2 y_s^4 \cos(2 (\theta_{12} - \theta_{23}))\\
    &+ 6 y_d^4 y_t^2 \cos(2 (\theta_{12} - \theta_{23})) - 6 y_s^4 y_t^2 \cos(2 (\theta_{12} - \theta_{23}))\\
    &+ y_c^2 y_d^4 \cos(2 (\theta_{12} - \theta_{13} - \theta_{23}))\\
    &- y_c^2 y_s^4 \cos(2 (\theta_{12} - \theta_{13} - \theta_{23})) - y_d^4 y_t^2 \cos(2 (\theta_{12} - \theta_{13} - \theta_{23}))\\
    & + y_s^4 y_t^2 \cos(2 (\theta_{12} - \theta_{13} - \theta_{23}))\\
    &- 4 y_b^4 y_c^2 \cos(2 (\theta_{13} - \theta_{23})) + 2 y_c^2 y_d^4 \cos(2 (\theta_{13} - \theta_{23})) + 2 y_c^2 y_s^4 \cos(2 (\theta_{13} - \theta_{23}))\\
    &+ 4 y_b^4 y_t^2 \cos(2 (\theta_{13} - \theta_{23})) - 2 y_d^4 y_t^2 \cos(2 (\theta_{13} - \theta_{23})) - 2 y_s^4 y_t^2 \cos(2 (\theta_{13} - \theta_{23}))\\
    &+ y_c^2 y_d^4 \cos(2 (\theta_{12} + \theta_{13} - \theta_{23})) - y_c^2 y_s^4 \cos(2 (\theta_{12} + \theta_{13} - \theta_{23}))\\
    &- y_d^4 y_t^2 \cos(2 (\theta_{12} + \theta_{13} - \theta_{23})) + y_s^4 y_t^2 \cos(2 (\theta_{12} + \theta_{13} - \theta_{23})) - 8 y_b^4 y_c^2 \cos(2 \theta_{23})\\
    &+ 4 y_c^2 y_d^4 \cos(2 \theta_{23}) + 4 y_c^2 y_s^4 \cos(2 \theta_{23}) + 8 y_b^4 y_t^2 \cos(2 \theta_{23}) - 4 y_d^4 y_t^2 \cos(2 \theta_{23})\\
    &- 4 y_s^4 y_t^2 \cos(2 \theta_{23}) - 6 y_c^2 y_d^4 \cos(2 (\theta_{12} + \theta_{23})) + 6 y_c^2 y_s^4 \cos(2 (\theta_{12} + \theta_{23})) \\
    &+ 6 y_d^4 y_t^2 \cos(2 (\theta_{12} + \theta_{23})) - 6 y_s^4 y_t^2 \cos(2 (\theta_{12} + \theta_{23})) + y_c^2 y_d^4 \cos(2 (\theta_{12} - \theta_{13} + \theta_{23}))\\
    &- y_c^2 y_s^4 \cos(2 (\theta_{12} - \theta_{13} + \theta_{23})) - y_d^4 y_t^2 \cos(2 (\theta_{12} - \theta_{13} + \theta_{23}))\\
    &+ y_s^4 y_t^2 \cos(2 (\theta_{12} - \theta_{13} + \theta_{23}))\\
    &- 4 y_b^4 y_c^2 \cos(2 (\theta_{13} + \theta_{23})) + 2 y_c^2 y_d^4 \cos(2 (\theta_{13} + \theta_{23})) + 2 y_c^2 y_s^4 \cos(2 (\theta_{13} + \theta_{23})) \\
    &+ 4 y_b^4 y_t^2 \cos(2 (\theta_{13} + \theta_{23})) - 2 y_d^4 y_t^2 \cos(2 (\theta_{13} + \theta_{23})) - 2 y_s^4 y_t^2 \cos(2 (\theta_{13} + \theta_{23})) \\
    &+ y_c^2 y_d^4 \cos(2 (\theta_{12} + \theta_{13} + \theta_{23})) - y_c^2 y_s^4 \cos(2 (\theta_{12} + \theta_{13} + \theta_{23}))\\
    &- y_d^4 y_t^2 \cos(2 (\theta_{12} + \theta_{13} + \theta_{23})) + y_s^4 y_t^2 \cos(2 (\theta_{12} + \theta_{13} + \theta_{23}))\\
    &+ 2 y_c^2 y_d^4 \sin(\delta - 2 \theta_{12} - \theta_{13} - 2 \theta_{23}) - 2 y_c^2 y_s^4 \sin(\delta - 2 \theta_{12} - \theta_{13} - 2 \theta_{23})\\
    &- 2 y_d^4 y_t^2 \sin(\delta - 2 \theta_{12} - \theta_{13} - 2 \theta_{23}) + 2 y_s^4 y_t^2 \sin(\delta - 2 \theta_{12} - \theta_{13} - 2 \theta_{23})\\
    &- 2 y_c^2 y_d^4 \sin(\delta + 2 \theta_{12} - \theta_{13} - 2 \theta_{23}) + 2 y_c^2 y_s^4 \sin(\delta + 2 \theta_{12} - \theta_{13} - 2 \theta_{23})\\
    &+ 2 y_d^4 y_t^2 \sin(\delta + 2 \theta_{12} - \theta_{13} - 2 \theta_{23}) - 2 y_s^4 y_t^2 \sin(\delta + 2 \theta_{12} - \theta_{13} - 2 \theta_{23})\\
    &- 2 y_c^2 y_d^4 \sin(\delta - 2 \theta_{12} + \theta_{13} - 2 \theta_{23}) + 2 y_c^2 y_s^4 \sin(\delta - 2 \theta_{12} + \theta_{13} - 2 \theta_{23})\\
    &+ 2 y_d^4 y_t^2 \sin(\delta - 2 \theta_{12} + \theta_{13} - 2 \theta_{23}) - 2 y_s^4 y_t^2 \sin(\delta - 2 \theta_{12} + \theta_{13} - 2 \theta_{23})\\
    &+ 2 y_c^2 y_d^4 \sin(\delta + 2 \theta_{12} + \theta_{13} - 2 \theta_{23}) - 2 y_c^2 y_s^4 \sin(\delta + 2 \theta_{12} + \theta_{13} - 2 \theta_{23}) \nonumber
\end{aligned}
\end{equation}
\begin{equation}
\begin{aligned}
    &- 2 y_d^4 y_t^2 \sin(\delta + 2 \theta_{12} + \theta_{13} - 2 \theta_{23}) + 2 y_s^4 y_t^2 \sin(\delta + 2 \theta_{12} + \theta_{13} - 2 \theta_{23}) \\
    &- 2 y_c^2 y_d^4 \sin(\delta - 2 \theta_{12} - \theta_{13} + 2 \theta_{23}) + 2 y_c^2 y_s^4 \sin(\delta - 2 \theta_{12} - \theta_{13} + 2 \theta_{23}) \\ 
    &+ 2 y_d^4 y_t^2 \sin(\delta - 2 \theta_{12} - \theta_{13} + 2 \theta_{23}) - 2 y_s^4 y_t^2 \sin(\delta - 2 \theta_{12} - \theta_{13} + 2 \theta_{23}) \\
    &+ 2 y_c^2 y_d^4 \sin(\delta + 2 \theta_{12} - \theta_{13} + 2 \theta_{23}) - 2 y_c^2 y_s^4 \sin(\delta + 2 \theta_{12} - \theta_{13} + 2 \theta_{23}) \\
    &- 2 y_d^4 y_t^2 \sin(\delta + 2 \theta_{12} - \theta_{13} + 2 \theta_{23}) + 2 y_s^4 y_t^2 \sin(\delta + 2 \theta_{12} - \theta_{13} + 2 \theta_{23})\\
    &+ 2 y_c^2 y_d^4 \sin(\delta - 2 \theta_{12} + \theta_{13} + 2 \theta_{23}) - 2 y_c^2 y_s^4 \sin(\delta - 2 \theta_{12} + \theta_{13} + 2 \theta_{23}) \\
    &- 2 y_d^4 y_t^2 \sin(\delta - 2 \theta_{12} + \theta_{13} + 2 \theta_{23}) + 2 y_s^4 y_t^2 \sin(\delta - 2 \theta_{12} + \theta_{13} + 2 \theta_{23}) \\
    &- 2 y_c^2 y_d^4 \sin(\delta + 2 \theta_{12} + \theta_{13} + 2 \theta_{23}) + 2 y_c^2 y_s^4 \sin(\delta + 2 \theta_{12} + \theta_{13} + 2 \theta_{23})  \\
    &+ 2 y_d^4 y_t^2 \sin(\delta + 2 \theta_{12} + \theta_{13} + 2 \theta_{23}) - 2 y_s^4 y_t^2 \sin(\delta + 2 \theta_{12} + \theta_{13} + 2 \theta_{23})).
\end{aligned}
\end{equation}
\vspace{12pt}

\begin{equation}
\begin{aligned}
    I_{42}= & y_c^4 y_d^2 \cos^2(\theta_{23}) \sin^2(\theta_{12}) + y_b^2 y_u^4 \sin^2(\theta_{13})\\
& + y_s^2 y_t^4 \cos^2(\delta) \cos^2(\theta_{23}) \sin^2(\theta_{12}) \sin^2(\theta_{13}) \\
& + y_s^2 y_t^4 \cos^2(\theta_{23}) \sin^2(\delta) \sin^2(\theta_{12}) \sin^2(\theta_{13}) \\
& + y_d^2 y_t^4 \sin^2(\theta_{12}) \sin^2(\theta_{23}) \\
& + y_c^4 y_s^2 \cos^2(\delta) \sin^2(\theta_{12}) \sin^2(\theta_{13}) \sin^2(\theta_{23}) \\
& + y_c^4 y_s^2 \sin^2(\delta) \sin^2(\theta_{12}) \sin^2(\theta_{13}) \sin^2(\theta_{23}) \\
& + \cos^2(\theta_{13}) (y_b^2 y_t^4 \cos^2(\theta_{23}) + y_s^2 y_u^4 \sin^2(\theta_{12}) \\
& + y_b^2 y_c^4 \sin^2(\theta_{23})) \\
& + \frac{1}{2} \cos^2(\theta_{12}) (2 y_d^2 y_u^4 \cos^2(\theta_{13}) \\
& + (2 y_c^4 y_s^2 + y_d^2 y_t^4 - y_d^2 y_t^4 \cos(2 \theta_{13})) \cos^2(\theta_{23}) \\
& + (y_c^4 y_d^2 + 2 y_s^2 y_t^4 - y_c^4 y_d^2 \cos(2 \theta_{13})) \sin^2(\theta_{23})) \\
& + (y_d^2 - y_s^2) (y_c^4 - y_t^4) \cos(\delta) \cos(\theta_{12}) \sin(\theta_{12}) \sin(\theta_{13}) \sin(2 \theta_{23}) 
\end{aligned}
\end{equation}

\newpage
\begin{align}
    I_{44}= & \frac{1}{32} ( 8 y_b^4 y_c^4 + 12 y_c^4 y_d^4 + 12 y_c^4 y_s^4 + 8 y_b^4 y_t^4 + 12 y_d^4 y_t^4 + 12 y_s^4 y_t^4 + 16 y_b^4 y_u^4  \nonumber\\
    &+ 8 y_d^4 y_u^4 + 8 y_s^4 y_u^4 \nonumber\\
    & - 4 (y_d^4 - y_s^4) (y_c^4 + y_t^4 - 2 y_u^4) \cos(2 \theta_{12}) \nonumber\\
    & - 2 (y_d^4 - y_s^4) (y_c^4 + y_t^4 - 2 y_u^4) \cos(2 (\theta_{12} - \theta_{13})) \nonumber\\
    & + 8 y_b^4 y_c^4 \cos(2 \theta_{13}) - 4 y_c^4 y_d^4 \cos(2 \theta_{13}) - 4 y_c^4 y_s^4 \cos(2 \theta_{13}) + 8 y_b^4 y_t^4 \cos(2 \theta_{13}) \nonumber\\
    & - 4 y_d^4 y_t^4 \cos(2 \theta_{13}) - 4 y_s^4 y_t^4 \cos(2 \theta_{13}) - 16 y_b^4 y_u^4 \cos(2 \theta_{13}) + 8 y_d^4 y_u^4 \cos(2 \theta_{13}) \nonumber\\
    & + 8 y_s^4 y_u^4 \cos(2 \theta_{13}) - 2 y_c^4 y_d^4 \cos(2 (\theta_{12} + \theta_{13})) + 2 y_c^4 y_s^4 \cos(2 (\theta_{12} + \theta_{13})) \nonumber\\
    & - 2 y_d^4 y_t^4 \cos(2 (\theta_{12} + \theta_{13})) + 2 y_s^4 y_t^4 \cos(2 (\theta_{12} + \theta_{13})) + 4 y_d^4 y_u^4 \cos(2 (\theta_{12} + \theta_{13})) \nonumber\\
    & - 4 y_s^4 y_u^4 \cos(2 (\theta_{12} + \theta_{13})) - 6 y_c^4 y_d^4 \cos(2 (\theta_{12} - \theta_{23})) + 6 y_c^4 y_s^4 \cos(2 (\theta_{12} - \theta_{23})) \nonumber\\
    & + 6 y_d^4 y_t^4 \cos(2 (\theta_{12} - \theta_{23})) - 6 y_s^4 y_t^4 \cos(2 (\theta_{12} - \theta_{23})) + y_c^4 y_d^4 \cos(2 (\theta_{12} - \theta_{13} - \theta_{23})) \nonumber\\
    & - y_c^4 y_s^4 \cos(2 (\theta_{12} - \theta_{13} - \theta_{23})) - y_d^4 y_t^4 \cos(2 (\theta_{12} - \theta_{13} - \theta_{23})) \nonumber\\
    & + y_s^4 y_t^4 \cos(2 (\theta_{12} - \theta_{13} - \theta_{23})) - 4 y_b^4 y_c^4 \cos(2 (\theta_{13} - \theta_{23})) + 2 y_c^4 y_d^4 \cos(2 (\theta_{13} - \theta_{23})) \nonumber\\
    & + 2 y_c^4 y_s^4 \cos(2 (\theta_{13} - \theta_{23})) + 4 y_b^4 y_t^4 \cos(2 (\theta_{13} - \theta_{23})) - 2 y_d^4 y_t^4 \cos(2 (\theta_{13} - \theta_{23})) \nonumber\\
    & - 2 y_s^4 y_t^4 \cos(2 (\theta_{13} - \theta_{23})) + y_c^4 y_d^4 \cos(2 (\theta_{12} + \theta_{13} - \theta_{23})) \nonumber\\
    & - y_c^4 y_s^4 \cos(2 (\theta_{12} + \theta_{13} - \theta_{23})) - y_d^4 y_t^4 \cos(2 (\theta_{12} + \theta_{13} - \theta_{23})) \nonumber\\
    & + y_s^4 y_t^4 \cos(2 (\theta_{12} + \theta_{13} - \theta_{23})) - 8 y_b^4 y_c^4 \cos(2 \theta_{23}) + 4 y_c^4 y_d^4 \cos(2 \theta_{23}) \nonumber\\
    & + 4 y_c^4 y_s^4 \cos(2 \theta_{23}) + 8 y_b^4 y_t^4 \cos(2 \theta_{23}) - 4 y_d^4 y_t^4 \cos(2 \theta_{23}) - 4 y_s^4 y_t^4 \cos(2 \theta_{23}) \nonumber\\
    & - 6 y_c^4 y_d^4 \cos(2 (\theta_{12} + \theta_{23})) + 6 y_c^4 y_s^4 \cos(2 (\theta_{12} + \theta_{23})) + 6 y_d^4 y_t^4 \cos(2 (\theta_{12} + \theta_{23})) \nonumber\\
    & - 6 y_s^4 y_t^4 \cos(2 (\theta_{12} + \theta_{23})) + y_c^4 y_d^4 \cos(2 (\theta_{12} - \theta_{13} + \theta_{23})) \nonumber\\
    & - y_c^4 y_s^4 \cos(2 (\theta_{12} - \theta_{13} + \theta_{23})) - y_d^4 y_t^4 \cos(2 (\theta_{12} - \theta_{13} + \theta_{23})) \nonumber\\
    & + y_s^4 y_t^4 \cos(2 (\theta_{12} - \theta_{13} + \theta_{23})) - 4 y_b^4 y_c^4 \cos(2 (\theta_{13} + \theta_{23})) + 2 y_c^4 y_d^4 \cos(2 (\theta_{13} + \theta_{23})) \nonumber\\
    & + 2 y_c^4 y_s^4 \cos(2 (\theta_{13} + \theta_{23})) + 4 y_b^4 y_t^4 \cos(2 (\theta_{13} + \theta_{23})) - 2 y_d^4 y_t^4 \cos(2 (\theta_{13} + \theta_{23})) \nonumber\\
    & - 2 y_s^4 y_t^4 \cos(2 (\theta_{13} + \theta_{23})) + y_c^4 y_d^4 \cos(2 (\theta_{12} + \theta_{13} + \theta_{23})) \nonumber\\
    &  - y_c^4 y_s^4 \cos(2 (\theta_{12} + \theta_{13} + \theta_{23})) - y_d^4 y_t^4 \cos(2 (\theta_{12} + \theta_{13} + \theta_{23}))  \nonumber\\
    &+ y_s^4 y_t^4 \cos(2 (\theta_{12} + \theta_{13} + \theta_{23}))\nonumber\\
    & + 2 y_c^4 y_d^4 \sin(\delta - 2 \theta_{12} - \theta_{13} - 2 \theta_{23}) - 2 y_c^4 y_s^4 \sin(\delta - 2 \theta_{12} - \theta_{13} - 2 \theta_{23})\nonumber\\
    & - 2 y_d^4 y_t^4 \sin(\delta - 2 \theta_{12} - \theta_{13} - 2 \theta_{23}) + 2 y_s^4 y_t^4 \sin(\delta - 2 \theta_{12} - \theta_{13} - 2 \theta_{23}) \nonumber\\
    & - 2 y_c^4 y_d^4 \sin(\delta + 2 \theta_{12} - \theta_{13} - 2 \theta_{23}) + 2 y_c^4 y_s^4 \sin(\delta + 2 \theta_{12} - \theta_{13} - 2 \theta_{23})\nonumber\\
    & + 2 y_d^4 y_t^4 \sin(\delta + 2 \theta_{12} - \theta_{13} - 2 \theta_{23}) - 2 y_s^4 y_t^4 \sin(\delta + 2 \theta_{12} - \theta_{13} - 2 \theta_{23}) \nonumber\\
    & - 2 y_c^4 y_d^4 \sin(\delta - 2 \theta_{12} + \theta_{13} - 2 \theta_{23}) + 2 y_c^4 y_s^4 \sin(\delta - 2 \theta_{12} + \theta_{13} - 2 \theta_{23}) \nonumber\\
    & + 2 y_d^4 y_t^4 \sin(\delta - 2 \theta_{12} + \theta_{13} - 2 \theta_{23}) - 2 y_s^4 y_t^4 \sin(\delta - 2 \theta_{12} + \theta_{13} - 2 \theta_{23})\nonumber\\
    & + 2 y_c^4 y_d^4 \sin(\delta + 2 \theta_{12} + \theta_{13} - 2 \theta_{23}) - 2 y_c^4 y_s^4 \sin(\delta + 2 \theta_{12} + \theta_{13} - 2 \theta_{23}) \nonumber\\
    \end{align}
    \begin{align}
    & - 2 y_d^4 y_t^4 \sin(\delta + 2 \theta_{12} + \theta_{13} - 2 \theta_{23}) + 2 y_s^4 y_t^4 \sin(\delta + 2 \theta_{12} + \theta_{13} - 2 \theta_{23}) \nonumber\\
    & - 2 y_c^4 y_d^4 \sin(\delta - 2 \theta_{12} - \theta_{13} + 2 \theta_{23}) + 2 y_c^4 y_s^4 \sin(\delta - 2 \theta_{12} - \theta_{13} + 2 \theta_{23})\nonumber\\
    & + 2 y_d^4 y_t^4 \sin(\delta - 2 \theta_{12} - \theta_{13} + 2 \theta_{23}) - 2 y_s^4 y_t^4 \sin(\delta - 2 \theta_{12} - \theta_{13} + 2 \theta_{23}) \nonumber\\
    & + 2 y_c^4 y_d^4 \sin(\delta + 2 \theta_{12} - \theta_{13} + 2 \theta_{23}) - 2 y_c^4 y_s^4 \sin(\delta + 2 \theta_{12} - \theta_{13} + 2 \theta_{23}) \nonumber\\
    & - 2 y_d^4 y_t^4 \sin(\delta + 2 \theta_{12} - \theta_{13} + 2 \theta_{23}) + 2 y_s^4 y_t^4 \sin(\delta + 2 \theta_{12} - \theta_{13} + 2 \theta_{23}) \nonumber\\
    & + 2 y_c^4 y_d^4 \sin(\delta - 2 \theta_{12} + \theta_{13} + 2 \theta_{23}) - 2 y_c^4 y_s^4 \sin(\delta - 2 \theta_{12} + \theta_{13} + 2 \theta_{23}) \nonumber\\
    & - 2 y_d^4 y_t^4 \sin(\delta - 2 \theta_{12} + \theta_{13} + 2 \theta_{23}) + 2 y_s^4 y_t^4 \sin(\delta - 2 \theta_{12} + \theta_{13} + 2 \theta_{23}) \nonumber\\
    & - 2 y_c^4 y_d^4 \sin(\delta + 2 \theta_{12} + \theta_{13} + 2 \theta_{23}) + 2 y_c^4 y_s^4 \sin(\delta + 2 \theta_{12} + \theta_{13} + 2 \theta_{23}) \nonumber\\
    & + 2 y_d^4 y_t^4 \sin(\delta + 2 \theta_{12} + \theta_{13} + 2 \theta_{23}) - 2 y_s^4 y_t^4 \sin(\delta + 2 \theta_{12} + \theta_{13} + 2 \theta_{23})).     
    \end{align}
    
\begin{align}
    I^{(-)}_{66} = & 2i (y_b^2 - y_d^2) (y_b^2 - y_s^2) (y_d^2 - y_s^2) (y_c^2 - y_t^2) (y_c^2 - y_u^2) (y_t^2 - y_u^2) \nonumber\\
& \times \cos(\theta_{12}) \cos^2(\theta_{13}) \cos(\theta_{23}) \sin(\delta) \sin(\theta_{12}) \sin(\theta_{13}) \sin(\theta_{23})
\end{align}
    
\end{apendicesenv}
% ---

%% file: postextual/anexos.tex
% ----------------------------------------------------------
% Apêndices
% ----------------------------------------------------------

% ---
% Inicia os anexos
% ---
 \begin{anexosenv}

% % Imprime uma página indicando o início dos anexos
% \partanexos

% % ---
 \chapter{Ata-Defesa}

   

\section*{Supplementary material}

# A. Ata-Defesa

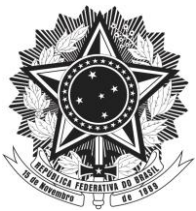

**MINISTÉRIO DA EDUCAÇÃO**
**Fundação Universidade Federal do ABC**
Avenida dos Estados, 5001 – Bairro Santa Terezinha – Santo André – SP
CEP 09210-580 · Fone: (11) 4996-0017

**Ata de Defesa de Dissertação de Mestrado**

No dia 11 de Março de 2024 às 15h00, no local: Sala 302 do Bloco B do Campus de Santo André da Universidade Federal do ABC, realizou-se a Defesa de Dissertação de Mestrado, que constou da apresentação do trabalho intitulado **"Hilbert series for flavor invariants and CP violation"** de autoria do candidato, **EDUARDO LOURENÇO FÁBIO DE LIMA**, RA nº 21202210105, discente do Programa de Pós-Graduação em FÍSICA da UFABC. Concluídos os trabalhos de apresentação e arguição, o candidato foi considerado _____**APROVADO**_____ pela Banca Examinadora.

E, para constar, foi lavrada a presente ata, que vai assinada pelos membros da Banca.

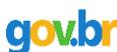

_____
**Prof.(a) ALYSSON FABIO FERRARI**
UNIVERSIDADE FEDERAL DO ABC - Membro Titular

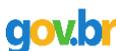

_____
**Prof.(a) ORLANDO LUIS GOULART PERES**
UNIVERSIDADE ESTADUAL DE CAMPINAS - Membro Titular

_____
**Prof.(a) CELIO ADREGA DE MOURA JUNIOR**
UNIVERSIDADE FEDERAL DO ABC - Membro Suplente

_____
**Prof.(a) PEDRO CUNHA DE HOLANDA**
UNIVERSIDADE ESTADUAL DE CAMPINAS - Membro Suplente

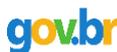

_____
**Prof.(a) CELSO CHIKAHIRO NISHI**
UNIVERSIDADE FEDERAL DO ABC - Presidente

\* Por ausência do membro titular, foi substituído pelo membro suplente descrito acima: nome completo, instituição e assinatura

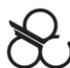
Universidade Federal do ABC

Ressalvas e sugestões da Banca examinadora:

_____

_____

_____

_____

_____

_____

_____

_____

_____

Os membros que participaram de modo remoto, foram:

Prof. Celso C. Nishi
_____

Prof. Orlando L. G. Peres
_____

_____

_____

_____

_____

\* Por ausência do membro titular, foi substituído pelo membro suplente descrito acima: nome completo, instituição e assinatura

Universidade Federal do ABC

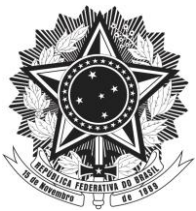

**MINISTÉRIO DA EDUCAÇÃO**
**Fundação Universidade Federal do ABC**
Avenida dos Estados, 5001 – Bairro Santa Terezinha – Santo André – SP
CEP 09210-580 · Fone: (11) 4996-0017

Por sugestão da Banca Examinadora, o novo título passa a ser (em letra de forma e legível):

Indique o **idioma** da dissertação/tese: (  ) **Português**  (✗) **Inglês**  (  ) _____

a) Novo Título em **Português** (preenchimento obrigatório):

Série de Hilbert para invariantes de sabor de quarks e violação de CP
_____

_____

_____

_____

b) Novo Título em **Inglês** (preenchimento obrigatório):

Hilbert series for quark flavor invariants and CP violation
_____

_____

_____

_____

c) Novo Título em **outro idioma**, conforme o indicado acima:

_____

_____

_____

_____

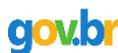

_____
**Assinatura do Presidente da Banca**

* Por ausência do membro titular, foi substituído pelo membro suplente descrito acima: nome completo, instituição e assinatura

**Universidade Federal do ABC**


  % \includepdf[pages=-]{Ata/folha2.pdf}
  % \includepdf[pages=-]{Ata/folha3.pdf}
  % \includepdf[pages=-]{Ata/folha4.pdf}

 \end{anexosenv}